\documentclass[reprint,amsmath,amssymb,aps,superscriptaddress]{revtex4-2}
\usepackage{dcolumn}
\usepackage{bm}
\usepackage{amsmath,amssymb}
\usepackage{amsthm}
\usepackage{mathrsfs}
\usepackage{indentfirst}
\usepackage{comment}
\usepackage{float}
\usepackage{braket}
\usepackage[utf8]{inputenc}
\usepackage{bm}
\usepackage{graphicx}
\usepackage{tikz}
\usepackage{physics}
\usetikzlibrary{arrows, shapes.gates.logic.US, calc}
\usetikzlibrary{angles,quotes}
\tikzstyle{branch}=[fill, shape=circle, minimum size=3pt, inner sep=0pt]
\usepackage{blochsphere}
\usepackage{mathtools}
\usepackage{color,graphicx} 
\newcommand\D{\!\operatorname{d}\!}
\usepackage{algorithm}
\usepackage[algo2e]{algorithm2e}
\usepackage[noend]{algpseudocode}
\usepackage{bbold}

\usetikzlibrary{quantikz}

\usepackage{hyperref}
\hypersetup{
 colorlinks=true,
 citecolor=red,
 linkcolor=blue,
 urlcolor=blue,
 pdfpagemode=UseNone,
 pdfstartview=FitH}
\usepackage{subfiles}

\newcommand{\rr}[1]{\mathrm{#1}}
\newcommand{\cl}[1]{\mathcal{#1}}
\newcommand{\orcidicon}[1]{\href{https://orcid.org/#1}{\includegraphics[height=\fontcharht\font`\B]{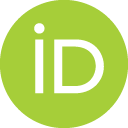}}}

\footnotetext{These authors contributed equally}

\begin{document}
\preprint{APS/123-QED}
\title{A novel approach to noisy gates for simulating quantum computers}

\author{Giovanni Di Bartolomeo\,\orcidicon{0000-0002-1792-7043}$^\diamond$}
\email{giovanni.dibartolomeo@phd.units.it}
\affiliation{
Department of Physics, University of Trieste, Strada Costiera 11, 34151 Trieste, Italy}
\affiliation{Istituto Nazionale di Fisica Nucleare, Trieste Section, Via Valerio 2, 34127 Trieste, Italy}


\author{Michele Vischi\,\orcidicon{0000-0002-5724-7421}$^\diamond$}

\email{michele.vischi@phd.units.it}
\affiliation{
Department of Physics, University of Trieste, Strada Costiera 11, 34151 Trieste, Italy}
\affiliation{Istituto Nazionale di Fisica Nucleare, Trieste Section, Via Valerio 2, 34127 Trieste, Italy}

\author{Francesco~Cesa}
\email{francesco.cesa@phd.units.it}
\affiliation{
Department of Physics, University of Trieste, Strada Costiera 11, 34151 Trieste, Italy}
\affiliation{Istituto Nazionale di Fisica Nucleare, Trieste Section, Via Valerio 2, 34127 Trieste, Italy}
\author{Roman~Wixinger\,\orcidicon{000-0001-7113-4370}}
\affiliation{
Institute of Particle Physics and Astrophysics, ETH Zürich, Zürich, Switzerland}

%
\author{Michele Grossi\,\orcidicon{0000-0003-1718-1314}}
\email{michele.grossi@cern.ch}
\affiliation{European Organization for Nuclear Research (CERN), Geneva 1211, Switzerland}
\author{Sandro Donadi\,\orcidicon{0000-0001-6290-5065}}
\affiliation{Istituto Nazionale di Fisica Nucleare, Trieste Section, Via Valerio 2, 34127 Trieste, Italy}
\author{Angelo Bassi\,\orcidicon{0000-0001-7500-387X}}
\affiliation{
Department of Physics, University of Trieste, Strada Costiera 11, 34151 Trieste, Italy}
\affiliation{Istituto Nazionale di Fisica Nucleare, Trieste Section, Via Valerio 2, 34127 Trieste, Italy}

\begin{abstract}
We present a novel method for simulating the noisy behaviour of quantum computers, which allows to efficiently incorporate  environmental effects in the driven evolution implementing the gates acting on the qubits. We show how to modify the noiseless gate executed by the computer to include any Markovian noise, hence resulting in what we will call a noisy gate. We compare our method with the IBM Qiskit simulator, and show that it follows more closely both the analytical solution of the Lindblad equation as well as the behaviour of a real quantum computer, where we ran algorithms involving up to 18 qubits; as such, our protocol offers a more accurate simulator for NISQ devices. The method is flexible enough to potentially describe any noise, including non-Markovian ones. The noise simulator based on this work is available as a python package at this \href{https://pypi.org/project/quantum-gates/}{link}.
\end{abstract}
\maketitle

\section{Introduction}\label{introduction}
Quantum computers are on the way; currently they manage between dozens and hundreds of qubits
\cite{chow2021ibm,kielpinski2002architecture,arute2019quantum,wu2021strong,zhong2020quantum}, which does not sound as an impressive number, yet it is already good enough to perform interesting tasks \cite{bharti2022noisy,cerezo2021variational}. As powerful as they promise to be, quantum computers are far from being ideal: since, as for any quantum system, they can hardly be isolated from the surrounding environment, they are prone to errors, which limit their capabilities. Like in the classical case, error correcting schemes have been developed \cite{steane2003overhead,fowler2009high,lidar2013quantum} and first tests have been performed \cite{google2023suppressing,sivak2022real}, but to be implemented they require to the least thousands qubits, which are not available; for the time being, we have to cope with errors.   

This stage of development is referred to as Noisy Intermediate-Scale Quantum (NISQ)~\cite{bharti2022noisy,preskill2018quantum} era; the major aim of the research during this near-term period is to maximize the computational power of current devices in view of the long-term goal of fault-tolerant quantum computation \cite{chow2021ibm}.

It is clear that NISQ computers require a good understanding of how noises affect  quantum circuits and, in order to do so, a proper modeling of the noises is needed. This requires essentially two major tasks: understanding the major sources of noise affecting the qubits, and writing better algorithms for simulating a given noise model on a classical computer. The present work deals with this second task.

To date, the simulations of noisy digital gate-based quantum computers is implemented by adding appropriate quantum operations  before and after each ideal gate \cite{breuer2002theory,nielsen2000quantum,benenti2019principles}: schematically, and working with the density matrix formalism, if an ideal  (unitary)  gate $G$ is supposed to be executed, the noises affecting it are modeled by adding appropriate operations $\mathcal{E}_1$ ($\mathcal{E}_2$) mimicking the noise, before (after) the gate:
\begin{equation}\label{QuantumMaps}
\begin{quantikz}
\lstick{$\rho$} & \gate{\mathcal{E}_1} &\gate{G} &\gate{\mathcal{E}_2}&\rstick{$\rho'$} \qw
\end{quantikz}.
\end{equation}
Such a modeling completely decouples the action of the controlled operation generating the gate $G$  from that of the environment. This approximation works well if $G$ acts almost instantaneously with respect to the noise, i.e. if the gate time $t_g$ required to implement the gate is much smaller than the characteristic time scales of the system-environment interaction.
For instance, in IBM's superconducting devices \cite{ibm_quantum_exp} $t_g\sim10^{-8}s$, while typical environmental effects such as relaxation and phase damping have characteristic times of order $T_1, T_2\sim10^{-4}s$. This justifies why this approach has been implemented by the majority of available noise simulators of NISQ computers (see appendix \ref{relevant_QC_frame}). 

Yet this approach has some limitations. By separating the action of the gate from that of the noise, it does not represent a faithful description of what happens inside a computer, where  the controlled action on the qubit(s) generating the gate and the environment act simultaneously and potentially affect each other. Therefore it is expected not to be fully accurate in describing a NISQ computer, especially when the number of gates and qubits is relatively large, which is actually the regime where simulations are more interesting.

In this article we propose an alternative approach, where the noise is {\it integrated} into the logical gates, in the sense that the resulting noisy gate is computed by solving for the dynamics generating it, with additional terms describing the noise added to it:
\begin{equation}\label{BigMap}
\begin{quantikz}
\lstick{$\rho$} &\gate{\mathcal{G}} &\rstick{$\rho'$} \qw
\end{quantikz},
\end{equation}
where in general $\cl{G} \neq \cl{E}_2 \circ G \circ \cl{E}_1$, and under standard assumptions (e.g., Markovianity) it gives an analytic expression for the solution of the Lindblad equation obtained with perturbative methods. Now $\cl{G}$ captures, within the limits of validity of Lindblad's equation, the entire physics occurring during the execution of each gate; not only it offers a more accurate description of the system and therefore a better protocol for circuit simulations, but also it helps to understand the different noises acting on the computer, especially in view of possible mitigation strategies. This new approach does not have any computational disadvantage with respect to~\eqref{QuantumMaps}.

As a note, Markovianity, which is the main physical assumption behind the Lindblad equation, and it is a very convenient working hypothesis, can be released in favour of more general noises \cite{breuer2002theory,adler2007collapse,bassi2003stochastic,maniscalco2006non,strunz1999quantum,gambetta2004non}; we will not touch on this possibility here, although the generalization of the approach here introduced is rather straightforward. 

Both approaches \eqref{QuantumMaps} and \eqref{BigMap} have a drawback if they are implemented at the density matrix level: the simulation will be slowed down quadratically as a function of the number of qubits. This drawback can be resolved for \eqref{QuantumMaps} by replacing the superoperations $\mathcal{E}_{1,2}$ acting on the density matrix with suitable stochastic operations acting on the state vector \cite{rigetti_noise,qiskit_notebook}; in this way, the noisy algorithm becomes random and each single run of the simulation can be seen as a single run of the algorithm on the noisy quantum computer. The same strategy can be adopted for \eqref{BigMap}; one writes 
\begin{equation}
\begin{quantikz}
\lstick{$\ket{\psi}$} &\gate{\mathcal{G}_{\bm{\xi}}} &\rstick{$|\psi'_{\bm{\xi}}\rangle$} \qw
\end{quantikz},
\end{equation}
where $\mathcal{G}_{\bm{\xi}}$ is a stochastic gate, solution of a stochastic Schr\"odinger equation, incorporating both the controlled action generating the (otherwise ideal) gate $G$ and the noise. Here $\bm{\xi}$ denotes a set of stochastic gaussian variables, and stresses the fact that $\mathcal{G}_{\bm{\xi}}$, and hence $|\psi'_{\bm{\xi}}\rangle$, are random; we will omit to indicate $\bm{\xi}$ in the rest of the paper. Physical quantities are obtained by averaging over the noise.

The general procedure therefore is the following. Given a noiseless algorithm, the corresponding noisy one is obtained by replacing each ideal gate with a noisy gate. The resulting noisy algorithm, which is stochastic, is repeated for different realizations of the random variables, as if they were different runs on a physical quantum computer. This produces a statistics of outcomes, to be compared with those of a real computer, or to be used to predict the behavior of a future NISQ device. 

As such, as already mentioned, a first application of our approach is to predict the behaviour of NISQ devices, their potentialities and limitations. But its use goes beyond the NISQ-era horizon: by offering a more accurate modeling of the noise, it allows to better understand the physics underlying the functioning of a quantum computer and to enforce appropriate error mitigation schemes \cite{van2022probabilistic,temme2017error,zhang2020error}. 

The rest of the paper details this program. We present the noisy gates method by designing it on the IBM superconducting computers \cite{ibm_quantum_exp} although the approach is general and can be used to describe any NISQ quantum platform, once the native gate set and the proper noise model are defined.

The paper is organized as follows. In Sec. \ref{noise_models} we review the main noises affecting superconducting qubits, and how they are described within the Lindblad's formalism; in Sec. \ref{General_derivation}, \ref{single_qubit} and \ref{two_qubit} we present the general derivation of the noisy gates, specializing it to the native single and two-qubits noisy gates of IBM devices. In Sec. \ref{algorithms_comp} we compare the structure of our algorithm with that of IBM Qiskit. 

In Sec. \ref{simulations} we present the results of the simulations, which test our algorithm against that of  Qiskit in reproducing the solution of the Lindlbad equation, as well as the outcomes from current IBM quantum computers: the simulations show that the proposed method is more accurate and precise compared to that of Qiskit in reproducing the Lindblad equation, with an average improvement between 50\% and 90\% and more. 

The improvement in simulating the real device fluctuates between 10\% and 30\%, because the underlying noise model is not accurate enough, and also because the devices are not really stable; for a large number of qubits it becomes even lower because  the number of runs of the device, which are necessary to recover a good statistic, is too high. In both cases, this is not a limitation of our algorithm, but of the physical model describing the computer. 

We conclude our paper with some general remarks and an outlook.

\section{Review of the noise model}\label{noise_models}
The noises which are more relevant in the functioning of superconducting devices have already been characterized in literature \cite{krantz2019quantum,nielsen2000quantum,benenti2019principles}; in this section we briefly present them. With good approximation they are described by a Lindblad dynamics \cite{gorini1976completely,lindblad1976generators}:
\begin{equation}\label{Lindblad_equation}
    \frac{\rr{d}\rho_s}{\rr{d}s}=-\frac{i}{\hbar}\big[\rr{H}_s, \rho_s\big] + \mathfrak{D}(\rho_s);
\end{equation}
here, $\rr{H}_s$ is the Hamiltonian of the system which implements the ideal gate, and $\mathfrak{D}(\rho)$ is a Lindblad term describing the effect of the environment. For convenience, we will describe the evolution with a time schedule $s\in[0,1]$, defined as $s=t/t_g$, where $t_g$ is the duration of a gate.  

Apart from state preparation and measurement (SPAM) errors, which happen at the very beginning and very end, during the execution of an algorithm there are two main sources of noise, namely, depolarization and relaxation \cite{krantz2019quantum,georgopoulos2021modeling}. The first, which can be ascribed to the imperfections of the device, tends to bring the state towards the totally mixed one, $\mathbb{1}/\sqrt{N}$, where $N=2^n$ and $n$ is the number of qubits; for the single qubit, this can be modeled by the following Lindblad term 
\cite{nielsen2000quantum,benenti2019principles},
\begin{equation}
    \mathfrak{D}_d(\rho)=\gamma_d\sum_{k=1}^3\big[\sigma^k\rho\sigma^k-\rho\big],
\end{equation}
where $\sigma^1=\rr{X}$, $\sigma^2=\rr{Y}, \sigma^3=\rr{Z}$ are the standard Pauli matrices and $\gamma_d\geq 0$ is the rate at which depolarization occurs.

The second type of noise is due to the interaction of the physical qubits with the surrounding environment; in particular, due the thermalization towards an equilibrium with the environment, energy exchanges occur. In the scenario of interest, this induces the decay of a qubit towards the ground state $\ket{0}$, an effect which is also known as amplitude damping \cite{benenti2019principles,nielsen2000quantum}. This damping is characterized by a relaxation time $T_1$, which identifies the scales at which the initial state decays towards $\ket{0}$; it causes also a damping of the off-diagonal elements of the density matrix in terms of dephasing, which (if only amplitude damping is acting) has a characteristic time $2T_1$. However, at the same time also a contribution of pure dephasing must be taken in account, resulting in an effective dephasing rate $1/T_2\geq 1/2T_1$. When also $T_1\geq T_2$ holds (and this is the case of interest to us), the combined action of these two effects, that from now on we will refer to as relaxation or amplitude and phase damping, can be described by the following Lindblad term,
\begin{equation}
   \mathfrak{D}_r(\rho)=\gamma_1\big[ \sigma^+\rho\sigma^- -\frac{1}{2}\big\{ \mathbb{P}^{(1)},\rho \big\}  \big] + \gamma_z\big[ \rr{Z}\rho\rr{Z}-\rho \big],
\end{equation}
where we use the convention $\sigma^\pm=(\rr{X}\pm i\rr{Y})/2$ and $\mathbb{P}^{(1)}=\ket{1}\bra{1}$ is the projector onto $\ket{1}$; the coefficients are related to the characteristic times as $\gamma_1=t_gT_1^{-1}$ and $\gamma_z=t_g(2T_1-T_2)/4T_1T_2$. 

We will consider both sources of noise together, meaning that the Lindblad term is $\mathfrak{D}(\rho)=\mathfrak{D}_d(\rho)+\mathfrak{D}_R(\rho)$, which can be diagonalized in the canonical Lindblad form by standard procedures. Eventually one obtains the Lindblad term
\begin{equation}\label{noisyterm}
    \mathfrak{D}(\rho)=\epsilon^2\sum_{k=1}^3\big[ \rr{L}_k\rho\rr{L}_k^\dag -\frac{1}{2}\big\{ \rr{L}_k^\dag\rr{L}_k, \rho \big\} \big],
\end{equation}
where the non normalized Lindblad operators are
\begin{equation}
    \rr{L}_1=\sqrt{\frac{\lambda_1}{\lambda}}\sigma^-,\;\;\;\rr{L}_2=\sqrt{\frac{\lambda_2}{\lambda}}\sigma^+,\;\;\;\rr{L}_3=\sqrt{\frac{\lambda_3}{\lambda}}\rr{Z};
\end{equation}
here, we set $\lambda_1=2\gamma_d$, $\lambda_2=2\gamma_d+\gamma_1$, $\lambda_3=\gamma_d+\gamma_z$ and $\lambda=\lambda_1+\lambda_2+\lambda_3$, and we defined the parameter $\epsilon=\sqrt{\lambda}$. As mentioned in the Introduction, in the case of IBM's superconducting devices the typical order of magnitude of the decoherence times is $\sim 10^{-4}$ s; by contrast, the typical order of magnitude of the time to execute a gate is $t_g\sim 10^{-8}$ s, which is small compared to $T_{1,2}$; in particular, one has $\gamma_d, \gamma_1,\gamma_z \ll 1$, which leads to $\epsilon=\sqrt{\lambda} \ll 1$. This justifies the perturbative expansion we will implement later.

While terms of the form \eqref{noisyterm} describe the dissipation occurring at the single qubit level, one straightforward generalization to the multi-qubit case (the one we will consider in this work) is obtained via the direct sum 
\begin{equation}\label{direct_sum}
    \mathfrak{D}(\rho)=\bigoplus_{k=1}^n \mathfrak{D}^{(k)}(\rho),
\end{equation}
where the upper index $(k)$ indicates that the Lindblad  term \eqref{noisyterm} acts on the $k-$th qubit. Such a generalization is based on the assumption that single qubit noises are dominating, therefore neglecting cross talks and correlated noises~\cite{ash2020analysis}; they can straightforwardly be implemented in our noisy framework,  and they will be the subject of future research. We stress that through Eq.~\eqref{direct_sum} we already account for the fact that (for instance, on IBM's devices) multiple-qubit operations are more faulty than single qubit manipulations: when entangling gates are performed, single qubit noises act together, and errors therefore amplify.

Before proceeding, one further comment is in order. Casting the behaviour of a real quantum device in a theoretical model is a hard task, and the more accurate the model, the less general it is. As remarked in the Introduction, the purpose of this work is not that of finding the best  noise model for a given quantum computer; rather, given a noise model, we are interested in the best way to simulate the device. The noise model we are considering here is therefore ultimately motivated by the fact that it is accurate enough to already give appreciable results in the simulations, but on the other hand it is also simple enough to efficiently enlighten our main points, and general enough to be readily extended to different platforms. It is understood that better results can be achieved only by specializing more  the analysis on physical device to be considered.

\section{General derivation of noisy gates}\label{General_derivation}
Let us consider the situation in which the computer executes a gate $\rr{U}_g$ on a set of $n$ qubits. This is achieved by driving the system with an Hamiltonian $\rr{H}_s$ for $s\in[0,1]$, which will induce some unitary evolution $\rr{U}_s$, defined by $i\hbar\rr{d}\rr{U}_s/\rr{d}s=\rr{H}_s\rr{U}_s$, and such that $\rr{U}_{s=1}=\rr{U}_g$. However, if noises and imperfections are taken in account, this coherent evolution is replaced by a partially non coherent one, which under the assumptions of Markovianity (and complete positivity) is described by a master equation of the form \eqref{Lindblad_equation} discussed in the previous section, with the Lindblad term given, in our case, by \eqref{direct_sum} and \eqref{noisyterm}, which needs to be solved in place of the Schr\"odinger equation. We recall here that  the coefficient $\epsilon$ is small, $\epsilon \ll 1$. 

In order to switch from the density matrix formalism to the state vector formalism, we perform a {\it linear stochastic unraveling} of the Lindbald equation~\cite{jacobs1998linear,caiaffa2017stochastic,bassi2003stochastic,jacobs2014quantum,wiseman2009quantum}; specifically we consider the following It\^o stochastic differential equation for the state vector: \cite{gardiner1985handbook}
\begin{equation}\label{ITO_eq}
    \rr{d}\ket{\psi_s}\!=\!\bigg[\!\!-\frac{i}{\hbar}\rr{H}_s\rr{d}s+\!\!\! \sum_{k=1}^{N^2-1}\!\!\Big[i\epsilon\rr{d}\rr{W}_{k,s}\rr{L}_{k}\! -\!\frac{\epsilon^2}{2}\rr{d}s \rr{L}^\dag_{k} \rr{L}_{k}\Big]\bigg]\!\ket{\psi_s},
\end{equation}
where $\rr{d}\rr{W}_{k,s}$ are differentials of standard independent Wiener processes, i.e. stochastic infinitesimal increments such that $\mathbb{E}\big[\rr{d}\rr{W}_{k,s}\big]=0$ and $\mathbb{E}\big[ \rr{d}\rr{W}_{k,s}\rr{d}\rr{W}_{k',s'}\big]=\delta_{k,k'}\rr{d}s $. Eq.~\eqref{ITO_eq} is an unraveling of the Lindblad equation in the sense that the density matrix obtained by averaging the pure states $\ket{\psi_s}\bra{\psi_s}$ over the noise:
\begin{equation}\label{averaged_solution}
    \rho_s=\mathbb{E}\Big[  \ket{\psi_s}\bra{\psi_s}  \Big], 
\end{equation}
is a solution of Eq.~\eqref{Lindblad_equation}. In this sense, Eqs.~\eqref{ITO_eq} and ~\eqref{Lindblad_equation} have the same physical content; the advantage of the stochastic unraveling is that it allows to work with Schr\"odinger-like equations for the state vector.


One key property of Eq. \eqref{ITO_eq} is that it is linear, and therefore it allows to write the solution as $\ket{\psi_{s=1}}=\rr{N}_g\ket{\psi_0}$, where $\rr{N}_g$ can be interpreted as a noisy random gate acting on the system. Since Eq. \eqref{ITO_eq} in general does not preserve the norm of the state vector, the associated gate $\rr{N}_g$ is not unitary; this is a consequence of the chosen unraveling: one could have chosen norm-preserving unravelings \cite{bassi2003dynamical,bassi2013models}, which however are not linear and therefore do not allow for a gate-like formulation. The lack of norm preservation is not  a problem since at the statistical level, i.e. when the average over the noise is taken as in \eqref{averaged_solution}, one recovers the Lindblad equation, which is trace preserving. 

In general, Eq. \eqref{ITO_eq} cannot be solved in a closed form \cite{gardiner1985handbook,arnold1974stochastic} except for few specific cases, for example when all operators commute. In Appendix \ref{derivations_approximate} we show how an approximate solution to order $\cl{O}(\epsilon^2)$ can be derived, which results in the following expression for the noisy version of a noiseless gate $\rr{U}_g$:
\begin{equation}\label{noisy_gate}
    \rr{N}_g=\rr{U}_g e^{\Lambda}e^{\Xi},
\end{equation}
where we defined the deterministic operator:
\begin{equation}\label{Lambda}
\Lambda:=-\frac{\epsilon^2}{2}\int_0^1\rr{d}s\sum_{k=1}^{N^2-1}\big[ \rr{L}^\dag_{k,s}\rr{L}_{k,s} -\rr{L}^2_{k,s} \big] 
\end{equation}
and the stochastic one:
\begin{equation}\label{Xi}
\Xi:=i\epsilon\sum_{k=1}^{N^2-1}\int_0^1\rr{d\rr{W}_{k,s}}\rr{L}_{k,s}.
\end{equation}
Note that in Eqs. \eqref{Lambda} and \eqref{Xi}, $\rr{L}_{k,s}=\rr{U}^\dag_s\rr{L}_k\rr{U}_s$ are the Lindblad operators in the interaction picture, therefore the noiseless part of the dynamics $\rr{U}_s$ and the noisy one given by the Lindblad operators $\rr{L}_k$ do not factorize, as it might look from a naive understanding of Eq.~\eqref{noisy_gate}. 

As explained in Appendix \ref{derivations_approximate}, we  omitted the additional term  $-(\epsilon^2/2) \sum_{k,l=1}^{N^2-1}\int_0^1\rr{d}\rr{W}_{k,s}\int_0^s\rr{d}\rr{W}_{l,s'}\big[ \rr{L}_{k,s},\rr{L}_{l,s'} \big]$ in Eq. \eqref{Xi}, which in principle should contribute to order $\epsilon^2$; this is legitimate because it is a nested It\^o integral of non anticipating functions \cite{gardiner1985handbook}, and hence its stochastic average is 0. For this reason, it drops from all final averaged quantities, and therefore we can neglect it from the start. 

Let us also point out that, in the cases of interest to us, the term \eqref{Lambda} can always be exponentiated, so that we will always be able to directly calculate $e^{\Lambda}$.

The only stochastic term entering the noisy gate $\rr{N}_g$ is $\Xi$ in Eq. \eqref{Xi}, which is a function of several random variables $\xi$ arising from the stochastic processes $\rr{W}_{k,s}$. Let us call $\rr{L}_{kij,s}=\rr{L}_{kij,s}^+ +i\rr{L}_{kij,s}^-$   the $ij$-th matrix element  of the jump operator $\rr{L}_{k,s}$ in the computational basis, divided in real  $(+)$ and imaginary $(-)$ part, respectively. Then, each entry of the stochastic matrix is of the form $\Xi_{ij}=i\epsilon\sum_{k=1}^{N^2-1}\big[\xi_{kij}^{+}+i\xi_{kij}^{-}\big]$, where we defined the random variables
\begin{equation}
    \xi_{kij}^+=\int_0^1\rr{d}\rr{W}_{k,s}\rr{L}_{kij,s}^{+}, \;\;\;    \xi_{kij}^-=\int_0^1\rr{d}\rr{W}_{k,s}\rr{L}_{kij,s}^{-},
\end{equation}
which, being It\^o integrals of deterministic functions, are all normally distributed with zero mean, $\mathbb{E}\big[\xi_{kij}^\pm\big]=0$, and variances $\mathbb{E}[(\xi_{kij}^\pm)^2]=\int_0^1\rr{d}s [\rr{L}_{kij}^\pm]^2$. Moreover, one can easily check that they are correlated with each other as
\begin{equation}
    \mathbb{E}\big[\xi_{kij}^\pm \xi_{k'i'j'}^\pm\big]=\delta_{k,k'}\int_0^1\rr{d}s \rr{L}_{kij,s}^\pm \rr{L}_{ki'j',s}^\pm. 
\end{equation}

The random variables giving $\Xi$ its stochastic character  may be defined in several other ways, and the best choice depends on the specific case of interest. In this section we presented one general strategy for defining them, but in practice this lead to an over estimation of the actual number of random variables needed. By straightforwardly counting, one has at most $2N^2(N^2-1)$ real gaussian random variables for a noisy gate acting on $n = \log_2 N$ qubits, each random variable being correlated with at most other $2N^2-1$ ones. In practice, however, we immediately point out that one shall expect neither the number of random variables, nor the number of correlations between them to really follow this scaling. This is mainly due to the fact that real quantum computers usually perform single and two qubit native gates, and single qubit noises are dominating. For instance, given \eqref{direct_sum}, one can  upper bound the number of random variables by $\sim 6 N^2\log_2 N$. In the following sections, as we go through the construction of the native set of noisy gates for IBM's quantum computers, we shall make this claim more clear. Note that our derivation works for any choice of the starting Lindblad master equation, meaning that any Markovian noise model can be treated. In particular, while in this paper we specialize on the noise model described in the previous section, one can add device-motivated modifications (such as correlated noises and leakages to upper levels in the case of IBM's platform); modifications of this kind are left to future research---as also the generalization of our derivation to non Markovian situations.

More details on the difference between our perturbative approximation and the one used in the standard approach \eqref{QuantumMaps} can be found in appendix \ref{comp}.

\section{Single qubit noisy gates}\label{single_qubit}
IBM's superconducting devices implement single qubits operations with unitaries of the form $\rr{U}(\theta,\phi)=e^{-i\theta\rr{R}_{xy}(\phi)/2}$, where we set $\rr{R}_{xy}(\phi)=\cos(\phi)\rr{X}+\sin(\phi)\rr{Y}$; such gates are achieved by driving the system with the Hamiltonian \cite{krantz2019quantum,mckay2017efficient}
\begin{equation}\label{single_qubit_Hamiltonian}
    \rr{H}(\theta,\phi)=\frac{\theta\hbar}{2}\rr{R}_{xy}(\phi)
\end{equation}
applied for a time $s = 1$~\footnote{The native single qubit gates chosen by IBM are $\rr{X}$ and $\rr{SX}$, which are rotations around the $\rr{x}-$axis obtained by fixing $\phi = 0$ in Eq.~\eqref{single_qubit_Hamiltonian}. 
Rotations around the $\rr{z}-$axis are implemented as \textit{virtual} gates, since they are mimicked by the software and are not associated to a physical action on the device~\cite{mckay2017efficient}.}. 
The Hamiltonian is driven by time-dependent pulses \cite{krantz2019quantum}, so that in Eq. \eqref{single_qubit_Hamiltonian} one should actually consider $\theta\rightarrow \omega_s$, and set $\int_0^1\rr{d}s\omega_s=\theta$. In this work we consider constant pulses for simplicity, being the generalization to general functions rather straightforward. It should be noted  that the functional form of $\omega_s$ affects the action of the noises on the system, meaning that different pulse shapes might lead to smaller noise effects, i.e. error mitigation; this is a question left for future research. 

The task now is to derive the noisy gates $\rr{N}(\theta,\phi)$ corresponding to the unitaries above, when depolarization and relaxation errors are both taken in account during the evolution. 

We begin by computing the evolution of the jump operators in the interaction picture, obtaining the expressions:
\begin{equation}
    \sigma^\pm_s(\theta,\phi)=\frac{e^{\pm i\phi}}{2}\Big[ \rr{R}_{xy}(\phi) \pm i\rr{R}(2\Bar{s\theta},\Bar{\phi}) \Big],
\end{equation}
and
\begin{equation}
    \rr{Z}_s=\rr{R}(2s\theta, \Bar{\phi}),
\end{equation}
where we defined $\rr{R}(\theta,\phi)=\cos(\theta/2)\rr{Z}+\sin(\theta/2)\rr{R}_{xy}(\phi)$ and for a generic angle $\alpha$ we set $\Bar{\alpha}=\alpha+\pi/2$. Then, based on Eq. \eqref{noisy_gate}, we compute the deterministic, non unitary term $\Lambda(\theta,\phi)$. Since in the interaction picture the evolution is unitary, one sees that the term corresponding to $k=3$ is always vanishing, and one has $\Lambda(\theta,\phi)=-\frac{1}{2}\int_0^1\rr{d}s[\epsilon_1^2\sigma^+_s\sigma^-_s+\epsilon_2^2\sigma^-_s\sigma^+_s]$, where we set $\epsilon_k^2\equiv \epsilon^2\lambda_k/\lambda$. Hence, we first calculate $\sigma^\pm_s\sigma^\mp_s=\rr{U}_s^\dag \sigma^\pm\sigma^\mp\rr{U}_s$, and after integration we get
\begin{equation}
    \int_0^1\rr{d}s \sigma^\pm_s\sigma^\mp_s = \frac{1}{2}\Big[ \mathbb{1}\pm\frac{\sin(\theta/2)}{\theta/2} \rr{R}(\theta,\Bar{\phi}) \Big],
\end{equation}
where $\rr{R}(\theta,\phi)=\cos(\theta/2)\rr{Z}+\sin(\theta/2)\rr{R}_{xy}(\phi)$ and $\Bar{\phi}=\phi+\pi/2$, so that one has
\begin{equation}
    \Lambda(\theta,\phi)=-\frac{\epsilon_1^2+\epsilon_2^2}{4}\mathbb{1}-\frac{\epsilon_1^2-\epsilon_2^2}{4}\frac{\sin(\theta/2)}{\theta/2}\rr{R}\big(\theta, \Bar{\phi}\big);
\end{equation}
such an expression can be readily exponentiated, leading to
\begin{equation}\label{deterministic_term_single}
    e^{\Lambda(\theta,\phi)}=e^{-\frac{\epsilon_1^2+\epsilon_2^2}{4}}\bigg[\cosh F(\theta)-\rr{R}\big(\theta,\Bar{\phi}\big)\sinh F(\theta)\bigg],
\end{equation}
where we defined $F(\theta)=\frac{\epsilon_1^2-\epsilon_2^2}{4}\frac{\sin(\theta/2)}{\theta/2}$. 

Next, we turn to investigating the stochastic term, $\Xi(\theta,\phi)$. Here, it is convenient to define the following real stochastic variables:
\begin{equation}
    \xi_{k,+}\!=\!\!\int_0^1\!\rr{d}\rr{W}_{k,s}\!\cos(s\theta), \;\;\;\xi_{k,-}\!=\!\!\int_0^1\!\rr{d}\rr{W}_{k,s}\!\sin(s\theta),
\end{equation}
whose variances are:
\begin{equation}
    \mathbb{E}\big[ \xi^2_{k,\pm}\big]=\frac{1}{2}\big[ 1\pm \frac{\sin(2\theta)}{2\theta} \big],
\end{equation}
while the correlations are:
\begin{equation}
    \mathbb{E}\big[\xi_{k,+}\xi_{j,-}\big]=\frac{1-\cos(2\theta)}{4\theta}\delta_{kj};
\end{equation}
moreover, we define 
\begin{equation}
    \xi_{k,w}=\int_0^1\rr{d}\rr{W}_{k,s},
\end{equation}
such that $\mathbb{E}\big[\xi_{k,w}^2\big]=1$, $\mathbb{E}\big[\xi_{k,+}\xi_{k,w}\big]=\sin(\theta)/\theta$ and $\mathbb{E}\big[\xi_{k,-}\xi_{k,w}\big]=\big[1-\cos(\theta)\big]/\theta$. \\
Summing all terms and re-arranging them conveniently, we arrive at the following expression
\begin{equation}\label{noisy_term_single}
    \Xi(\theta,\phi)=if_0\rr{Z}+if_1\rr{R}_{xy}(\phi)+if_2\rr{R}_{xy}(\Bar{\phi}),
\end{equation}
where we defined the following set of complex stochastic coefficients:
\begin{align}
    f_0= & \epsilon_3\xi_{3,+}-i\frac{e^{i\phi}\epsilon_2\xi_{2,-}-e^{-i\phi}\epsilon_1\xi_{1,-}}{2},  \\
    f_1 = &  \frac{e^{i\phi}\epsilon_2\xi_{2,w}+ e^{-i\phi}\epsilon_1\xi_{1,w}}{2}, \\
    f_2 = & \epsilon_3\xi_{3,-} +i\frac{e^{i\phi}\epsilon_2\xi_{2,+}- e^{-i\phi}\epsilon_1\xi_{1,+}}{2}.
\end{align}
Since these quantities are all combinations of gaussian random variables with the correlations previously discussed, they can be efficiently sampled with known algorithms; then, the stochastic matrix \eqref{noisy_term_single} can be assembled and numerically exponentiated. Multiplication by the deterministic term \eqref{deterministic_term_single} and then by the noiseless gate $\rr{U}(\theta,\phi)$ eventually lead to the noisy gate $\rr{N}(\theta,\phi)$ for the single qubit, which, as shown only depends on 8 correlated gaussian variables.


\section{Two-qubit noisy gates}\label{two_qubit}
On IBM's quantum chips, two qubit gates are implemented by  a driven cross resonance \cite{krantz2019quantum,mckay2017efficient,rigetti2010fully}; labeling with an upper index the qubit each operator acts on, this consists in the execution of the unitary $\rr{U}^{(1,2)}(\theta,\phi)=e^{-i\theta\rr{Z}^{(1)}\otimes\rr{R}_{xy}^{(2)}(\phi)/2}$, which can be realised by driving the composite system with the Hamiltonian
\begin{equation}\label{cross_resonance}
    \rr{H}^{(1,2)}(\theta,\phi)=\frac{\hbar\theta}{2}\rr{Z}^{(1)}\otimes\rr{R}_{xy}^{(2)}
\end{equation}
for a duration $s = 1$, where, from now on, the tensor product symbol will be dropped, unless otherwise specified. In the proposed approach we take in consideration only noises acting on single qubits, so that the Lindblad term reads
\begin{equation}
    \mathfrak{D}^{(1,2)}(\rho)=\epsilon^{2}\sum_{i\in\{1,2\}}\sum_{k=1}^3 \big[ \rr{L}_k^{(i)}\rho\rr{L}_k^{(i)\dag} -\frac{1}{2}\big\{ \rr{L}_k^{(i)\dag}\rr{L}_k^{(i)}, \rho \big\} \big],
\end{equation}
where now $\rho$ is the two-qubit statistical operator. The procedure for calculating the noisy gates is the same as in the single qubit case.

First, we compute the Lindblad operators on the first qubit ($i=1$) in the  interaction picture:
\begin{equation}
    \sigma_s^{\pm(1)}=e^{\pm is\theta\rr{R}_{xy}^{(2)}(\phi)}\sigma^{\pm(1)},
\end{equation}
while $\rr{Z}_s^{(1)}=\rr{Z}^{(1)}$ remains constant as it commutes with the Hamiltonian. For the second qubit ($i=2$), one has
\begin{equation}
    \sigma_s^{\pm(2)}=\frac{e^{\pm i\phi}}{2}\big[\rr{R}_{xy}^{(2)}(\phi)\pm i \rr{Z}^{(1)}\cl{R}(2s\Bar{\theta}, \Bar{\phi})\big], 
\end{equation}
and $\rr{Z}_s^{(2)}=\cl{R}(2s\theta, \Bar{\phi})$,
where we defined for convenience
\begin{equation}
    \cl{R}(\theta, \phi)=\cos(\theta/2)\rr{Z}^{(2)}+\sin(\theta/2)\rr{Z}^{(1)}\rr{R}_{xy}^{(2)}(\phi).
\end{equation}
The deterministic term $\Lambda(\theta, \phi)$, see Eq.\eqref{Lambda}, can be calculated straightforwardly, leading to
\begin{equation}
    \Lambda(\theta, \phi)= -\frac{\epsilon_1^2+\epsilon_2^2}{2}\mathbb{1}-\frac{\epsilon_1^2-\epsilon_2^2}{4}\Big[\rr{Z}^{(1)}+\frac{\sin(\theta/2)}{\theta/2}\cl{R}(\theta, \Bar{\phi})\Big];
\end{equation}
notice that again this term can be exponentiated analytically as all the terms involved commute; in particular, one has 
\begin{align}
    e^{\Lambda(\theta, \phi)}\!&=e^{\!-\!\frac{\epsilon_1^2+\epsilon_2^2}{2}}\Big[\!\cosh\big(\frac{\epsilon_1^2-\epsilon_2^2}{4}\big)\mathbb{1}\!-\!\rr{Z}^{(1)}\!\sinh\!\big(\frac{\epsilon_1^2-\epsilon_2^2}{4}\big)\!\Big] \times \nonumber \\
    & \times \Big[\cosh F(\theta)-\cl{R}(\theta, \Bar{\phi})\sinh F(\theta)\Big],
\end{align}
where $F(\theta)$ is the same function defined in the single qubit case.

In order to efficiently write the stochastic term $\Xi(\theta, \phi)$, it is convenient to define, in analogy with the single qubit case, the gaussian random variables
\begin{equation}
    \xi^{(i)}_{k,+}\!=\!\!\int_0^1\!\rr{d}\rr{W}^{(i)}_{k,s}\!\cos(s\theta),\;\;\;\xi^{(i)}_{k,-}\!=\!\!\int_0^1\!\rr{d}\rr{W}^{(i)}_{k,s}\!\sin(s\theta),
\end{equation}
and
\begin{equation}
    \xi_{k,w}^{(i)}=\int_0^1\rr{d}\rr{W}^{(i)}_{k,s},
\end{equation}
whose correlations are straightforward to calculate and mimic those already seen in Sec. \ref{single_qubit}. Then, we can separate $\Xi(\theta,\phi)$ in two parts as $\Xi^{(1)}(\theta,\phi)+\Xi^{(2)}(\theta,\phi)$; the first is equal to 
\begin{align}
    \Xi^{(1)}(\theta,\phi) & =\epsilon_3 \xi_{3,w}^{(1)}\rr{Z}^{(1)} + \epsilon_1\big[ \xi^{(1)}_{1,+}+i\xi^{(1)}_{1,-}\rr{R}_{xy}^{(2)}(\phi) \big]\sigma^{-(1)}+ \nonumber \\
    & + \epsilon_2\big[ \xi^{(1)}_{2,+}+i\xi^{(1)}_{2,-}\rr{R}_{xy}^{(2)}(\phi) \big]\sigma^{+(1)},
\end{align}
while the second part reads
\begin{align}
    \Xi^{(2)}(\theta,\phi) & = if_w\rr{R}^{(2)}_{xy}(\phi)-f_-\rr{Z}^{(1)}\rr{Z}^{(2)}+ f_+\rr{R}_{xy}(\Bar{\phi})+\nonumber \\
    & + i\epsilon_3\xi^{(2)}_{3,+}\rr{Z}^{(2)}+i\epsilon^{(2)}_{3,+}\rr{Z}^{(1)}\rr{R}_{xy}(\Bar{\phi}),
\end{align}
where we defined
\begin{equation}
    f_w=\frac{1}{2}\big[\epsilon_1e^{-i\phi}\xi^{(2)}_{1,w}+\epsilon_2e^{i\phi}\xi^{(2)}_{2,w}\big]
\end{equation}
and
\begin{equation}
    f_\pm=\frac{1}{2}\big[\epsilon_1e^{-i\phi}\xi^{(2)}_{1,\pm}-\epsilon_2e^{i\phi}\xi^{(2)}_{2,\pm}\big].
\end{equation}
Again, as in the single qubit case, the stochastic matrix $\Xi(\theta,\phi)$ can be assembled by combining gaussian random variables, and hence it can be efficiently sampled and numerically exponentiated; this, combined with the term $\rr{U}(\theta,\phi)e^{\Lambda(\theta,\phi)}$, gives the noisy gate for two qubits. 

\section{Comparison of the algorithms}\label{algorithms_comp}
It is instructive to compare the structure of our approach to noise simulation with that of noise simulators based on the standard approach in Eq. \eqref{QuantumMaps}. As shown in appendix \ref{relevant_QC_frame} all relevant quantum computing frameworks implement such standard approach, and we chose IBM's Qiskit as term of comparison since it is the most developed one; in the next section we will compare also their performances in simulating the Lindblad equation as well as a real quantum computer.

Both methods rely on the state vector formulation, with important differences though. 
According to the Qiskit documentation \cite{qiskit,qiskit_notebook} the noises  are implemented by  Kraus maps, which in the density matrix formalism read:
\begin{equation}
    \mathcal{E}(\rho) = \sum_i \rr{K}_i \rho \rr{K}_i^{\dagger},
\end{equation}
where $\sum_i \rr{K}_i^{\dagger}\rr{K}_i = \mathbb{1}$. The map can be unraveled as a stochastic map on the state vector  by imposing that, at a given time, $\ket{\psi}$ changes randomly as follows:
\begin{equation}
\ket{\psi'} = \frac{1}{\sqrt{p_j}}\rr{K}_j \ket{\psi},
\end{equation}
with probability:
\begin{equation}\label{prob_kraus}
p_j = |\bra{\psi}\rr{K}_j^{\dagger}\rr{K}_j\ket{\psi}|^2.
\end{equation}
The associate pseudo code is reported in Alg. \ref{alqiskit}.
\begin{algorithm}[H]
\caption{\textsc{Qiskit Simulation}}
\label{alqiskit}
\KwIn{Initial state $\ket{\psi_{0}}$, a noiseless circuit $\rr{C} =\{\rr{U}^{(1)},...,\rr{U}^{(n_{g})}\}$ composed by $n_{g}$ gates $\rr{U}^{(i)}$ and number of samples $N_{s}$} 
\For{$0 \leq$ k $\leq N_{s}$} 
{
\While {$1 \leq$ i $\leq n_{g}$}{
compute $\ket{\psi_{k}}^{(i)} = \rr{U}^{(i)} \ket{\psi_k}^{(i-1)}$\;
\BlankLine
compute $p_j = |\bra{\psi_{k}}^{(i)}\rr{K}_j^{\dagger}\rr{K}_j\ket{\psi_{k}}^{(i)}|^2$\;
\BlankLine
sample $K_{j}$ operator from $\{p_{j}\}$\;
\BlankLine
update the state to $\ket{\psi_{k}}^{(i)} = \frac{1}{\sqrt{p_j}}\rr{K}_j\ket{\psi_{k}}^{(i)}$\;}

compute $\rho_{k} = \ket{\psi_{k}}^{(n_{g})}\bra{\psi_{k}}^{(n_{g})}$}\;
\KwOut{$\rho_{f}=\frac{1}{N_{s}}\sum_{k=1}^{N_{s}}\rho_{k}$}
\end{algorithm}

The time complexity of Alg.\ref{alqiskit} is primarily determined by the matrix vector multiplication step, exhibiting a complexity of $\mathcal{O}(2^{2n})$, where $n$ is the number of qubits. The space complexity is dominated by the storage of the state vector and it scales as $\mathcal{O}(2^{n})$.
It has to be noted that when the Kraus operators are not unitary, as for relaxation, one needs to store the intermediate state vectors, which are necessary in order to compute the probabilites in Eq.(\ref{prob_kraus}). This operation has the same time and space complexity as those of the the previous step.
(This can be avoided for mixed unitary error channels: probabilities are known and independent of the current state.)

Our noisy gates simulation instead is based on the  algorithm summarized in Alg. \ref{alNG}.

\begin{algorithm}[H]
\caption{\textsc{Noisy Gates Simulation}}
\label{alNG}
\KwIn{Initial state $\ket{\psi_{0}}$, a noiseless circuit $\rr{C} =\{\rr{U}^{(1)},...,\rr{U}^{(n_{g})}\}$ composed by $n_{g}$ gates $\rr{U}^{(i)}$ and number of samples $N_{s}$} 
\For{$0 \leq$ k $\leq N_{s}$} 
{
map a noisy circuit $\rr{\tilde{C}} = \{\rr{N}^{(1)},...,\rr{N}^{(n_{g})}\}$ on $\rr{C}$\;
\BlankLine
sample stochastic processes $\xi$ inside noisy gates $\rr{N}^{(i)}$\;
\BlankLine
compute $\ket{\psi_{k}} = \rr{N}^{(n_{g})}\dots \rr{N}^{(1)}\ket{\psi_0}$\;
\BlankLine
compute $\rho_{k} = \ket{\psi_{k}}\bra{\psi_{k}}$\;
}
\KwOut{$\rho_f=\frac{1}{N_{s}}\sum_{k=1}^{N_{s}}\rho_{k}$}
\end{algorithm}
The time complexity of Alg.~\ref{alNG} is again $\mathcal{O}(2^{2n})$, determined by the matrix vector multiplication step. Analogously, the space complexity is $\mathcal{O}(2^{n})$. We notice that in Alg.~\ref{alNG} there is no need to perform the scalar product in Eq. \eqref{prob_kraus}. Moreover, all optimization to reduce the time complexity that are possible for the first step of Alg.~\ref{alqiskit} are also possible for Alg.~\ref{alNG}. Finally, both algorithms perform samples of random numbers, but this operation has a constant scaling.

\section{Simulations}\label{simulations}
\begin{figure*}[htp]
\centering
\begin{minipage}{0.328\textwidth}
\includegraphics[width=\textwidth]{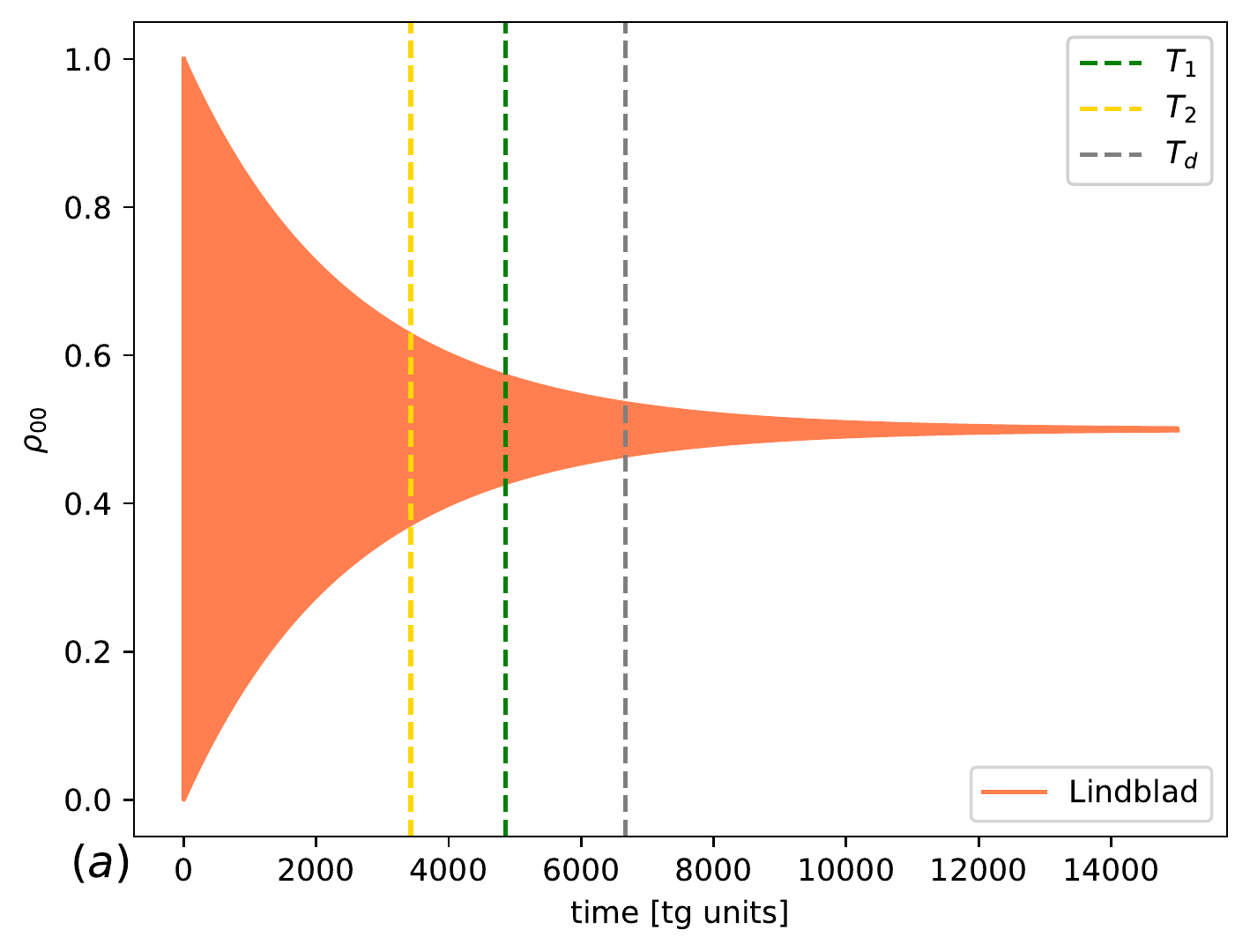}
\end{minipage}
\begin{minipage}{0.328\textwidth}
\includegraphics[width=\textwidth]{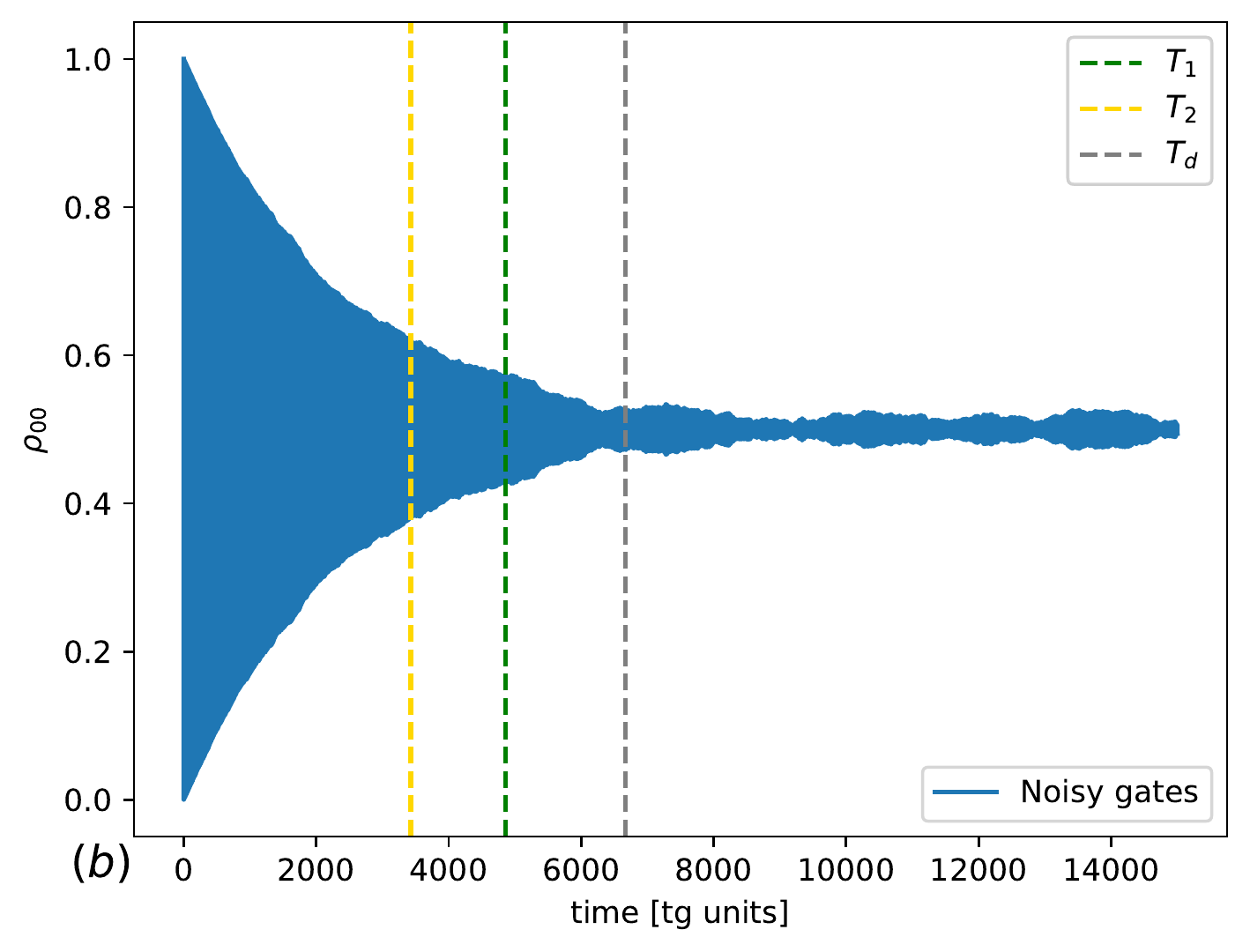}
\end{minipage}
\begin{minipage}{0.328\textwidth}
\includegraphics[width=\textwidth]{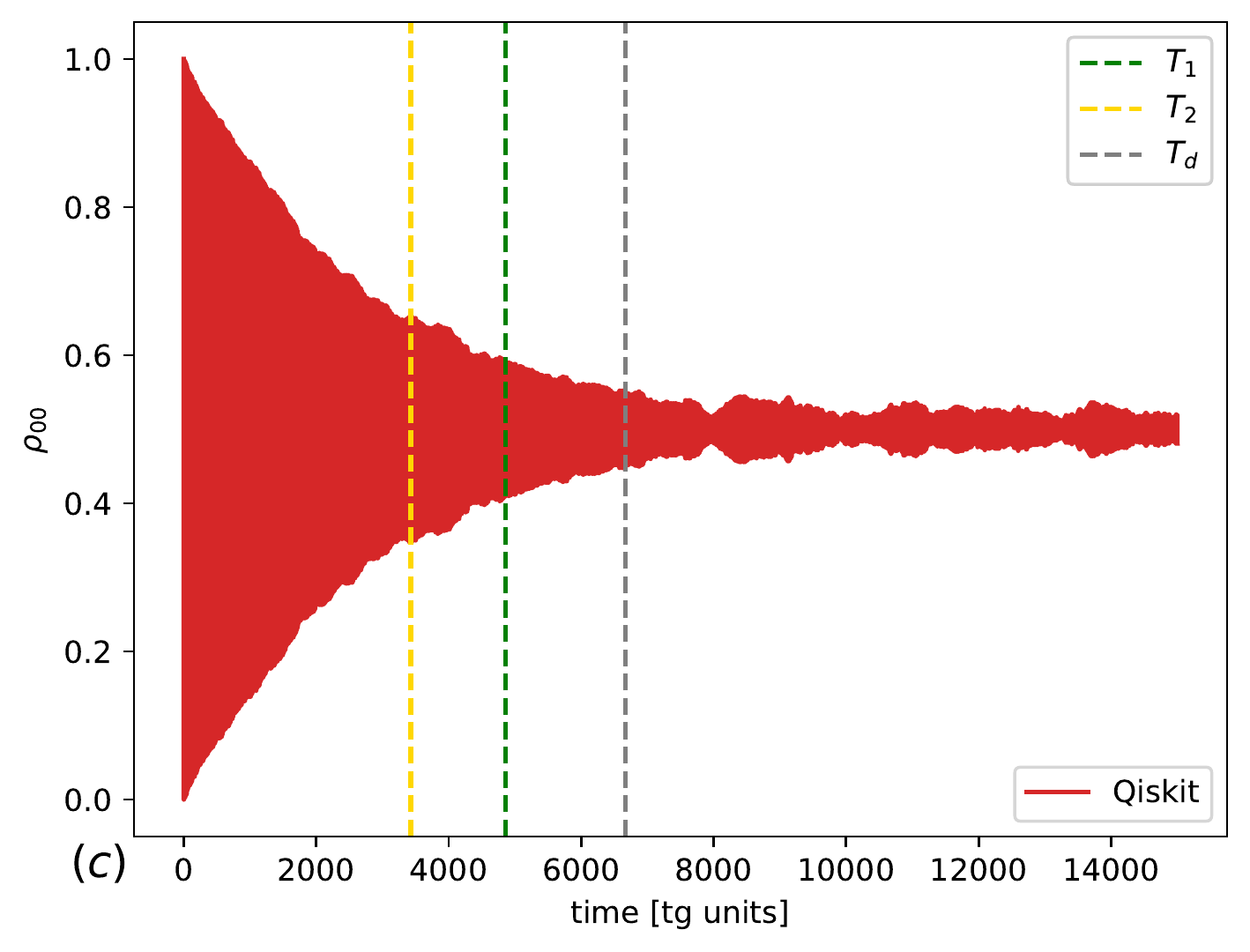}
\end{minipage}
\begin{minipage}{0.328\textwidth}
\includegraphics[width=\textwidth]{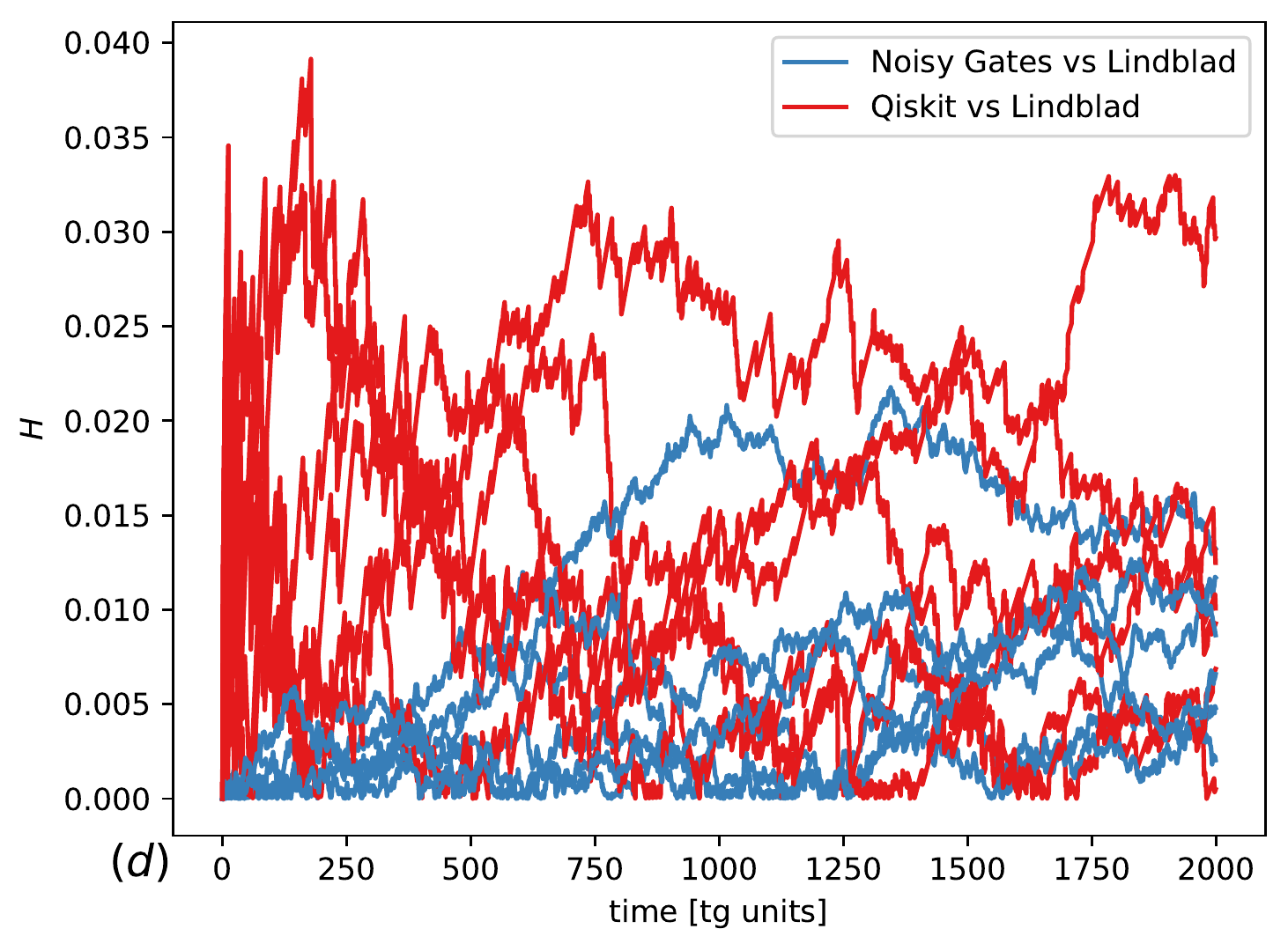}
\end{minipage}
\begin{minipage}{0.328\textwidth}
\includegraphics[width=\textwidth]{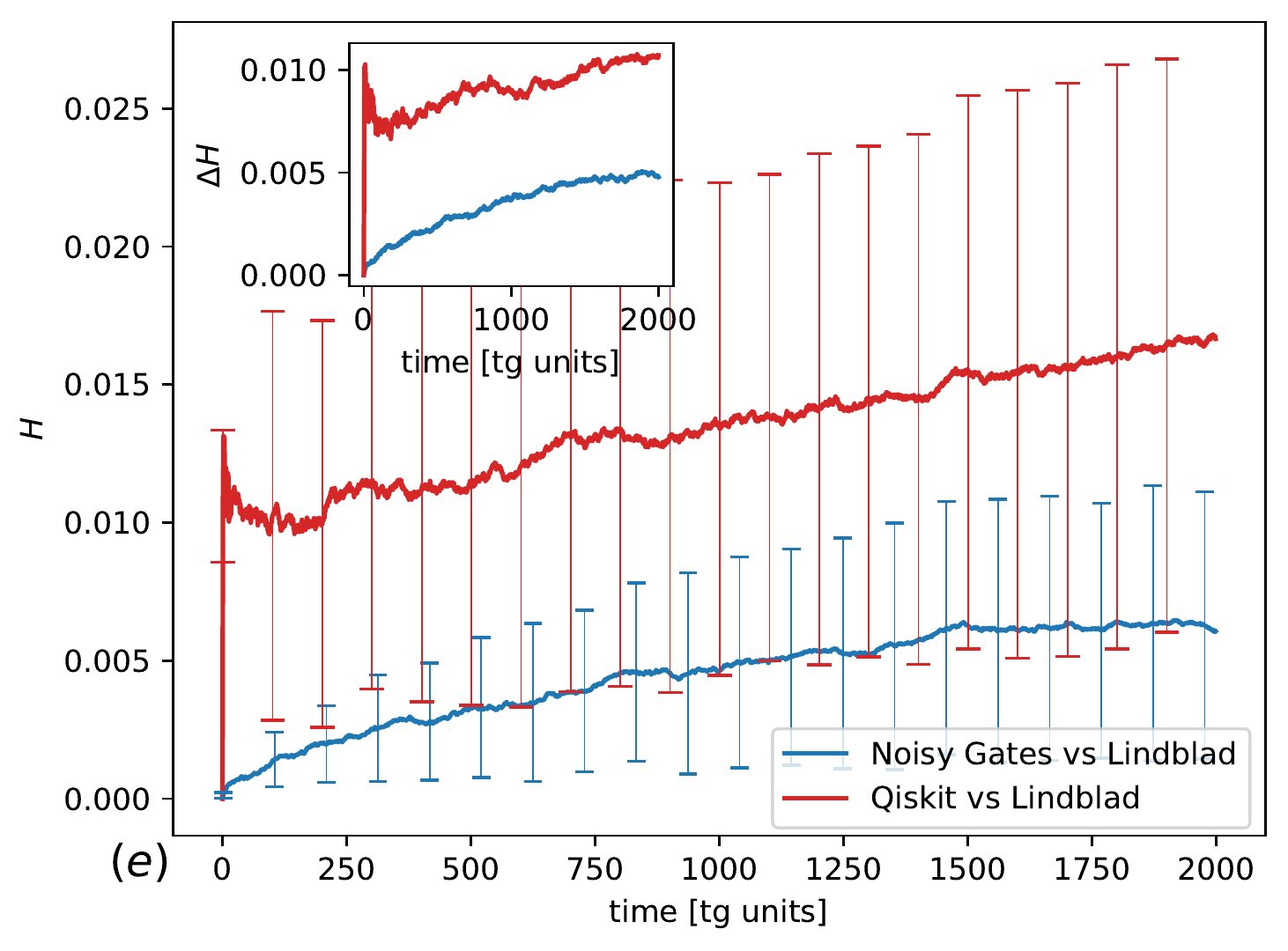}
\end{minipage}
\begin{minipage}{0.328\textwidth}
\includegraphics[width=\textwidth]{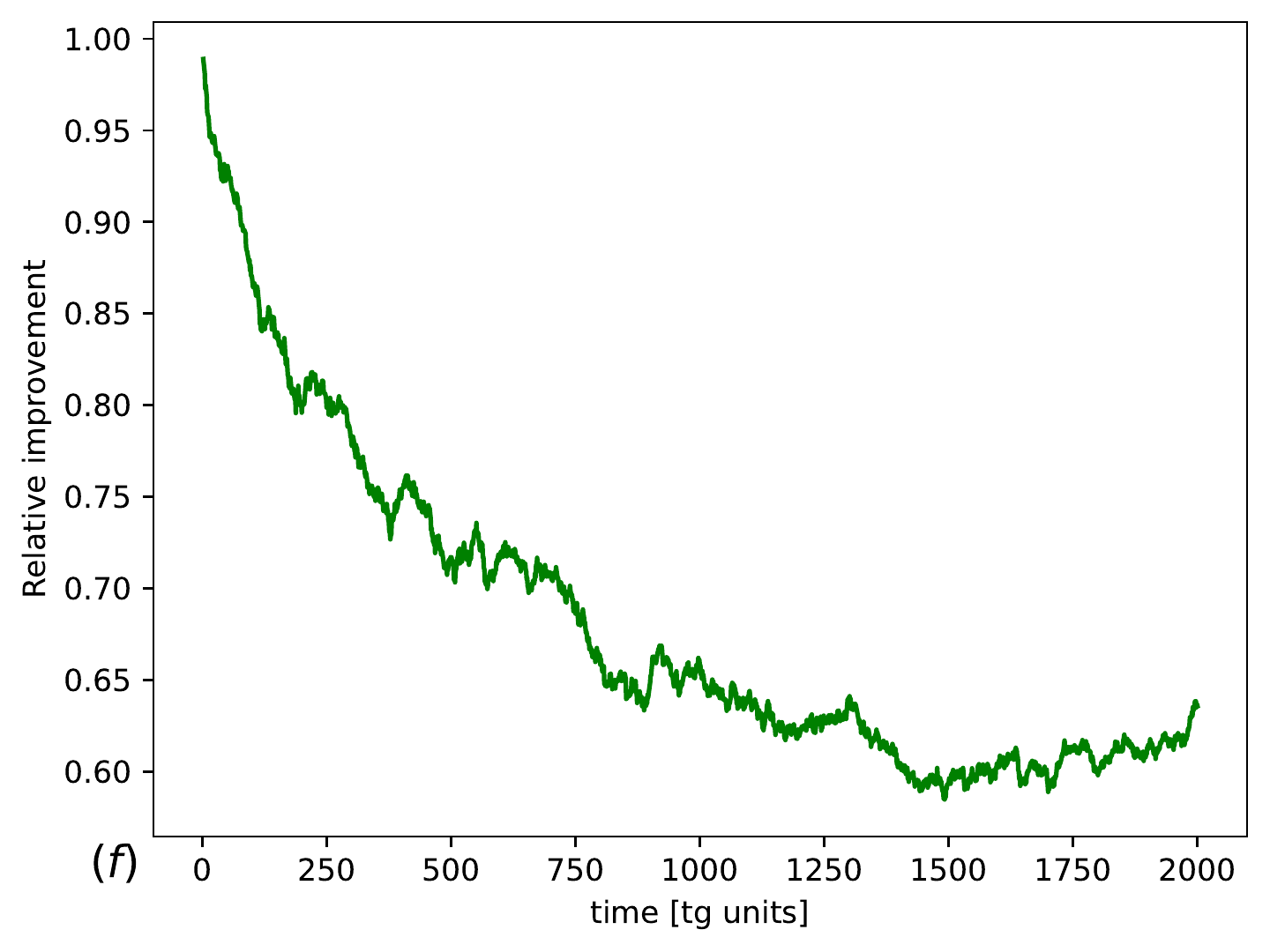}
\end{minipage}

\caption{Repetition of X gates. The three upper panels (a), (b), (c) show the time evolution of the $\rho_{00} = \langle 0 | \rho | 0 \rangle$ entry of the density matrix. The numerical solution of the Lindblad equation is displayed in orange (a), that of the noisy gates simulation in blue (b), and that of the Qiskit simulation in red (c). The noisy gates and Qiskit simulations are obtained with 1000 samples, and qualitatively they reproduce the time evolution of the Lindblad equation. Vertical dashed lines in the three top panels represent the time scales of relaxation $T_1$ (green), $T_2$ (yellow) and depolarization $T_d$ (grey). Panel (d) shows the Hellinger distances $\cl{H}^{\text{ng}}_{\sigma}$, in blue, and $\cl{H}^{\text{ibm}}_{\sigma}$, in red, as a function of time. Different curves are obtained from 100 independent runs of the two methods (for better readability only five are shown), where each simulation is obtained by averaging over 1000 samples. Panel (e) shows the mean of the Hellinger distances $\Bar{\cl{H}}^{\text{ng}}_{\sigma}$, and $\Bar{\cl{H}}^{\text{ibm}}_{\sigma}$, obtained from the $100$ independent runs, and vertical error bars show their standard deviations $\Delta\cl{H}^{\text{ng}}_{\sigma}$, $\Delta\cl{H}^{\text{ibm}}_{\sigma}$. The inset displays $\Delta\cl{H}^{\text{ng}}_{\sigma}$ and $\Delta\cl{H}^{\text{ibm}}_{\sigma}$ as functions of time. Panel (f) shows the relative improvement of the distance $\Bar{\cl{H}}^{\text{ng}}_{\sigma}$ with respect to $\Bar{\cl{H}}^{\text{ibm}}_{\sigma}$, calculated as $|\Bar{\cl{H}}^{\text{ibm}}_{\sigma}-\Bar{\cl{H}}^{\text{ng}}_{\sigma}|/\Bar{\cl{H}}^{\text{ibm}}_{\sigma}$. The fact that noises drive the system towards the maximally mixed state is the reason why the improvement decreases in time. The noisy gates and the standard approaches lead to the same predictions when one is close to decoherence times, as the noise is dominant over the unitary evolution. In the interesting regime $[0,2000\cdot t_g]$ before decoherence dominates, our improvement is always above $60\%$.}
\label{hellinger_X}
\end{figure*}
We now study the performances of our noisy gates method, and compare them with those of Qiskit's simulator \cite{qiskit}. First, in subsection \ref{numerical_comp} we test the two approaches against the  solution of Lindblad equation \eqref{Lindblad_equation}, by studying a repeated application of IBM's native gate set. Then, in subsection \ref{device_comp} we compare the predictions of both methods with the behaviour of an actual quantum computer, by running the inverse QFT algorithm on the IBM's quantum processors $\text{ibmq\_kolkata}$ and $\text{ibmq\_oslo}$. In appendix \ref{GHZ} we perform the same analysis by running the GHZ algorithm on $\text{ibm\_oslo}$. All  simulations are performed by using the noise model described in section \ref{noise_models} (see also appendices \ref{Krausmaps} and \ref{spam_relax}). The implementation of the work proposed in this paper is open source and available as a python package at this \href{https://pypi.org/project/quantum-gates/}{link}. It allows the user to run noisy simulations.

\subsection{Comparison with the numerical solution of Lindblad equations}\label{numerical_comp}
First, let us compare our method with the one implemented in the Qiskit simulator for the task of simulating the Lindblad equation. To this purpose, we simulate the same Lindblad equation with both methods, obtaining the density matrix $\rho^{\text{ng}}$, from the noisy gates simulation, and the density matrix $\rho^{\text{ibm}}$ from the Qiskit simulation. We then benchmark the results with the density matrix $\sigma$ obtained by directly solving numerically the Lindblad equation with Mathematica \cite{Mathematica}. We compare these density matrices by computing the Hellinger distances $\cl{H}^{\text{ng}}_{\sigma} = \cl{H}(\rho^{\text{ng}},\sigma)$, $\cl{H}^{\text{ibm}}_{\sigma}=\cl{H}(\rho^{\text{ibm}},\sigma)$  where the Hellinger distance is defined by
\begin{equation}
    \cl{H}(\rho,\sigma)=\frac{1}{\sqrt{2}}\sqrt{\sum_{k=1}^{N}\big(\sqrt{\rho_{kk}}-\sqrt{\sigma_{kk}}\big)^2}\, ,
\end{equation} with $\rho_{kk}$ ($\sigma_{kk}$) the diagonal elements of $\rho$ ($\sigma$). Note that the Hellinger distance is a classical measure of the distance between the readout probability distributions: while it cannot be interpreted as a distance between quantum states (it does not take in account the coherences), it directly compares the concrete outputs of the real device, which are classical (the oucomes of $\rr{Z}$ measurements).    
In appendix \ref{fidelities} we also compute the fidelities $\cl{F}^{\text{ng}}_{\sigma} =\cl{F}(\rho^{\text{ng}},\sigma)$ and $\cl{F}^{\text{ibm}}_{\sigma} = \cl{F}(\rho^{\text{ibm}},\sigma)$.
We run the simulations on both single and two qubit gates. Considering the native gate set of IBM's quantum computers, $\{\rr{R}_z(\phi),\rr{X},\rr{SX},\rr{CNOT}\}$, we remind that $\rr{R}_z(\phi)$ are implemented as virtual gates \cite{krantz2019quantum,mckay2017efficient}, i.e. they are noiseless, and the CNOT gates are implemented by combining single qubit gates in Eq. \eqref{single_qubit_Hamiltonian} and CR gates in Eq. \eqref{cross_resonance} \cite{krantz2019quantum,mckay2017efficient,alexander2020qiskit}. Moreover, X and SX gates are both rotations around the X-axis for different values of $\theta$, see Eq. \eqref{single_qubit_Hamiltonian}. Thus for our purposes, it is sufficient to simulate the X, CR and CNOT gates affected by noises.

\textit{Single qubit simulations.}
\begin{figure*}[htp]
    \begin{minipage}{0.330\textwidth}
        \includegraphics[width=\textwidth]{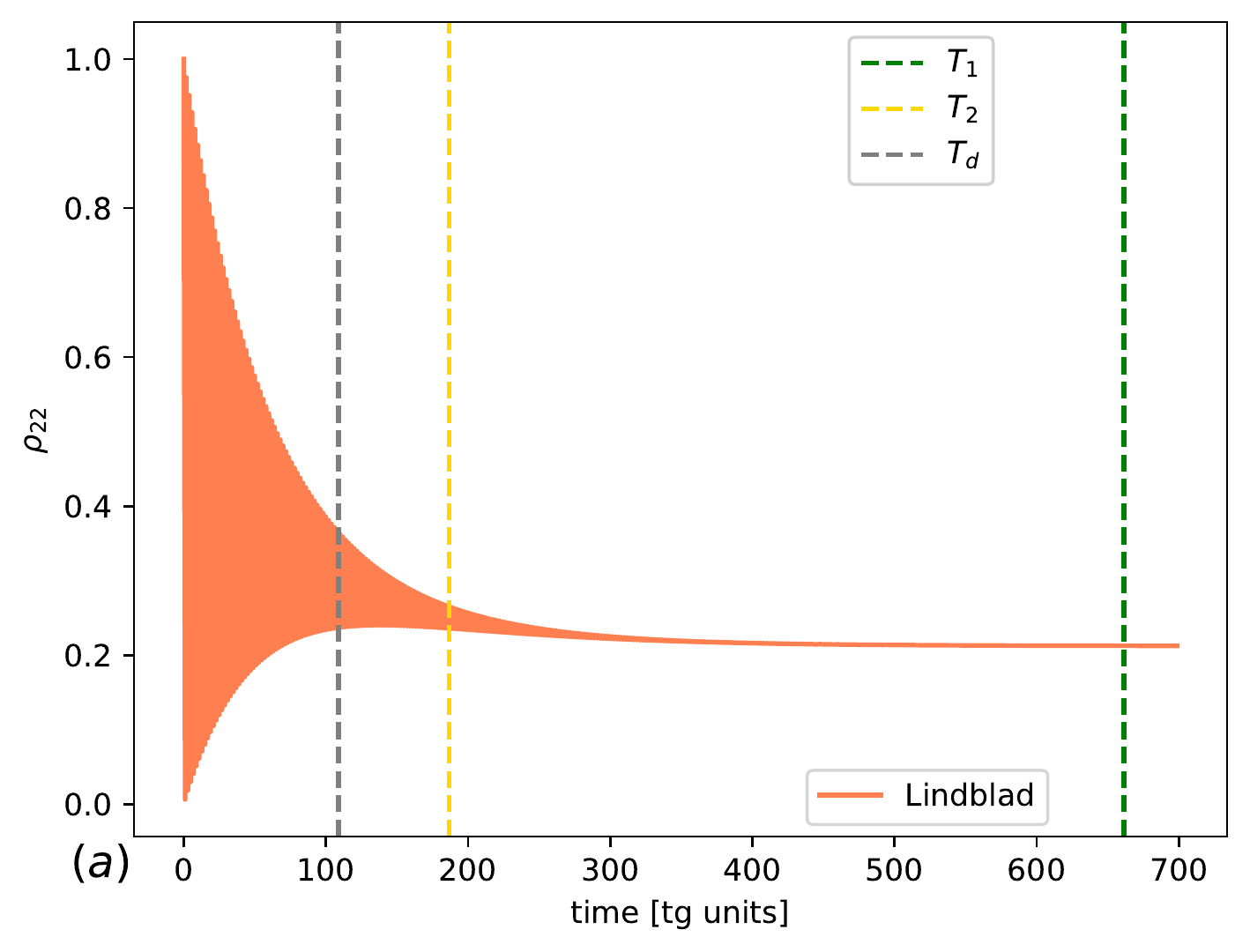}
    \end{minipage}%
    \hfill%
    \begin{minipage}{0.330\textwidth}
        \includegraphics[width=\textwidth]{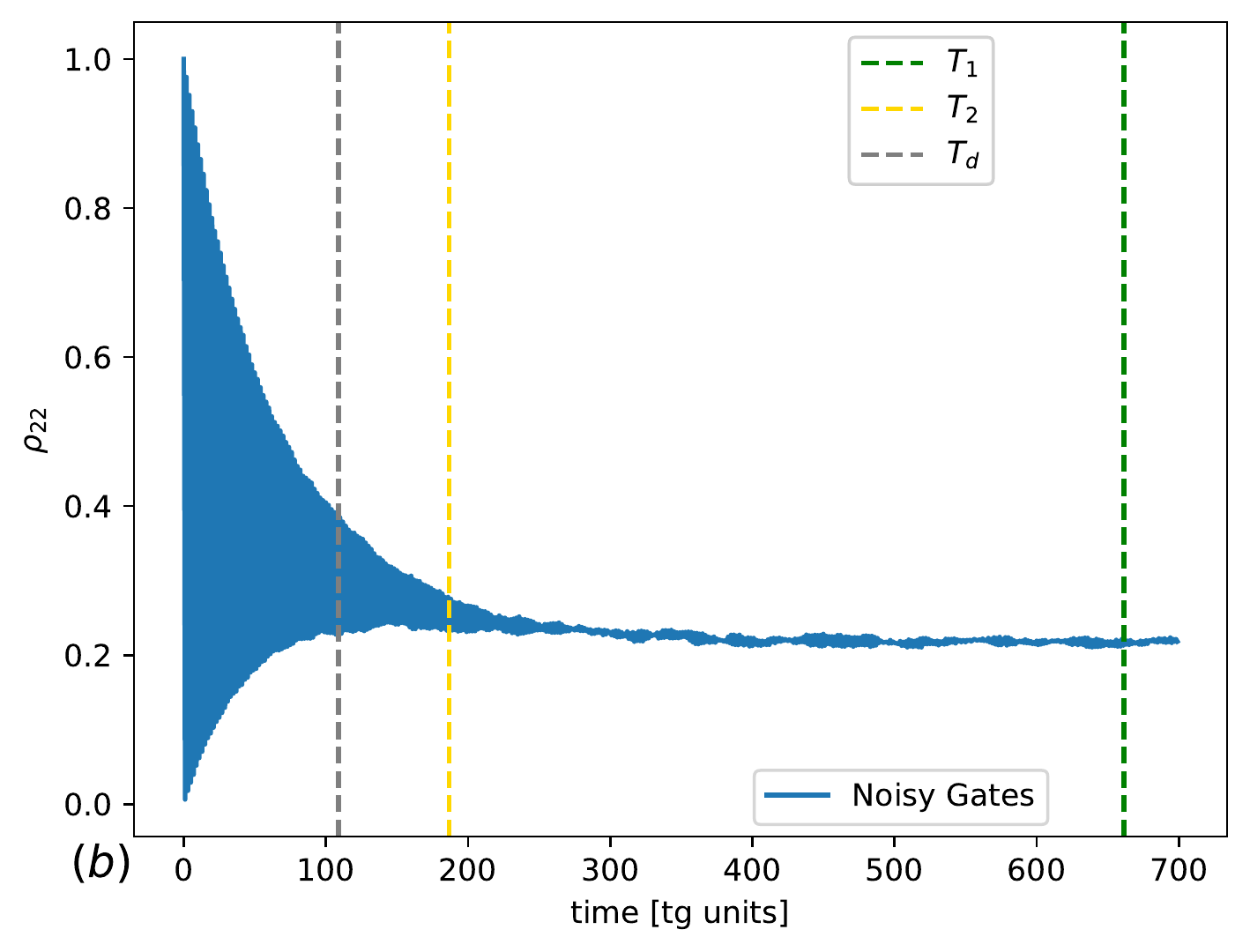}
    \end{minipage}
    \hfill%
    \begin{minipage}{0.330\textwidth}
        \includegraphics[width=\textwidth]{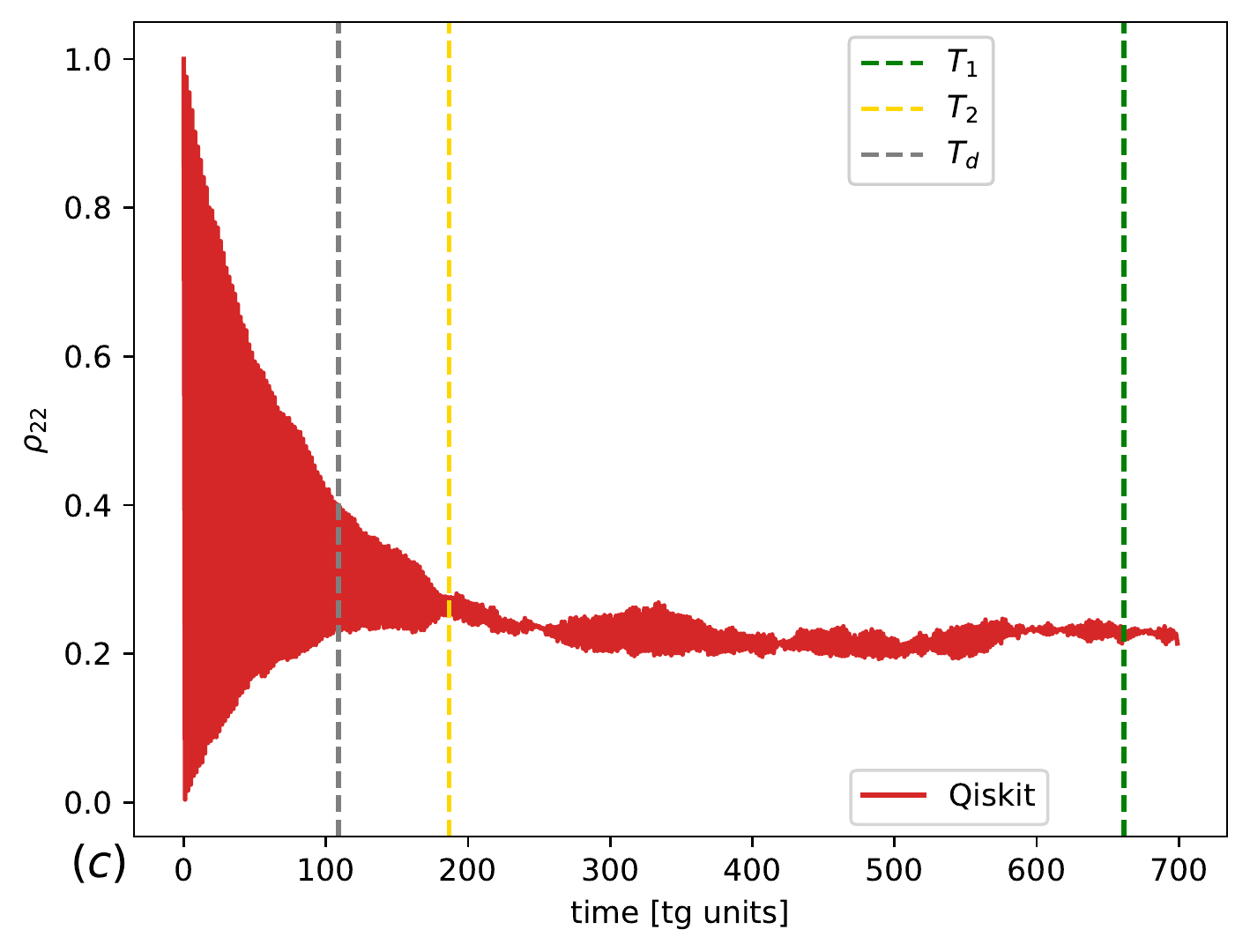}
    \end{minipage}
    \begin{minipage}{0.330\textwidth}
        \includegraphics[width=\textwidth]{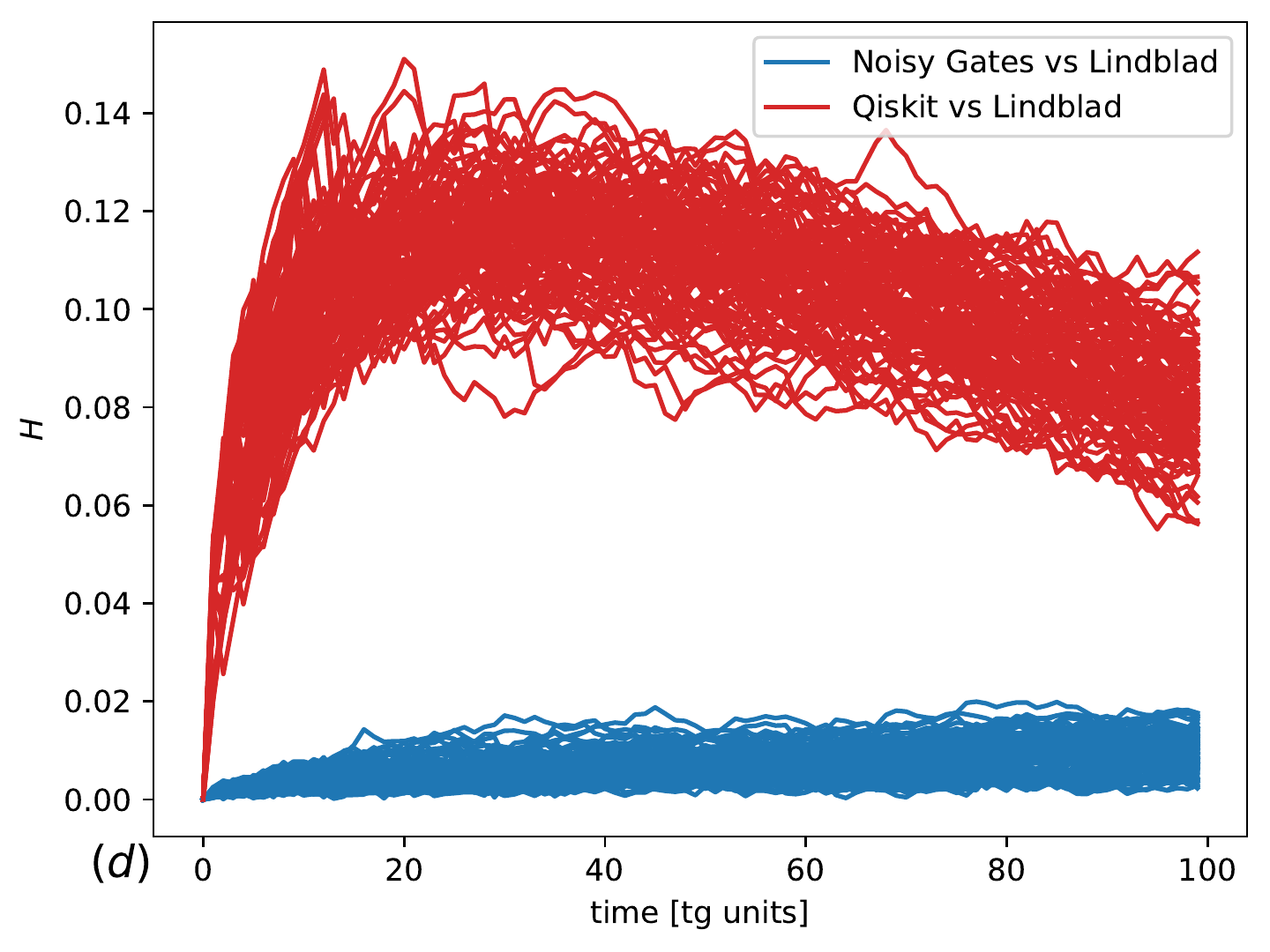}     
    \end{minipage}%
    \hfill%
    \begin{minipage}{0.330\textwidth}
        \includegraphics[width=\textwidth]{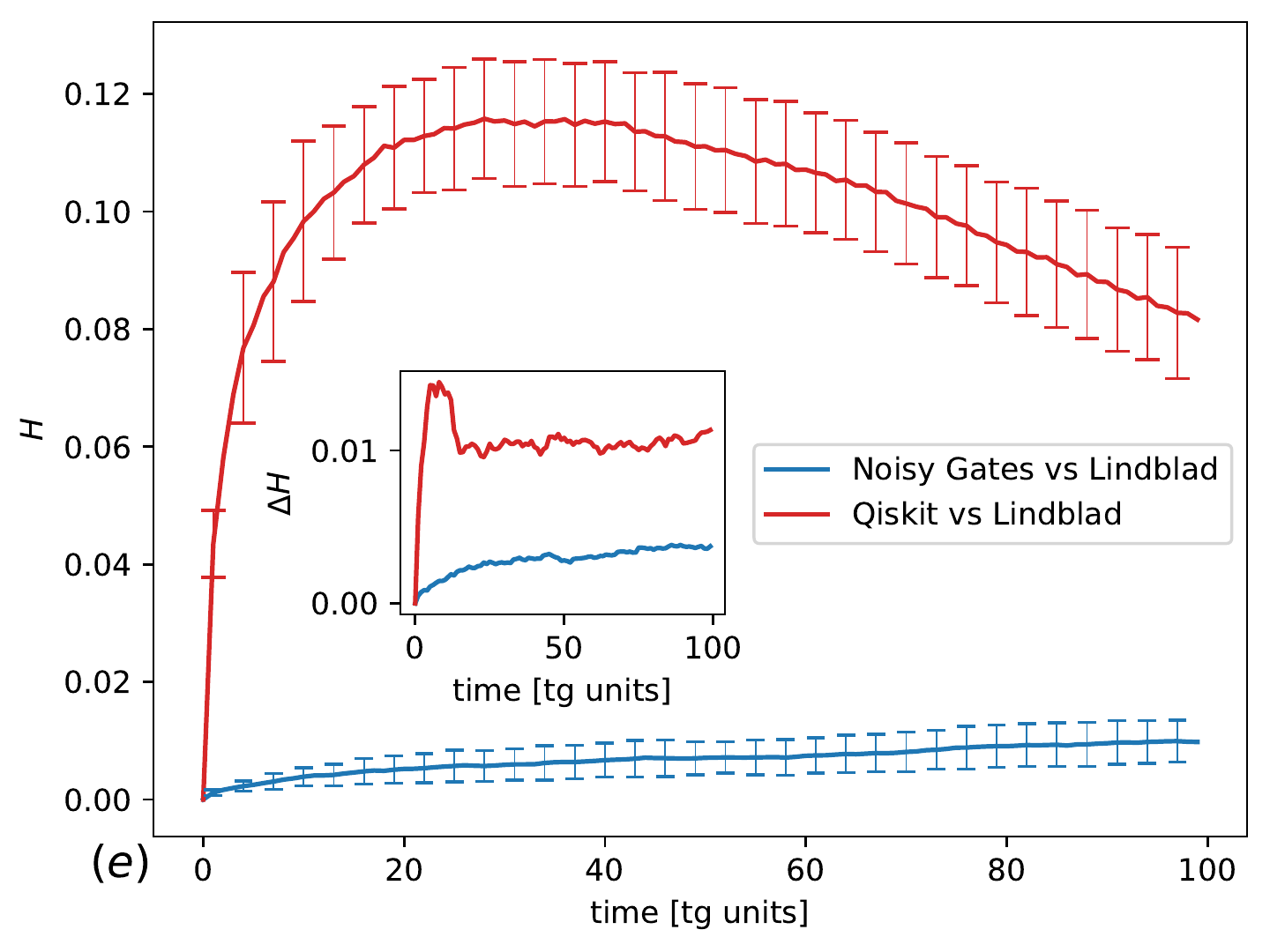}      
    \end{minipage}
    \hfill%
    \begin{minipage}{0.330\textwidth}
         \includegraphics[width=\textwidth]{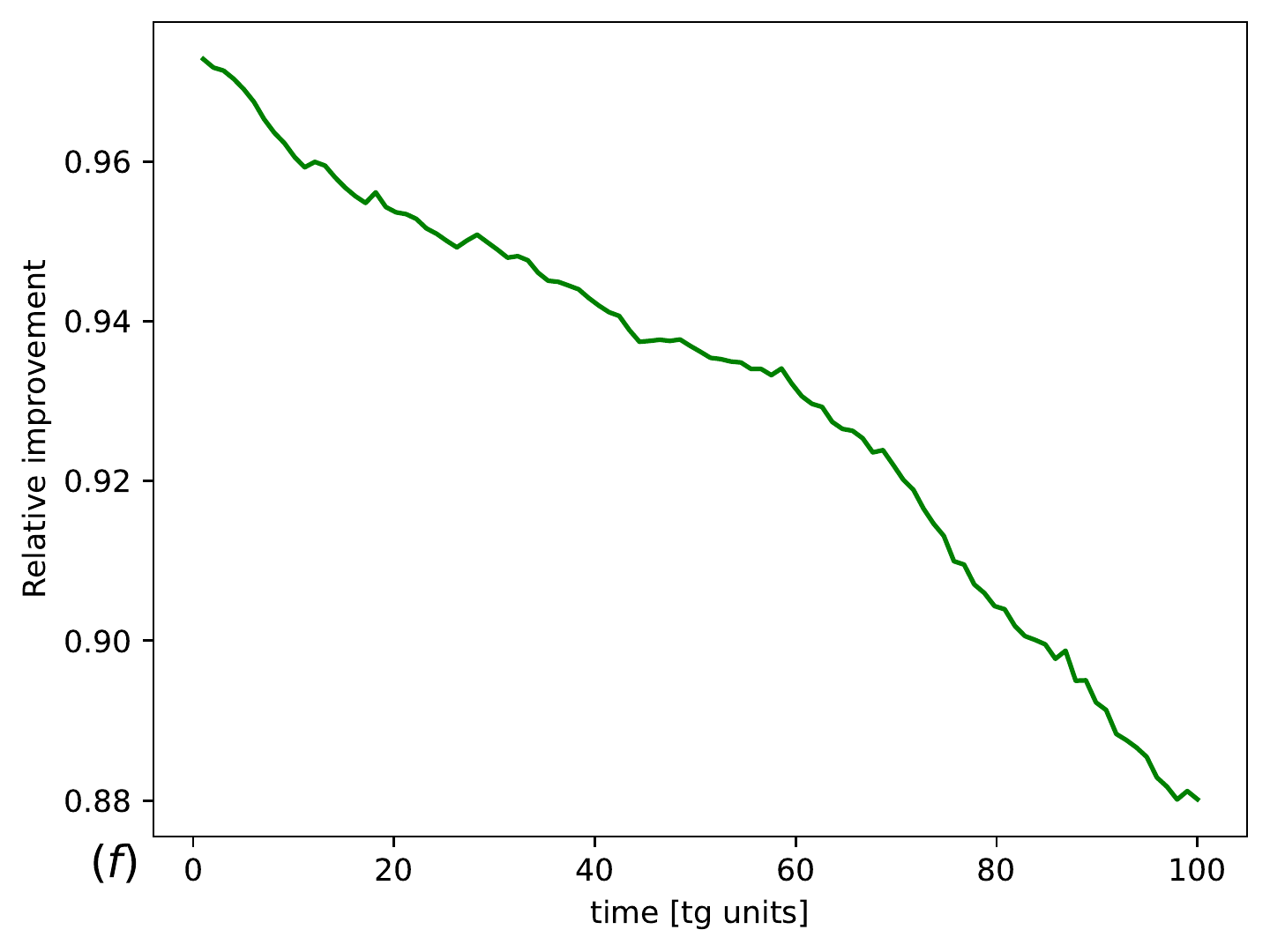}
    \end{minipage}
\caption{Repetition of CR gates. The three upper panels (a), (b), (c) show the time evolution of the $\rho_{22}$ entry of the density matrix for the $\rr{CR}$ gate with $\theta=\pi$ and $\phi=0$. Colors have the same meaning as for Fig.~\ref{hellinger_X}. Vertical dashed lines represent the time scales of relaxation, $T_1$ (in green) and $T_2$ (in yellow) of the target qubit, and depolarization $T_d$ (grey). The noisy gates simulations reproduce qualitatively better the time evolution obtained from the direct numerical solution of the Lindblad equation. Panels (d) and (e) display the Hellinger distances $\cl{H}^{\text{ng}}_{\sigma}$, in blue, and $\cl{H}^{\text{ibm}}_{\sigma}$, in red, as a function of time, for a repetition of CR gates. The plots have the same meaning as for Fig.\ref{hellinger_X}. Panel (f) shows the relative improvement of the distance $\Bar{\cl{H}}^{\text{ng}}_{\sigma}$ with respect to $\Bar{\cl{H}}^{\text{ibm}}_{\sigma}$, calculated as $|\Bar{\cl{H}}^{\text{ibm}}_{\sigma}-\Bar{\cl{H}}^{\text{ng}}_{\sigma}|/\Bar{\cl{H}}^{\text{ibm}}_{\sigma}$. The fact that noises drive the system towards the maximally mixed state is the reason why the improvement decreases in time. The noisy gates and the standard approaches lead to the same predictions when one is close to decoherence times, as the noise is dominant over the unitary evolution. In the interesting regime $[0,100\cdot t_g]$ our improvement is always above $88\%$.}
\label{hellinger_CR}
\end{figure*}
We first simulate a repetition of $\rr{X}$ gates, each of which can be obtained by setting $\theta=\pi$ and $\phi=0$ in Eq. \eqref{single_qubit_Hamiltonian}; we initialize the qubit in $\ket{0}$ and we use the qubit noise parameters of $\text{ibmq\_manila}$ (more details on the device can be found in appendix \ref{Device_parameters}). We evolve the state of the qubit for a time $T=\cl{N}t_g$, with $\cl{N}= 15000$. In the upper panels of Fig. \ref{hellinger_X} we plot the time evolution of the population of the ground state, $\rho_{00} =\langle 0 | \rho | 0 \rangle $, as obtained with the three methods. In the noiseless case, $\rho_{00}$ should oscillate between $0$ and $1$ with period $2t_g$, as at each step of $t_g$ a complete $\rr{X}$ rotation is performed; in the presence of noises, the oscillations are damped due to the relaxation of the qubit, while the depolarization drives probabilities towards the asymptotic value $\rho_{00}\rightarrow 0.5$. 

Both our simulation and that obtained using Qiskit's simlulator qualitatively reproduce this behaviour. In Fig. \ref{hellinger_X} we have also highlighted with dashed vertical lines the characteristic times of relaxation and depolarization (see the caption); for times approaching these values the state is not a reliable quantum state anymore, as the density matrix becomes completely mixed. Given this consideration, in the lower plots we stop at  $\cl{N}=2000$. 

In order to inspect which of the two models reproduces more accurately and precisely the Lindblad evolution, we have run 100 independent simulations with both the noisy gates simulator and the Qiskit simulator, computing for each run the Hellinger distances $\cl{H}^{\text{ng}}_{\sigma}$, $\cl{H}^{\text{ibm}}_{\sigma}$. We computed the means over the 100 independent simulations, $\Bar{\cl{H}}^{\text{ng}}_{\sigma}$, $\Bar{\cl{H}}^{\text{ibm}}_{\sigma}$ and the standard deviations $\Delta\cl{H}^{\text{ng}}_{\sigma}$, $\Delta\cl{H}^{\text{ibm}}_{\sigma}$. These quantities are shown in the lower panels of Fig. \ref{hellinger_X}. During the relevant time interval $[0,T]$ the Hellinger distance of the noisy gates simulator is closer to zero, than that obtained with the Qiskit simulator. Both results are compatible within the error bars, however the standard deviations associated to the noisy gates simulations are significantly smaller than those associated to the Qiskit simulations, as also highlighted in the inset of Fig. \ref{hellinger_X} (e). We notice that the difference between $\Bar{\cl{H}}^{\text{ng}}_{\sigma}$ and $\Bar{\cl{H}}^{\text{ibm}}_{\sigma}$ is of the order $\sim 10^{-3} - 10^{-2}$, and this corresponds to a relative improvement, calculated as $|\Bar{\cl{H}}^{\text{ibm}}_{\sigma}-\Bar{\cl{H}}^{\text{ng}}_{\sigma}|/\Bar{\cl{H}}^{\text{ibm}}_{\sigma}$, in the range from $90\%$ to $60\%$ as time increases. The relative improvement is shown in Fig. \ref{hellinger_X} (f). The fact that noises drive the system towards the maximally mixed state is the reason why the improvement decreases over time. The noisy gates and the standard approaches lead to the same predictions when one is close to decoherence times. Indeed after such times the strength of the noise is dominant over the unitary evolution, or the Hamiltonian contribution is negligible with respect to the Lindblad term (see Eq.\eqref{Lindblad_equation}), which is the same in the two approaches. In the interesting regime $[0,T]$ our improvement is always above $60\%$. In appendix \ref{fidelities} we repeat a similar analysis for the fidelities.
\begin{figure*}[htp]
    \begin{minipage}{0.330\textwidth}
        \includegraphics[width=\textwidth]{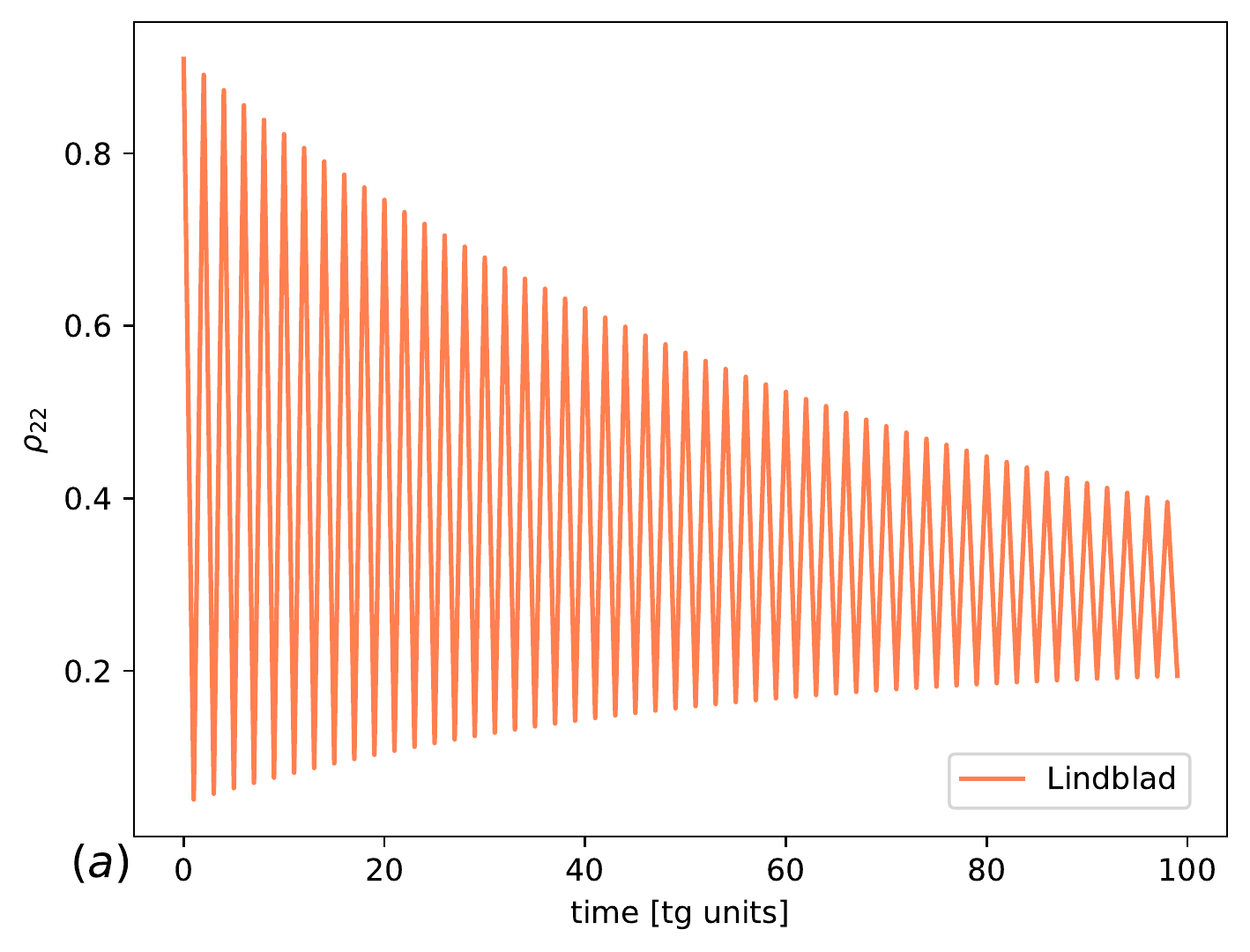}
    \end{minipage}%
    \hfill%
    \begin{minipage}{0.330\textwidth}
        \includegraphics[width=\textwidth]{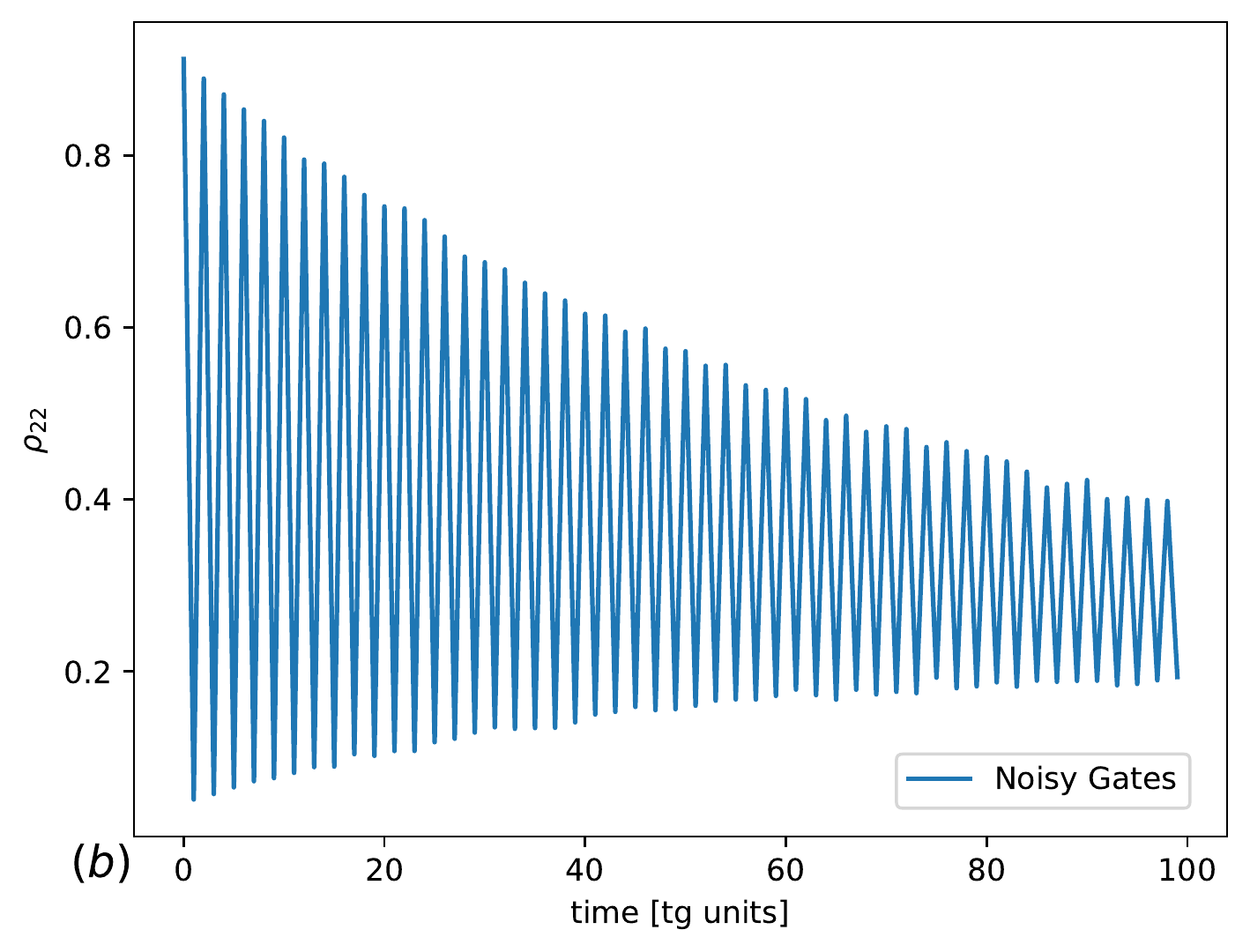}
    \end{minipage}
    \hfill%
    \begin{minipage}{0.330\textwidth}
        \includegraphics[width=\textwidth]{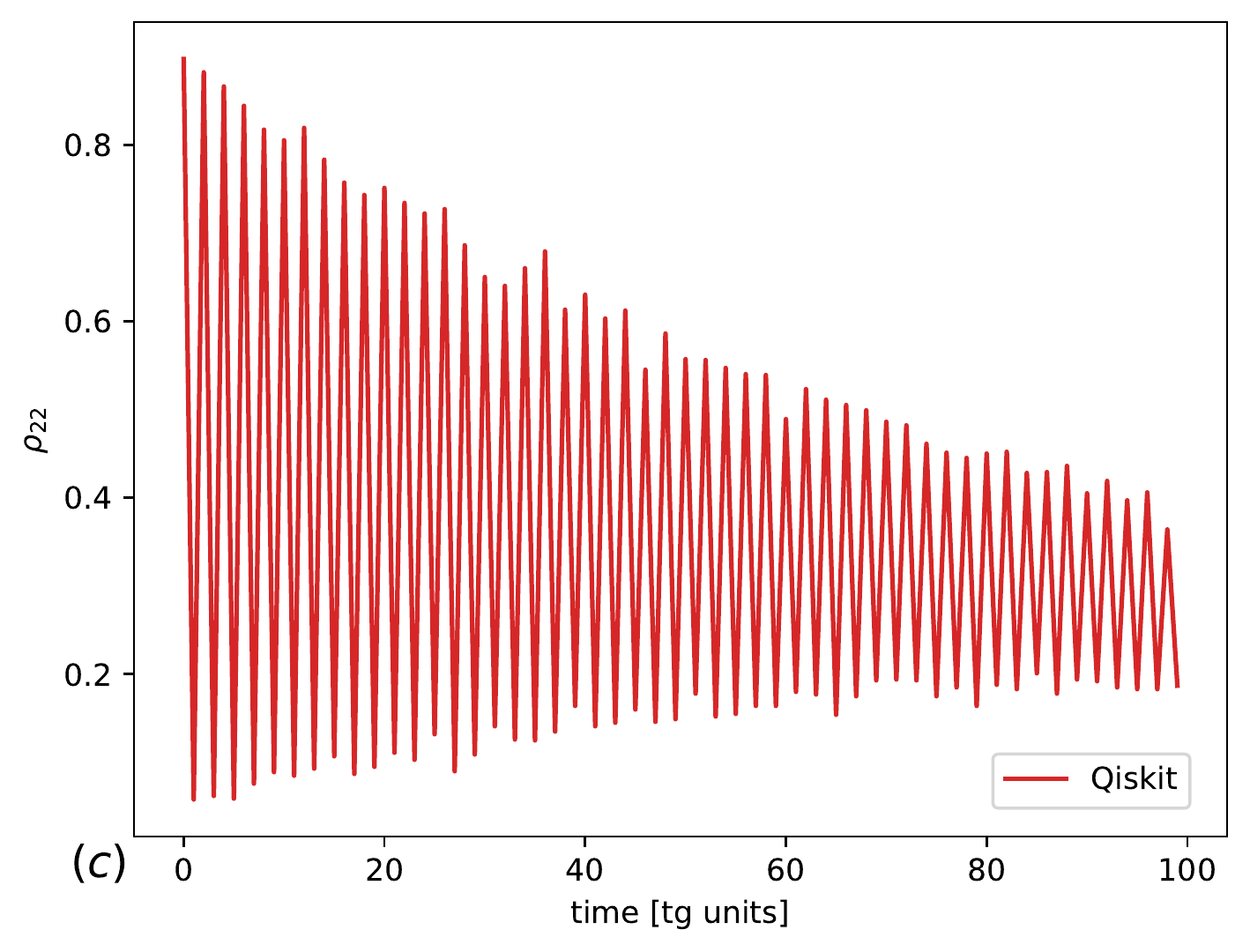}
    \end{minipage}
    \begin{minipage}{0.330\textwidth}
        \includegraphics[width=\textwidth]{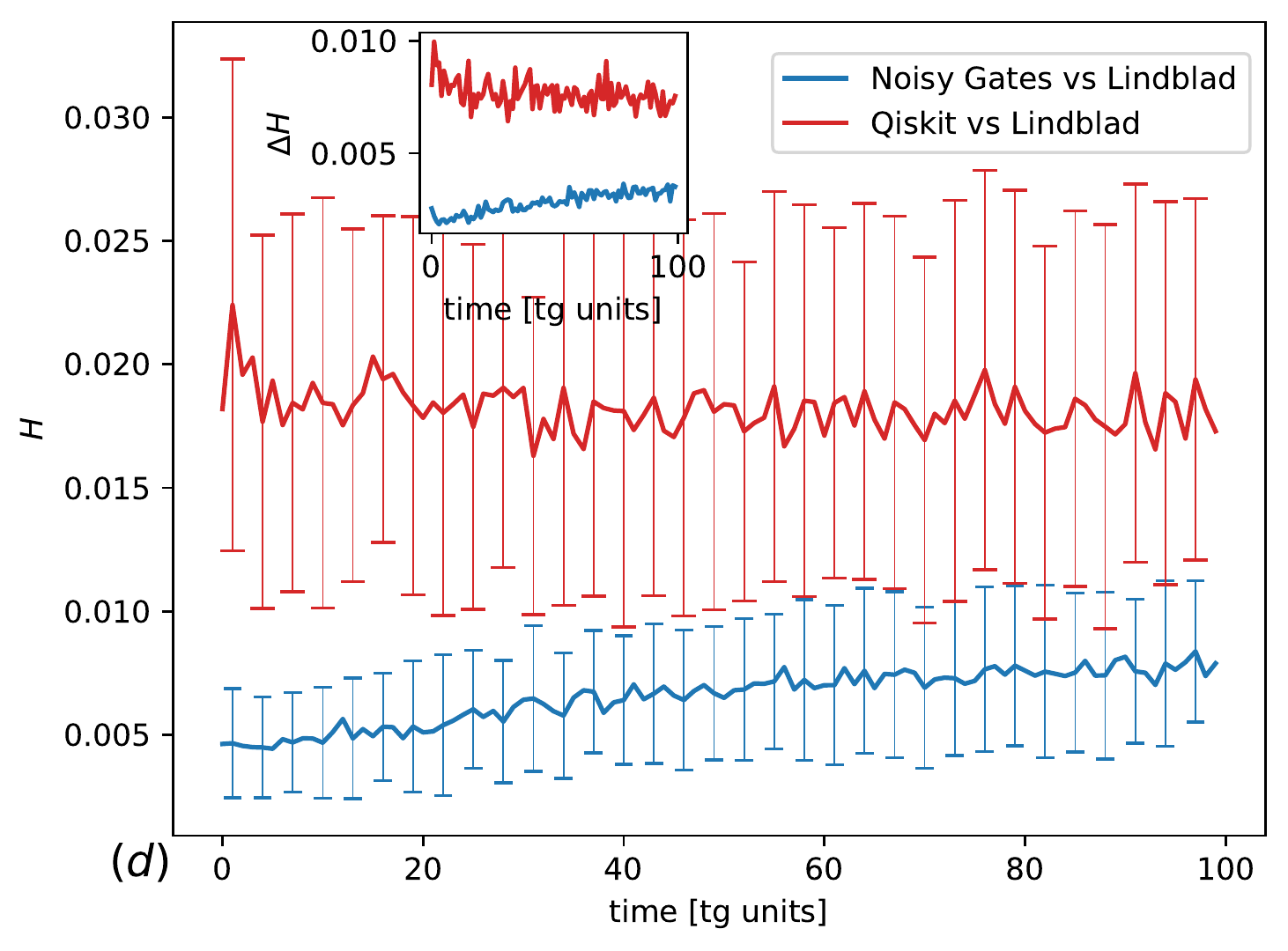}     
    \end{minipage}%
    \hfill%
    \begin{minipage}{0.330\textwidth}
        \includegraphics[width=\textwidth]{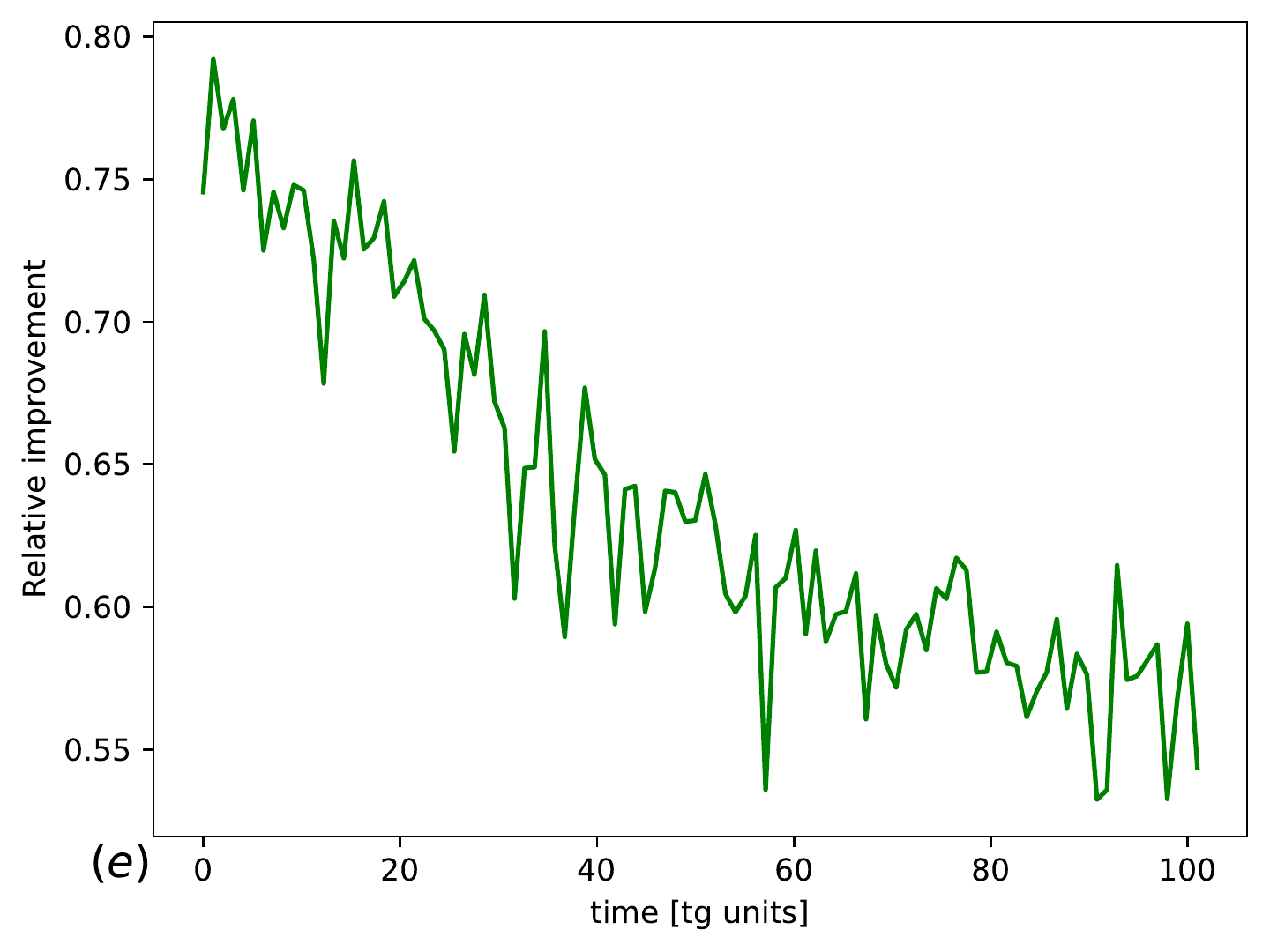}      
    \end{minipage}
    \hfill%
    \begin{minipage}{0.330\textwidth}
         \includegraphics[width=\textwidth]{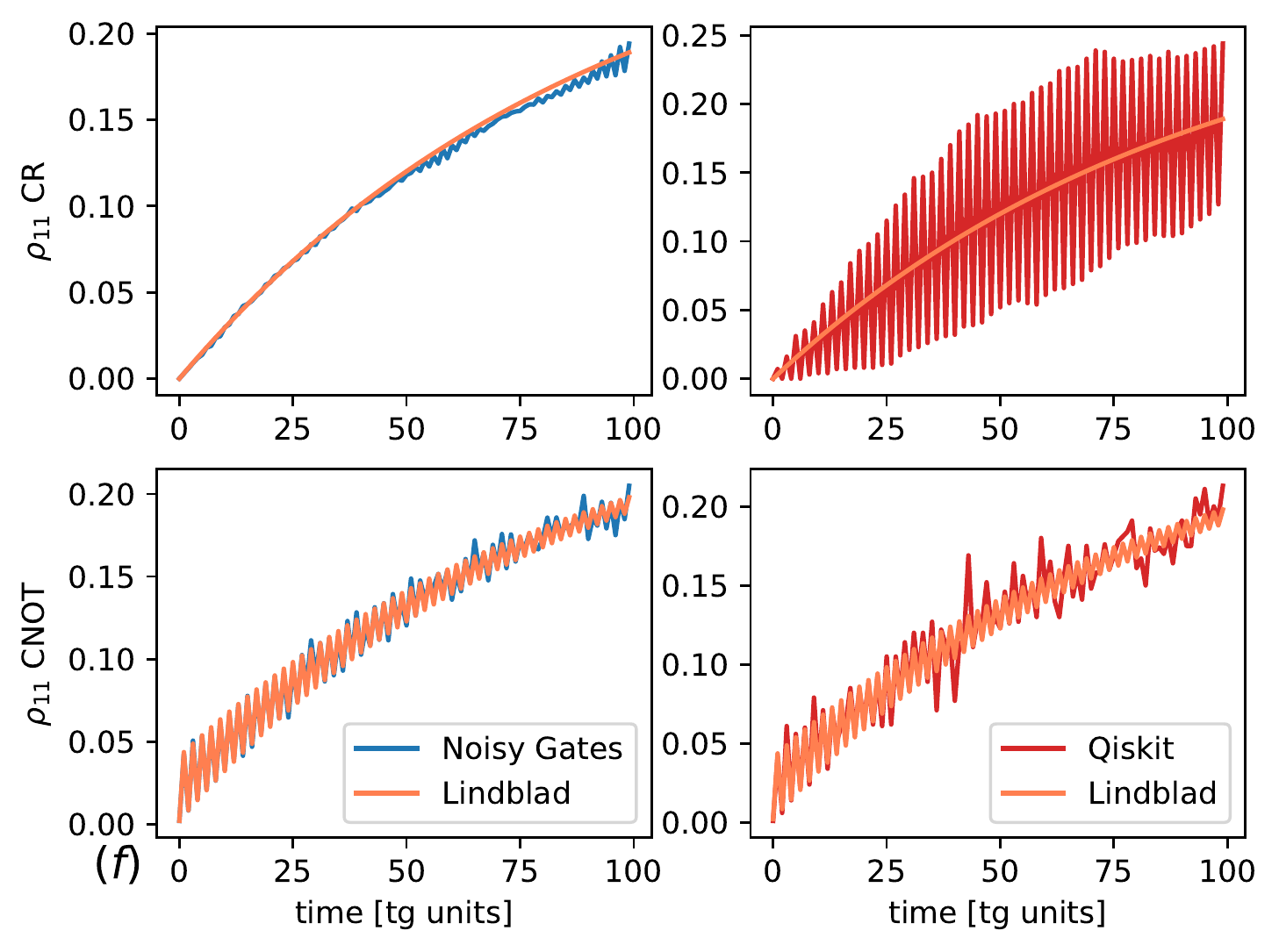}
    \end{minipage}
\caption{Repetition of CNOT gates. The three upper panels (a), (b), (c) show the time evolution of the $\rho_{22}$ entry of the density matrix for the $\rr{CNOT}$ gate. Colors have the same meaning as for Fig.~\ref{hellinger_X}. The noisy gates simulations reproduce qualitatively better the time evolution obtained from the direct numerical solution of the Lindblad equation. Panel (d) displays mean of the Hellinger distances $\Bar{\cl{H}}^{\text{ng}}_{\sigma}$, in blue, and $\Bar{\cl{H}}^{\text{ibm}}_{\sigma}$, in red, and their standard deviations as functions of time. Panel (e) shows the relative improvement. The fact that noises drive the system towards the maximally mixed state is again the reason why the improvement decreases in time: the noisy gates and the standard approaches lead to the same predictions when one is close to decoherence times, as the noise is dominant over the unitary evolution. In the interesting regime $[0,100\cdot t_g]$ our improvement is always above $55\%$. The upper subplots of panel (f) show the time evolution of the $\rho_{11}$ entry of the density matrix for the same sequence of $\rr{CR}$ gates in Fig.~\ref{hellinger_CR} and the lower subplots show the time evolution of the $\rho_{11}$ entry of the density matrix for the sequence of $\rr{CNOT}$ gates. Colors have the same meaning as for Fig.~\ref{hellinger_X}.  For the $\rr{CR}$ gates, the Qiskit simulation of $\rho_{11}$ is visibly different from the Lindblad evolution, thus explaining the higher improvement of the noisy gates simulation in the Hellinger distance in Fig.~\ref{hellinger_CR}.}
\label{hellinger_CX}
\end{figure*}

\textit{Two qubits simulations.} Next, we simulate a repetition of Cross-resonance gates as defined in Eq. \eqref{cross_resonance}, where we choose $\phi=0$ and $\theta=\pi$. We initialize the system in the state $\ket{10}$ and we use the qubit noise parameters of $\text{ibmq\_manila}$. In the three upper panels of Fig. \ref{hellinger_CR} we show the time evolution of the entry $\rho_{22}=\langle 10 | \rho | 10 \rangle $; the x-axis is normalized in terms of the two-qubit gate time $t_g$. The two-qubit state goes asymptotically towards the completely mixed state as $\rho_{22}$ reaches the asymptotic value $0.25$. The probability $\rho_{22}$, which in the ideal case should flip between one and zero, is again damped over time by relaxation effects.
Again, we have highlighted with vertical dashed lines the characteristic time scales of the noises, showing only the $T_1$ and $T_2$ values of the target qubit as representative values. The depolarizing error is the dominant one, spoiling the quantum state already after $\sim100$ $\rr{CR}$ gates; for this reason, the the lower panels we  consider a total duration $\cl{N}\sim 100$. As before, we report the Hellinger distances, showing the different results of $100$ independent simulations together with their mean and standard deviation in the three lower panels of Fig. \ref{hellinger_CR}. 

As in the single qubit case, within the relevant time interval $[0,T]$ the Hellinger distances obtained with the noisy gates simulations are closer to zero than those obtained with the Qiskit simulator. However now, differently from the single qubit case, the two results are not compatible within error bars. Moreover the difference between $\Bar{\cl{H}}^{\text{ng}}_{\sigma}$ and $\Bar{\cl{H}}^{\text{ibm}}_{\sigma}$ is now of the order $\sim 10^{-1}$. This corresponds to a relative improvement in the range from $90\%$ to $88\%$ as time increases, shown in Fig. \ref{hellinger_CR} (f). In the interesting regime $[0,T]$ our improvement is always above $88\%$.
We notice that in Fig. \ref{hellinger_CR} (e) the value of $\Bar{\cl{H}}^{\text{ibm}}_{\sigma}$ approaches that of $\Bar{\cl{H}}^{\text{ng}}_{\sigma}$ for times close to 100 $\rr{CR}$ gate times. The reason why this happens is the same explained above for the single qubit case. In appendix \ref{fidelities} we repeat a similar analysis for the fidelities.

We then perform the analysis for a repetition of $\rr{CNOT}$ gates, for an initial state given by $\ket{10}$ and qubit noise parameters of $\text{ibmq\_quito}$ (see appendix \ref{Device_parameters}). We notice that in this simulation we implement each CNOT gate directly without expressing it as a combination of single qubit gates and CR gates, as it is done in IBM devices. We make this choice because in this way it is easier to solve numerically the target Lindblad equation. At each time step of the evolution we simulate a circuit with an increasing number of CNOT gates and measurements at the end. Thus we add SPAM channels (see appendices \ref{Krausmaps} and \ref{spam_relax}) to model measurements errors. This allows to extend the analysis to runs on real hardware, that involve measurements, as we will show later in subsection \ref{device_comp}. In the three upper panels of Fig. \ref{hellinger_CX} we show the time evolution of the $\rho_{22}=\langle 10 | \rho | 10 \rangle $ entry of the density matrix. The relevant time interval is again given by a total duration of $\cl{N}\sim 100$ gates. Indeed the depolarizing error in this case spoils the quantum state after $\sim120$ $\rr{CNOT}$ gates. 
Fig. \ref{hellinger_CX} (d) shows the mean of the Hellinger distances $\Bar{\cl{H}}^{\text{ng}}_{\sigma}$ (in blue) and $\Bar{\cl{H}}^{\text{ibm}}_{\sigma}$ (in red) and their standard deviations $\Delta\cl{H}^{\text{ng}}_{\sigma}$ and $\Delta\cl{H}^{\text{ibm}}_{\sigma}$, also shown in the inset. Once more, within the relevant time interval $[0,T]$ the Hellinger distances obtained with the noisy gates simulations are closer to zero than those obtained with the Qiskit simulator and the difference between $\Bar{\cl{H}}^{\text{ng}}_{\sigma}$ and $\Bar{\cl{H}}^{\text{ibm}}_{\sigma}$ is of the order $\sim 10^{-2}$. This corresponds to a relative improvement in the range from $80\%$ to $55\%$ as time increases. This is shown in Fig. \ref{hellinger_CX} (e). In the interesting regime $[0,T]$ the relative improvement is always above $55\%$.

\begin{figure}[!htp]
\centering
\includegraphics[width=1.01\linewidth]{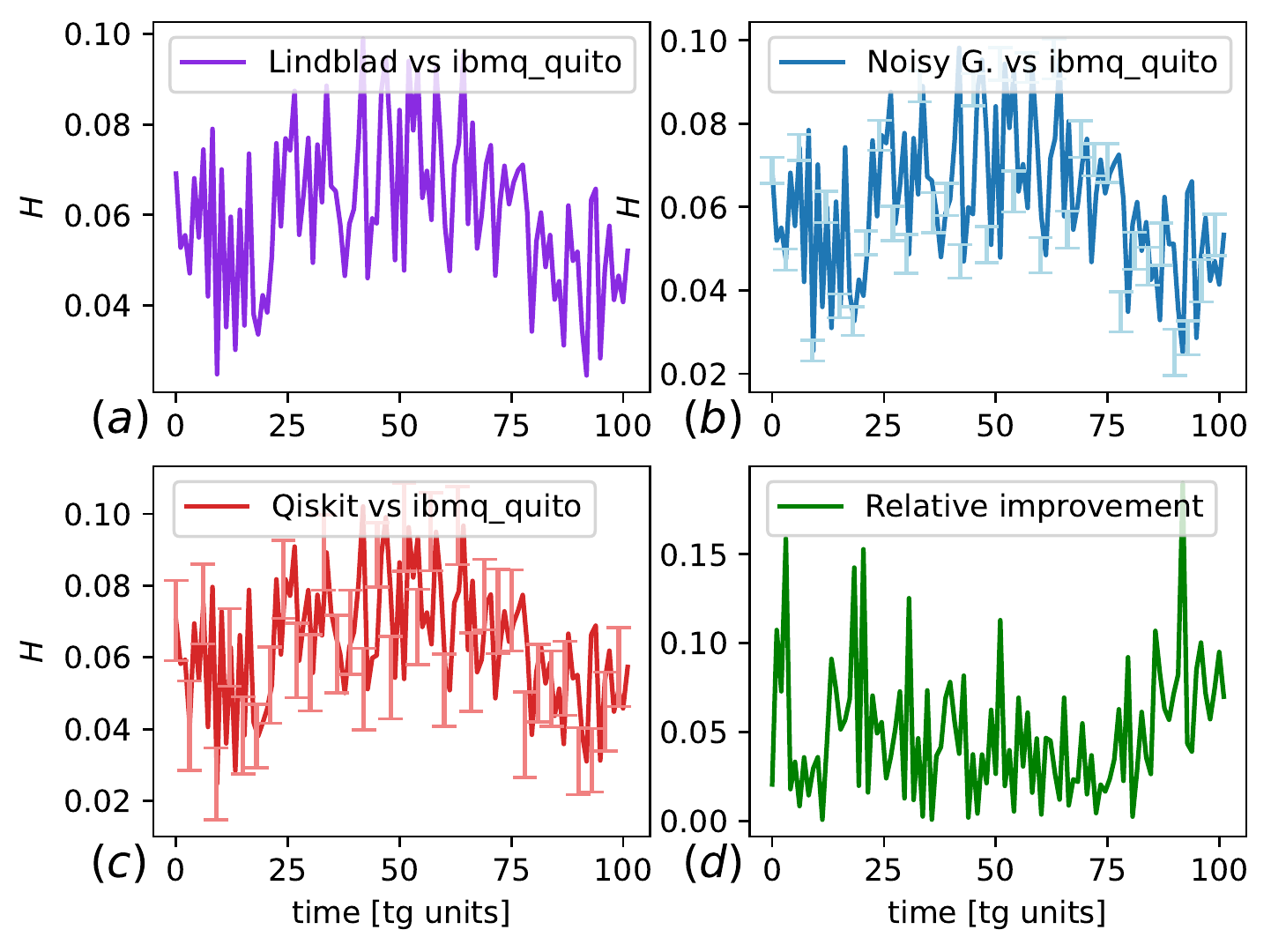} 
\caption{Repetition of CNOT gates. Panel (a) shows the Hellinger distance $\cl{H}^{\chi}_{\sigma}$ between the Lindblad evolution and $\text{ibmq\_quito}$ for the repetition of CNOT gates. Panel (b) shows the mean Hellinger distance $\Bar{\cl{H}}^{\text{ng}}_{\chi}$ and the standard deviations between the noisy gates simulation and $\text{ibmq\_quito}$. Panel (c) shows the mean Hellinger distance $\Bar{\cl{H}}^{\text{ibm}}_{\chi}$ and the standard deviations between the Qiskit simulation and $\text{ibmq\_quito}$. Panel (d) shows the relative improvement calculated as $|\Bar{\cl{H}}^{\text{ibm}}_{\chi}-\Bar{\cl{H}}^{\text{ng}}_{\chi}|/\Bar{\cl{H}}^{\text{ibm}}_{\chi}$. The relative improvement is around $10\%$. The smaller relative improvement with respect to those shown in the previous figures, is mainly due to  additional noises present in ibmq devices, i.e. crosstalks, correlated noises and coherent errors.} 
\label{hellinger_CX_all}
\end{figure}
By looking at Figs. \ref{hellinger_CR} (e) and \ref{hellinger_CX} (d), we notice that the improvement in the Hellinger distance gained by using the noisy gates approach is much higher for $\rr{CR}$ gates with respect to $\rr{CNOT}$ gates. The reason why this happens is clarified in Fig. \ref{hellinger_CX} (f). The panel consists of two upper subplots showing the time evolution of the $\rho_{11}$ entry of the density matrix for the $\rr{CR}$ gates and two lower subplots showing the time evolution of the $\rho_{11}$ entry of the density matrix for the $\rr{CNOT}$ gates. Similarly to the convention used above, orange curves are obtained with the numerical solution of the Lindblad equation, blue curves are obtained with the noisy gates simulations and red curves are obtained with Qiskit simulations. The noisy gates simulations make good predictions for both  gate sequences, as the blue curves follow closely the orange curves. On the other hand, the Qiskit simulation for the $\rr{CR}$ gates is visibly different from the numerical solution of the Lindblad equation. This might be due to the fact that the CR gate is a block diagonal matrix with $X(\theta)$ in the upper block and $X(-\theta)$ in the lower block while the CNOT gate is block diagonal with an identity in the upper block and $X(\theta)$ in the lower block. The identity in the CNOT might lead to a lower influence of noises on the $\rho_{00}$ and $\rho_{11}$ entries of the density matrix. These observations explain why the Hellinger distances obtained with the noisy gates in different simulations are very good and similar to each other, while the Hellinger distance obtained with Qiskit is better for the $\rr{CNOT}$ with respect to the $\rr{CR}$. Nevertheless, the noisy gates approach always outperforms the standard one by a significant amount, as shown by the relative improvements.
\subsection{Comparison with the behaviour of a real quantum computer}\label{device_comp}

Now, we inspect the performances of the noisy gates approach when trying to reproduce the behaviour of a real quantum computation. To this purpose, we first extend the analysis of the CNOT gates sequence in subsection \ref{numerical_comp}, and then we focus on the inverse Quantum Fourier Transform (QFT$^{\dagger}$). When dealing with a real hardware, we must take into account that the noise model we implement in this analysis (see section \ref{noise_models}) might not be accurate enough in describing the device, and that different quantum devices might behave very differently from one another. As we will show,  despite the choice of a simple noise model and the instability of ibmq devices, our approach is still able to outperform the standard one also when compared with the real hardware.

\textit{CNOT simulations.}
We run the sequence of CNOT gates of subsection \ref{numerical_comp} on $\text{ibmq\_quito}$, available on the cloud and comprising 7 superconducting transmon qubits \cite{koch2007charge} (see appendix \ref{Device_parameters}) and we reconstruct the density matrix  $\chi$ obtained from the physical device, to be compared with the density matrices $\rho^{\text{ng}}$, $\rho^{\text{ibm}}$ and $\sigma$ obtained for the CNOT simulations discussed in subsection \ref{numerical_comp}. We remark again that for the CNOT simulations of subsection \ref{numerical_comp}, we implemented each CNOT gate directly without expressing it as a combination of single qubit gates and CR gates, as it is done in IBM devices, because in this way it is easier to solve numerically the target Lindblad equation. We create a list of circuits, each consisting of an increasing number of CNOT gates, and measure each circuit $1000$ times to obtain the output probability distributions, thus deriving the evolution of the outcome probabilities as the number of gates increases. As noted above, since each circuit involves measurements we added a SPAM error to model measurement errors. 

The Hellinger distance $\cl{H}^{\chi}_{\sigma}=\cl{H}(\chi,\sigma)$ between the Lindblad evolution and the evolution obtained with $\text{ibmq\_quito}$ is shown in Fig. \ref{hellinger_CX_all} (a). This distance is three to tens time larger with respect to $\Bar{\cl{H}}^{\text{ng}}_{\sigma}$ and $\Bar{\cl{H}}^{\text{ibm}}_{\sigma}$ that are shown in Fig. \ref{hellinger_CX} (d). While the standard approach and the noisy gates approach have a certain level of agreement with the Lindblad equation, the latter is deviating from the quantum hardware by a significantly higher level. This is also the reason why it is not possible to appreciate the difference between the mean Hellinger distance $\Bar{\cl{H}}^{\text{ng}}_{\chi} = \Bar{\cl{H}}(\rho^{\text{ng}},\chi)$ of the noisy gates with $\text{ibmq\_quito}$ and the mean Hellinger distance $\Bar{\cl{H}}^{\text{ibm}}_{\chi}=\Bar{\cl{H}}(\rho^{\text{ibm}},\chi)$ of Qiskit with $\text{ibmq\_quito}$, as shown in Fig. \ref{hellinger_CX_all} (b) and Fig. \ref{hellinger_CX_all} (c). Fig. \ref{hellinger_CX_all} (d) shows the relative improvement with respect to the device, which is calculated as $|\Bar{\cl{H}}^{\text{ibm}}_{\chi}-\Bar{\cl{H}}^{\text{ng}}_{\chi}|/\Bar{\cl{H}}^{\text{ibm}}_{\chi}$. The relative improvement is around $10\%$. The smaller relative improvement with respect to those shown in the previous subsection is only to a small extend due to the fact that we do not decompose CNOT gates. The main reason, as we explain when discussing the  simulations of the QFT (see below), is that additional noises are present in ibmq devices, i.e. crosstalks, correlated noises and coherent errors \cite{wilen2021correlated,zhao2022quantum}. The simple noise model that we consider in this work  does not take  such noises into account.
\begin{figure*}[htp]
    \begin{minipage}{0.331\textwidth}
        \includegraphics[width=\textwidth]{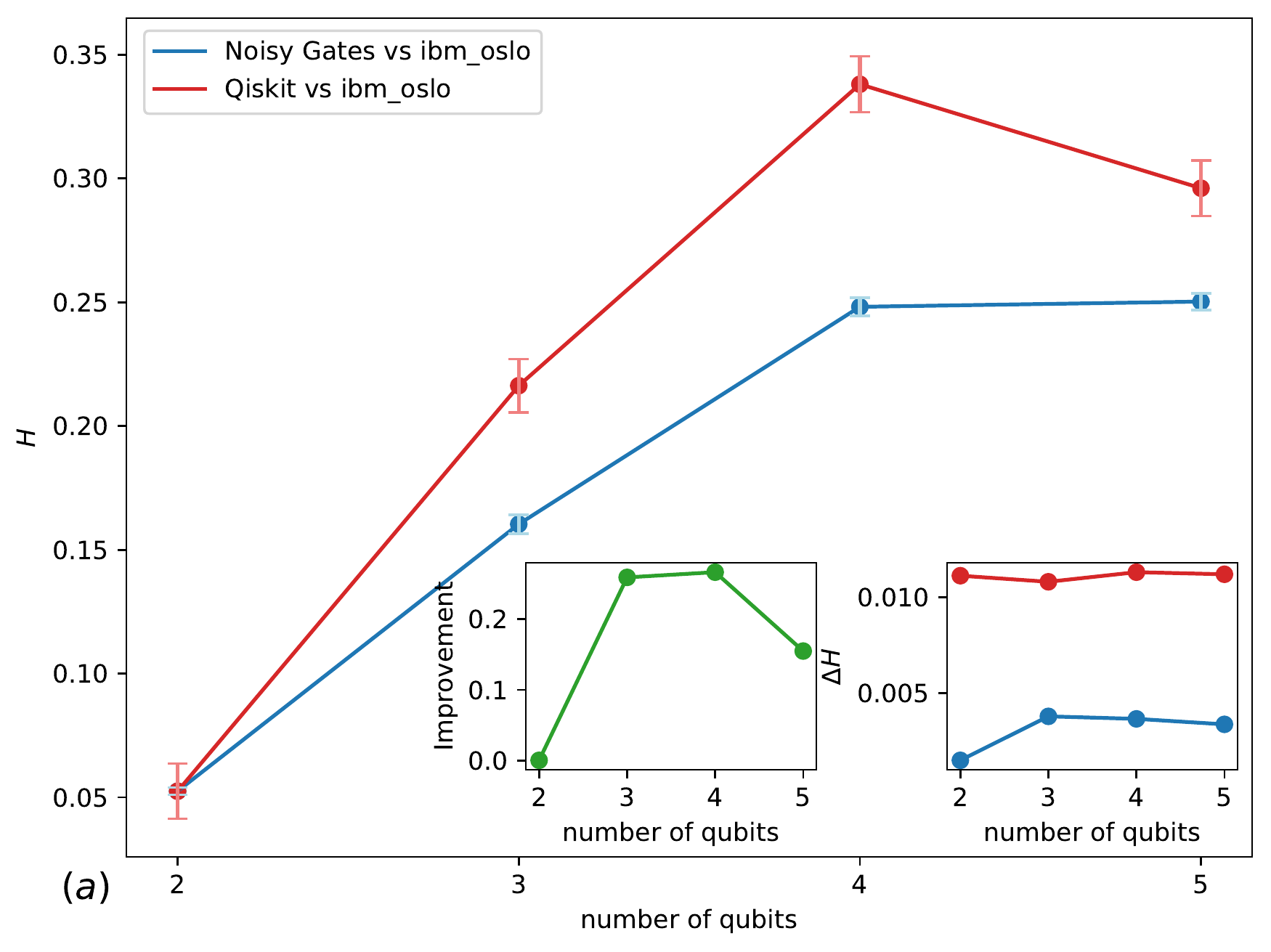}
    \end{minipage}%
    \hfill%
    \begin{minipage}{0.331\textwidth}
        \includegraphics[width=\textwidth]{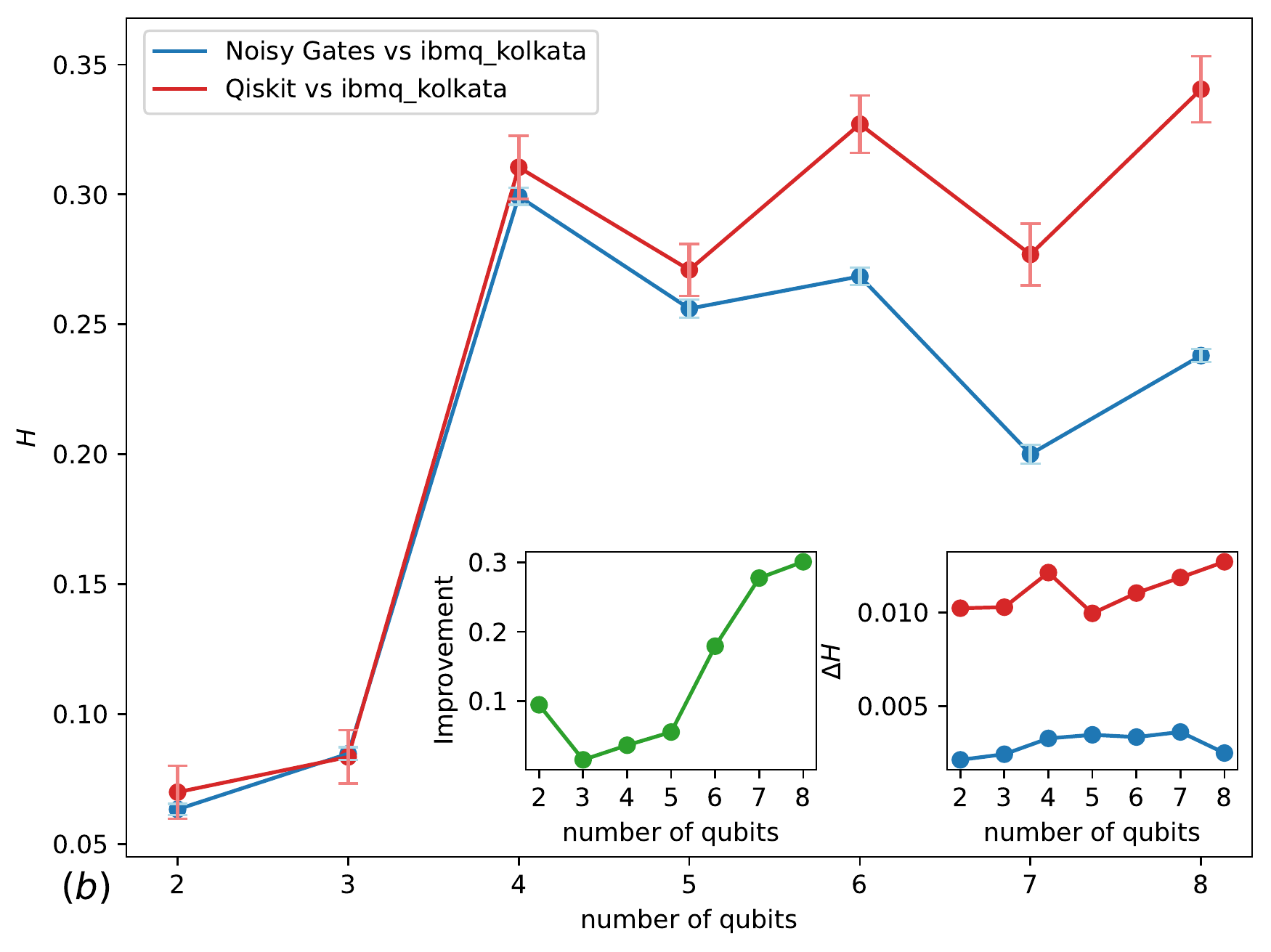}
    \end{minipage}
    \hfill%
    \begin{minipage}{0.331\textwidth}
        \includegraphics[width=\textwidth]{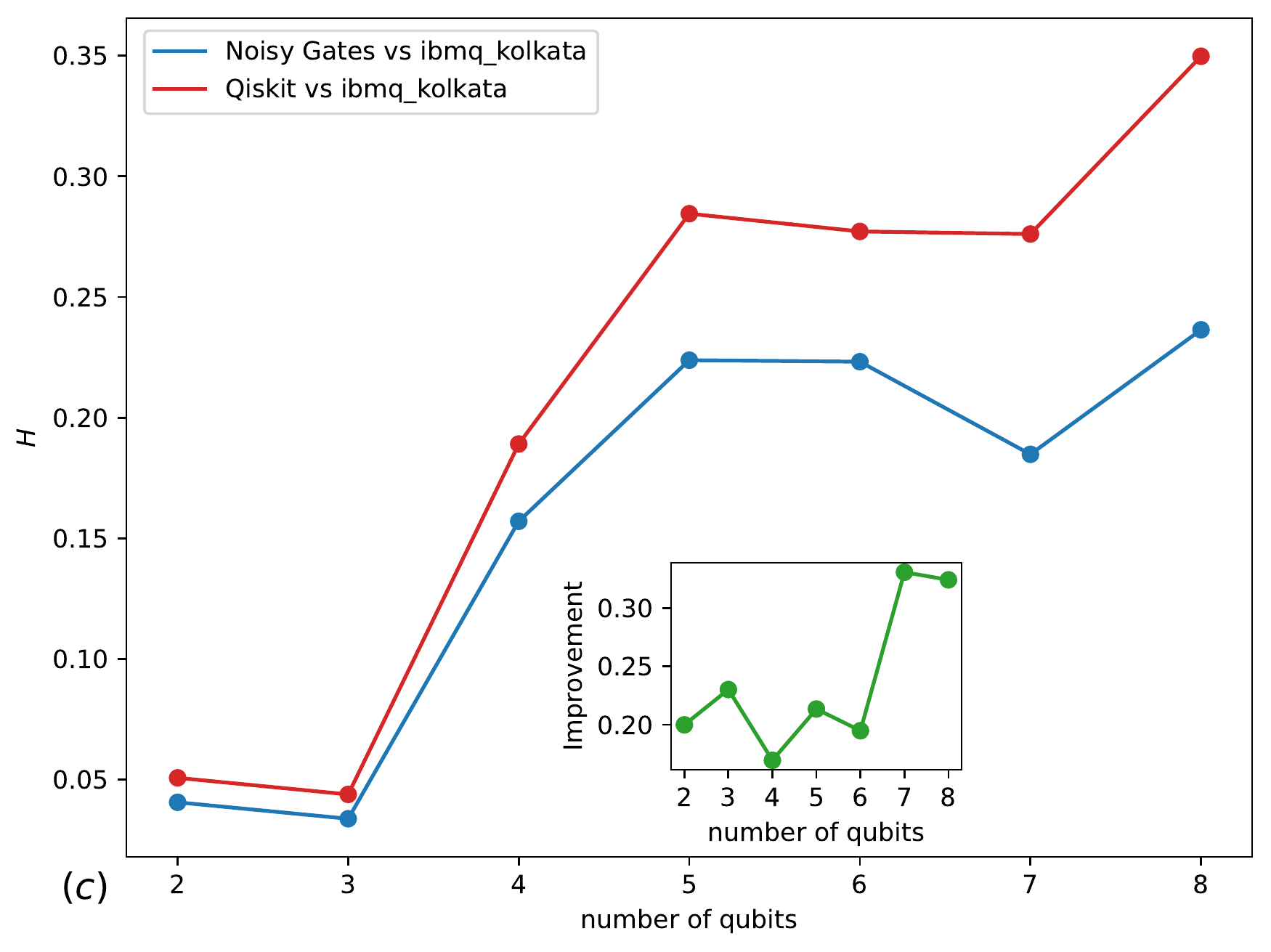}
    \end{minipage}
\caption{Quantum Fourier Transform. Panel (a) shows the Hellinger distances  between the noisy gate approach and $\text{ibmq\_oslo}$, and between the Qiskit simulator and $\text{ibmq\_oslo}$, when executing the QFT$^\dagger$ algorithm for $n = 2,\dots, 5$ qubits. Each value is the mean of $100$ independent simulations for the noisy gates, in blue, and for the Qiskit simulations, in red. The left inset shows the relative improvement, calculated as $|\Bar{\cl{H}}^{\text{ibm}}_{\chi}-\Bar{\cl{H}}^{\text{ng}}_{\chi}|/\Bar{\cl{H}}^{\text{ibm}}_{\chi}$, while the right inset shows the standard deviations as functions of the number of qubits. Panel (b) is the same as (a), with $\text{ibmq\_oslo}$ replaced by $\text{ibmq\_kolkata}$ and the number of qubit going up to 8. In panel (c) the comparison presented in (b) has been repeated a second time on $\text{ibmq\_kolkata}$; in this case,  only a single simulation of $1000$ samples is considered. The inset shows the relative improvement.}
\label{hellinger_QFT_oslo_kolkata}
\end{figure*}

\textit{QFT simulations}
The (QFT$^{\dagger}$) is a subroutine of many important quantum algorithms, as for example the Shor's algorithm \cite{shor1994algorithms,ruiz2017quantum}. An important feature of QFT$^{\dagger}$ is that the circuit for $n$ qubits is readily extendable to $n+1$ qubits; thus we can efficiently test the robustness of the method as the circuit's width and depth increase. We run QFT$^{\dagger}$ for $n = 2,\dots,5$ on $\text{ibmq\_oslo}$ and for $n = 2,\dots,18$ on $\text{ibmq\_kolkata}$. These devices are available on the cloud, comprising respectively 7 and 27 superconducting transmon qubits \cite{koch2007charge}, see appendix \ref{Device_parameters} for further details. We set as input of QFT$^{\dagger}$ the state $\ket{+}^{\otimes n}$, obtained by applying a layer of Hadamard gates on each qubit initialized in $\ket{0}$. In this way the ideal output of QFT$^{\dagger}$ should be $\ket{0}^{\otimes n}$. 
Runs on real quantum computers are performed by taking $1000$ shots, i.e. measurements. We also run the corresponding noisy gates and Qiskit simulations. (In appendix \ref{GHZ} we perform a similar analysis for the GHZ algorithm \cite{greenberger2009ghz}.) 

Implementing $\text{QFT}^{\dagger}$ circuit on ibmq devices requires to transpile the circuit into their native gate set. We have defined a custom noise model in Qiskit, by adding after each gate of the transpiled circuit the depolarizing and relaxation channels, and the SPAM channel before measurements, see appendix \ref{Krausmaps}. Similarly, in the noisy gates simulation each gate is replaced with its noisy version according to the noise model in section \ref{noise_models}. During idle-times of qubits we put the relaxation noise gates (see appendix \ref{spam_relax}) in order to take into account the stand-by times of the physical qubits; before measurements, we apply SPAM noise gates (see appendix \ref{spam_relax}) which accounts for read-out errors. In these simulations the CNOT gates inside the circuits are decomposed in terms of single-qubit and CR gates, as in ibmq devices. \\
\indent In order to measure the performance of different approaches in simulating the behaviour of the quantum computer, we look at their  distance with the outcomes of the real device; this is achieved by computing the Hellinger distance between the probability distributions, and it can be done without performing full tomography on the quantum states, which scales exponentially with the number of qubits and becomes unfeasible for the current simulations.


In Fig. \ref{hellinger_QFT_oslo_kolkata} (a) we plot the average values of $\cl{H}^{\text{ng}}_{\chi} = \cl{H}(\rho^{\text{ng}},\chi)$, $\cl{H}^{\text{ibm}}_{\chi}=\cl{H}(\rho^{\text{ibm}},\chi)$ as the number of qubits $n$ increases from $2$ to $5$, where now the diagonal elements of $\chi$ are the outcome probabilities of $\text{ibmq\_oslo}$. Fig. \ref{hellinger_QFT_oslo_kolkata} (b) displays again the average values of $\cl{H}^{\text{ng}}_{\chi} = \cl{H}(\rho^{\text{ng}},\chi)$, $\cl{H}^{\text{ibm}}_{\chi}=\cl{H}(\rho^{\text{ibm}},\chi)$ up to $8$ qubits, where now the diagonal elements of $\chi$ come from $\text{ibmq\_kolkata}$. As shown in Fig. \ref{hellinger_QFT_oslo_kolkata} (c) we compute again $\cl{H}^{\text{ng}}_{\chi}$, $\cl{H}^{\text{ibm}}_{\chi}$  to test the stability of $\text{ibmq\_kolkata}$ in different runs.

As in the previous section, we have run $100$ independent simulations, each including $1000$ samples, for both methods and for each $n$, in order to compute the standard deviations $\Delta\cl{H}^{\text{ng}}_{\chi}$, $\Delta\cl{H}^{\text{ibm}}_{\chi}$  shown in the insets of Fig. \ref{hellinger_QFT_oslo_kolkata}. Only for Fig. \ref{hellinger_QFT_oslo_kolkata} (c) we have run a single simulation of $1000$ samples, thus standard deviations are not present. We notice that for every $n$ we get $\Bar{\cl{H}}^{\text{ng}}_{\chi} < \Bar{\cl{H}}^{\text{ibm}}_{\chi}$ and $\Delta\cl{H}^{\text{ng}}_{\chi} < \Delta\cl{H}^{\text{ibm}}_{\chi}$. 
The relative improvement, shown in green in the insets of of Fig. \ref{hellinger_QFT_oslo_kolkata}, changes significantly between different devices and also for the same device but in different moments, namely with different noise parameters, meaning that the performances of such devices are not very stable. For example at $n = 3$, in the left panel the relative improvement is $\sim 25\%$, in the central panel it is $\sim 5\%$ and in the right panel it is $\sim 25\%$. The highest relative improvement obtained with the run on $\text{ibmq\_oslo}$ is $\sim 30\%$ and for runs on $\text{ibmq\_kolkata}$ is $\sim 35\%$.

The results show that our method is more accurate than existing ones. Actually, it reproduces the Lindblad dynamics better (Figures \ref{hellinger_X} and \ref{hellinger_CR}) than the  dynamics of the quantum devices. The reason, mentioned before, is that quantum devices are affected by additional and more complicated noises, which are not taken into account by the noise model we are using; we stress again that to find a better noise model is not the scope of this work, and will be subject of future research.
\begin{figure}[ht!]
\centering
\includegraphics[width=0.7\linewidth]{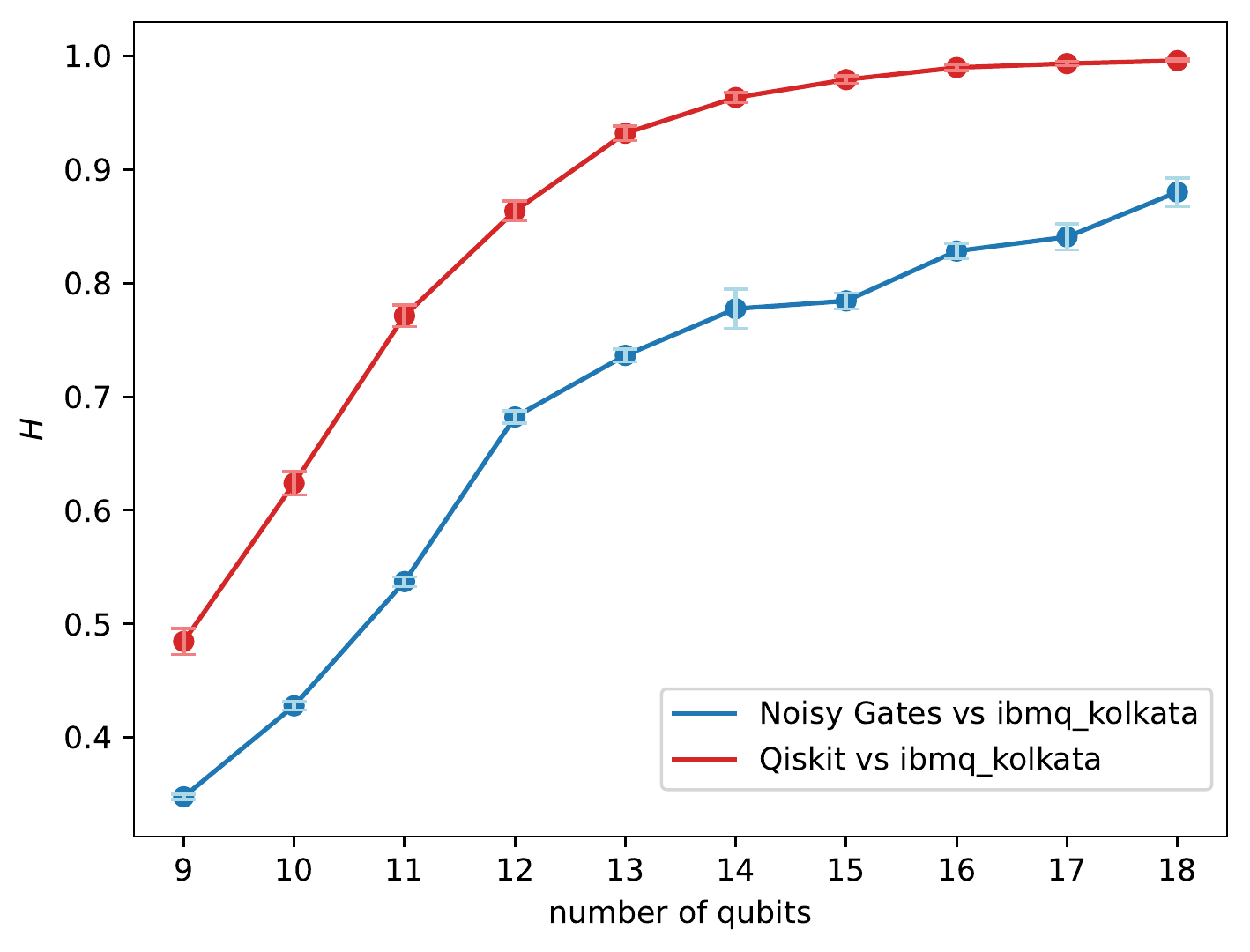} 
\caption{Mean Hellinger distances $\Bar{\cl{H}}^{\text{ng}}_{\chi}$ and $\Bar{\cl{H}}^{\text{ibm}}_{\chi}$ and their standard deviations $\Delta\cl{H}^{\text{ng}}_{\chi}$ and $ \Delta\cl{H}^{\text{ibm}}_{\chi}$ from $n = 9$ to $n = 18$ qubits for the QFT$^\dagger$ executed on $\text{ibmq\_kolkata}$. Since for $n \geq 9$ the depth of the circuit is such that noises make the resulting probability distribution very flat, $1000$ runs of the circuit on the quantum device, which returns a single computational basis state in each run, are not sufficient to reconstruct faithfully the probability distribution over the $2^n$ basis states; by increasing the number of qubits,  a 0 probability is associated to an increasing number of basis states. The simulator instead does does not suffer from this limitation: in each run, it returns a non-zero value for each possible output. Then the respective probability distributions differ more and more, and this is the reason why the  Hellinger distance rapidly increases. Despite the fact that the number of runs of the device is not sufficient to derive clear conclusions, we notice that the noisy gates approach still performs better than the standard one.} 
\label{hellinger_QFT_9_18}
\end{figure}

The simulations on $\text{ibmq\_kolkata}$ have been extended to $18$ qubits to test the computational scalability of the noisy gates simulator. In Fig. \ref{hellinger_QFT_9_18} we show the mean Hellinger distances $\Bar{\cl{H}}^{\text{ng}}_{\chi}$ and $\Bar{\cl{H}}^{\text{ibm}}_{\chi}$ and their standard deviations $\Delta\cl{H}^{\text{ng}}_{\chi}$ and $ \Delta\cl{H}^{\text{ibm}}_{\chi}$ from $n = 9$ to $n = 18$ qubits: simulations apparently become rapidly bad, since the Hellinger distance approaches 1, its maximum value. There is a clear reason behind that, which does not represent a limitation of our simulator. First of all, for such an high number of qubits, the depth of the transpiled circuit is so large that noises dominate~\footnote{e.g. for $n = 9$ around $50$ CNOTs are performed on almost every pair of connected qubits while for $n = 18$ more than $100$ CNOTs. As one can see in Figs. \ref{hellinger_CR} and \ref{hellinger_CX}, when more than $100$ CR or CNOT gates are applied, the total execution time is larger than the decoherence times. Thus, the state of the system approaches rapidly the maximally mixed state.} and the resulting probability distributions are very flat. Then, to recover a faithful  probability distribution over the $2^n$ possible outcomes by the quantum device, which returns a single outcome in each run,  the number of circuit runs must be significantly larger than $2^n$. Therefore, for $n \geq 9$ a number of runs equal to $1000$  is not sufficient (and increasing this number becomes soon impractical): the output distribution from the device is increasingly dominated by 0's, while our simulator returns (in general) a non-zero probability for each output state: this makes the Hellinger distances of Fig. \ref{hellinger_QFT_9_18} approach 1. Nevertheless, also in this case the noisy gates approach  performs better than the standard one, even if the number of runs of the quantum device are not enough to properly recover the full probability distribution.

As a final remark, we stress that we obtain better results with respect to Qiskit, despite the fact that we have chosen the simplest time dependent pulse shape in the Hamiltonians (see Eq. \eqref{single_qubit_Hamiltonian} and Eq. \eqref{cross_resonance}). 

\section{Conclusions and outlook}
We have developed a novel approach, called noisy quantum gates, to improve classical simulations of NISQ computers: it is based on integrating the noise into the gates, rather than keeping gates and noise as two separate dynamics. We have shown that our approach is very successful in simulating the Lindblad dynamics, with a relative improvement between 50\% and 90\% and more, 
compared with the standard gate-noise separation method. 

When compared against real quantum devices, the improvement fluctuates between 10\% and  30\%; this is largely due to the fact that the underlying noise model is too simple to accurately represent the dynamics of the device, as discussed in connection to the simulation of the CNOT gate. This is not a weakness of the noisy gate approach here presented, but of the underlying noise model, which we used since it is rather standard in the literature. 


There is a number of potential improvements that can be straightforwardly implemented; all of them require an update of the noise model, not of the simulation strategy, which is already very good. First of all, there are likely additional single-qubit errors which should be taken into account, for example those induced by the driving pulses. Secondly, in the present work we considered only non-correlated single-qubit errors, but the method can easily accommodate also correlated two-qubits errors \cite{wilen2021correlated,preskill2012sufficient} by introducing proper correlated noises into the stochastic equations. Another possible extension of the approach is to add in the Hamiltonians small interactions between adjacent qubits in order to mimic cross talk errors \cite{sarovar2020detecting,zhao2022quantum}. Last, the current version of the noisy gates approach relies on the Lindblad equation that works in the  Markovian limit; this is reflected in the fact that  we used stochastic equations based on white noises. The approach can  be generalized to non-Markovian dynamics by using colored noises, as already discussed in the literature in different contexts \cite{bassi2003stochastic,maniscalco2006non,strunz1999quantum,gambetta2004non}.

Furthermore, our approach is also useful for other purposes that go beyond plane error analysis. For example, the shape of the pulse in the driving Hamiltonians, (see Eq. \eqref{single_qubit_Hamiltonian} and Eq. \eqref{cross_resonance}), can affect the noise. In our work we chose for simplicity a rectangular shape, but usually in real devices different shapes can be used, for example Gaussian ones. Consequentially, a natural application of our approach is error mitigation \cite{cai2022quantum, endo2018practical}, by optimizing the parameters of the pulse in order to minimize the effect of the noise \cite{liang2022variational,ibrahim2022pulse,greenaway2022variational}; the optimization can be performed for example by exploiting machine learning techniques, to find the best pulse parameters, which can be  tested  on real quantum hardware.

In this work we specified our approach to the native gate set and noise model of IBM devices; clearly the approach is general and can be used to describe in principle any NISQ platform.
\begin{acknowledgments}
\noindent
G.D.B. and M.V. thank A. Gundhi, L. Pintucci and L.L. Viteritti for useful discussions. F.C., G.D.B. and M.V. acknowledge the financial support from University of Trieste and INFN; F.C. acknowledges also the financial support of PON Ricerca e Innovazione 2014-2020 (D.M. 1061, 10.08.21) and QTI (Quantum Telecommunications Italy).  M.G. is supported by CERN through the CERN Quantum Technology Initiative.  S.D. acknowledges the financial support from INFN. A.B. acknowledges financial support from the EIC Pathfinder project QuCoM (GA no. 101046973), the PNRR PE National Quantum Science and Technology Institute (PE0000023), the University of Trieste and INFN.
\end{acknowledgments}

\appendix

\section{Derivation of the approximate solution}\label{derivations_approximate}
In this appendix, let us show how the approximate solution in Eq. \eqref{noisy_gate} to Eq. \eqref{ITO_eq} can be rigorously derived to order $\cl{O}(\epsilon^2)$. We propose two different methods.
\subsection{Perturbative expansion in the interaction picture}
As a first proof, let us perform the stochastic unraveling in the interaction picture, hence defining the state quantum trajectory at any time as $\ket{\psi_s}=\rr{U}_s\ket{\phi_s}$, where the state vector $\ket{\phi_s}$ is the solution, at time $s$, of the It\^o equation
\begin{equation}
    \rr{d}\ket{\phi_s}=\bigg[ i\epsilon\sum_{k=1}^{N^2-1}\rr{d}\rr{W}_{k,s}\rr{L}_{k,s} -\frac{\epsilon^2}{2}\sum_{k=1}^{N^2-1}\rr{d}s \rr{L}^\dag_{k,s}\rr{L}_{k,s} \bigg]\ket{\phi_s};
\end{equation}
here, $\rr{d}\rr{W}_{k,s}$ are defined as in the main text, and we defined the jump operators in the interaction picture, $\rr{L}_{k,s}:=\rr{U}^\dag_s\rr{L}_k\rr{U}_s$. Then, by dividing the time interval $s\in[0,1]$ in infinitesimal steps of width $1/M$ and taking the limit $M\rightarrow \infty$, formally the solution to Eq. \eqref{ITO_eq} can be written as
\begin{equation}
\rr{N}=\rr{U}_g\lim_{M\rightarrow\infty}\rr{N}_M,
\end{equation}
where we defined $\rr{N}_M:=\prod_{m=0}^{M-1}\exp\big[ \epsilon \rr{B}_m +\frac{\epsilon^2}{2}\rr{A}_m \big]$, with 
\begin{equation}
    \rr{A}_m=-\frac{1}{M}\sum_{k=1}^{N^2-1} \bigg[ \rr{L}^\dag_{k,m/M}\rr{L}_{k,m/M}-\rr{L}^2_{k,m/M} \bigg]
\end{equation}
and
\begin{equation}
    \rr{B}_m=i\sum_{k=1}^{N^2-1}\rr{L}_{k,m/M}\int_{m/M}^{(m+1)/M}\rr{d}\rr{W}_{k,s}.
\end{equation}
For general purposes (and, in particular, for ours) $\rr{N}$ can not be calculated analytically; hence, we show how to obtain a general form to the second order in $\epsilon$ (i.e., to first order in $\lambda t_g$). First, let us prove that the following approximation holds:
\begin{equation}\label{claim}
\rr{N}_M=e^{\frac{\epsilon^2}{2}\mathcal{A}_M}e^{\epsilon\mathcal{B}_M+\frac{\epsilon^2}{2}\mathcal{C}_M}+\mathcal{O}(\epsilon^3),
\end{equation}
where we defined $\mathcal{B}_M=\sum_{k=0}^{M-1}\mathrm{B}_k$, $\mathcal{A}_M=\sum_{k=0}^{M-1}\mathrm{A}_k$, and
\begin{equation}
\mathcal{C}_M=\sum_{k=0}^{M-1}\sum_{j=0}^k\big[\mathrm{B}_k, \mathrm{B_j}\big]=\sum_{k=0}^{M-1}\big[\mathrm{B}_k, \mathcal{B}_k\big].
\end{equation}
The proof follows by induction. First, one can straightforwardly check that Eq. \eqref{claim} holds for $M=0$; then, suppose it holds for $\Tilde{M}=M-1$. Since by definition $\rr{N}_{M}=e^{\epsilon\mathrm{B}_{M}+\frac{\epsilon^2}{2}\mathrm{A}_{M}}\rr{N}_{M-1}$, applying the inductive hypothesis one can see that 
\begin{align}
\rr{N}_{M} & = 1+\epsilon\mathcal{B}_{M} + \frac{\epsilon^2}{2}\big[\mathcal{A}_{M}+\mathcal{B}_{M}^2+ \mathcal{C}_{M}	\big]+\mathcal{O}(\epsilon^3) \nonumber \\
& = e^{\frac{\epsilon^2}{2}\mathcal{A}_{M}}e^{\epsilon\mathcal{B}_{M}+\frac{\epsilon^2}{2}\mathcal{C}_{M}}+\mathcal{O}(\epsilon^3),
\end{align}
which concludes the proof. Then, inserting \eqref{claim} in the formal expression for $\rr{N}$, one can perform the limit $M\rightarrow \infty$, ending up with $\rr{N}=\rr{U}_g e^{\Lambda}e^{\Xi}$, where we defined
\begin{equation}
\Lambda=-\frac{\epsilon^2}{2}\int_0^1\rr{d}s\sum_{k=1}^{N^2-1}\big[ \rr{L}^\dag_{k,s}\rr{L}_{k,s} -\rr{L}^2_{k,s} \big] 
\end{equation}
and
\begin{equation}
    \Xi=i\epsilon\sum_{k=1}^{N^2-1}\int_0^1\rr{d}\rr{W}_{k,s}\rr{L}_{k,s}-\frac{\epsilon^2}{2}\cl{C};
\end{equation}
here, $\cl{C}=\sum_{k,l=1}^{N^2-1}\int_0^1\rr{d}\rr{W}_{k,s}\int_0^s\rr{d}\rr{W}_{l,s'}\big[ \rr{L}_{k,s},\rr{L}_{l,s'} \big]$. As explained in the main text, this term can actually be dropped at second order in $\epsilon$, leading to the expressions given in \eqref{Lambda} and \eqref{Xi}. 

\subsection{Small noise expansion}
A second approach makes use of a perturbative method known as \textit{small noise expansion} or \textit{asymptotic perturbative expansion} \cite{gardiner1985handbook}. For simplicity, let us consider the SDE with one single Lindblad operator,
 \begin{equation}\label{perturbeq}
    d\ket{\psi_s} =\bigg[-\frac{i}{\hbar}\rr{H}_s \rr{d} s + i\epsilon \rr{L} \rr{d} \rr{W}_s-\frac{\epsilon^2}{2} \rr{L}^\dag\rr{L} \rr{d} s\bigg]\ket{\psi_s},
\end{equation}
the generalization to $N^2-1$ Lindblad operators being straightforward, and let us set the following ansatz:
\begin{equation}
    \ket{\psi_s}= \ket{\psi_s^0}+\epsilon\ket{\psi_s^1}+\epsilon^2\ket{\psi_s^2}+\dots
\end{equation}
Substituting this ansatz into Eq.(\ref{perturbeq}) and equating terms with the same power of $\epsilon$, up to second order we get a system of SDEs:
\begin{align}
& \rr{d} \ket{\psi_s^0} = -\frac{i}{\hbar}\rr{H}_s\ket{\psi_s^0}\rr{d} s \nonumber\\
& \rr{d} \ket{\psi_s^1} = -\frac{i}{\hbar}\rr{H}_s\ket{\psi_s^1}\rr{d} s +i L\ket{\psi_s^0}\rr{d} \rr{W}_s\nonumber\\
& \rr{d} \ket{\psi_s^2} = -\frac{i}{\hbar}\rr{H}_s\ket{\psi_s^2}\rr{d} s +i L\ket{\psi_s^1}\rr{d} \rr{W}_s -\frac{1}{2}L^{\dag}L\ket{\psi_s^0}\rr{d} s,
\end{align}
which must be solved with the initial conditions $\ket{\psi_0^0}=\ket{\psi_0}$. The zeroth order differential equation is the deterministic one given by the Hamiltonian evolution alone, hence its solution is simply $\ket{\psi_s^0}=\rr{U}_s\ket{\psi_0}$. The first order SDE is an example of a time-dependent Ornstein-Uhlenbeck process \cite{gardiner1985handbook}: the solution is  
\begin{equation}
\ket{\psi_s^1}=i\rr{U}_s \rr{S}_s\ket{\psi_0},    
\end{equation}
 where we defined $\rr{S}_{s}=\int_0^s\rr{d}\rr{W}_\tau\rr{L}_\tau$. Finally, the solution to the second order SDE is 
\begin{equation}\label{secondordersol}
    \ket{\psi_s^2}=-\rr{U}_s\int_0^s\Big[\frac{1}{2}\rr{L}_s^\dag\rr{L}_s\rr{d}s+\rr{L}_s\rr{S}_s\rr{d}\rr{W}_s  \Big]\ket{\psi_0},
\end{equation}
where $\rr{L}_s=\rr{U}_s^\dag\rr{L}\rr{U}_s$. Then, the solution at order $\epsilon^2$ is given by $\ket{\psi_1}=\rr{N}\ket{\psi_0}+\cl{O}(\epsilon^3)$, where the evolution operator is $\rr{N}=\rr{U}_g\rr{N}'$, with
\begin{equation}\label{evolutionnn}
    \rr{N}'=\bigg[\mathbb{1}+\epsilon\rr{S}_1-\epsilon^2\int_0^1\Big[ \frac{1}{2} \rr{L}_s^\dag\rr{L}_s\rr{d}s + \rr{L}_s\rr{S}_s\rr{d}\rr{W}_s \Big]\bigg].
\end{equation}
In order to evaluate the solution in the form given in the main text, we make use of the following equality:
\begin{equation}
    \int_0^{\tau}\rr{d}\rr{W}_s \rr{L}_s\rr{S}_s = \frac{1}{2}\bigg[ \rr{S}^2_s+\int_0^{\tau} \rr{d}\rr{W}_s [\rr{L}_s,\rr{S}_s] -\int_0^{\tau} \rr{d} s \rr{L}^2_s\bigg]
\end{equation}
obtained by using the It\^o rule \cite{gardiner1985handbook} for each entry of the stochastic matrices. Substituting this expression into Eq.\eqref{evolutionnn}, we get to second order:
\begin{equation}
\begin{split}
    &\rr{N}'  =  \mathbb{1}+i\epsilon\rr{S}_1 -\frac{\epsilon^2}{2}\bigg[\rr{S}_1^2+\int_0^1\rr{d}s\Big[\rr{L}_s^\dag-\rr{L}_s\Big]\rr{L}_s +\cl{C} \bigg] = \nonumber \\
    & \;\;\;\;\;\;\;\; =  e^{\Lambda}e^{\Xi} + \cl{O}(\epsilon^3),
\end{split}    
\end{equation}
where $\Lambda$, $\Xi$ and $\mathcal{C}=\int_0^1\rr{d}\rr{W}_s\big[\rr{L}_s,\rr{S}_s\big]$ are the same quantities defined in the main text.

\section{Comparison of the approximations}\label{comp}

We focus on the main differences between the standard approximation made in error analysis against the one considered in the noisy gates approach. 

Given the following Lindblad master equation
\begin{equation}\label{mastereq_schr}
    \frac{\D}{\D t} {\rho}_{t} = -\frac{i}{\hbar}[H_{t},\rho_{t}] +\gamma\mathfrak{L}\left[\rho_{t}\right],
\end{equation}
where $\mathfrak{L}\left[\rho_{t}\right]= L\rho_{t}L^{\dagger} - \frac{1}{2}\{L^{\dagger}L,\rho_{t}\}$, let's move to the interaction picture by defining $\chi_{t} = U_{t,t_{0}}^{\dagger}\rho_{t}U_{t,t_{0}}$ and $\chi_{t_{0}} = \rho_{t_{0}}$. Then
\begin{equation}\label{mastereq_int}
    \frac{\D}{\D t} {\chi}_{t} = \gamma\mathfrak{L}(t)\left[\chi_{t}\right],
\end{equation}
where  $\mathfrak{L}(t)\left[\chi_{t}\right] = U_{t,t_{0}}^{\dagger}\mathfrak{L}\left[\rho_{t}\right]U_{t,t_{0}}$.
\\

The formal solution of Eq. \eqref{mastereq_int} is
\begin{equation}
\chi_{t} = \text{T}\biggl[e^{\gamma\int_{t_{0}}^{t} \D s\mathfrak{L}(s)}\biggr]\chi_{t_{0}},
\end{equation}
where $\text{T}\left[\cdot\right]$ is the time ordering. Thus in the Schr\"odinger picture we can write the formal solution of Eq. \eqref{mastereq_schr} as

\begin{equation}
\label{formal}
\rho_{t} = U_{t,t_{0}}\text{T}\biggl[e^{\gamma\int_{t_{0}}^{t} \D s\mathfrak{L}(s)}\biggr]\rho_{t_{0}}U_{t,t_{0}}^{\dagger}.
\end{equation}

- \emph{Standard approximation} - The main approximation that can be found in the literature is to separate the Hamiltonian dynamics from the noise one \cite{nielsen2000quantum,benenti2019principles}. This choice is based on the observation that in general in quantum devices $\omega >> \gamma$, where $\omega$ is the pulse frequency of the Hamiltonian. Thus, the noise dynamics can be seen as frozen with respect to the faster Hamiltonian one. It means that in Eq. \eqref{formal} one assumes
\begin{equation}
\mathfrak{L}(t) \simeq \mathfrak{L}
\end{equation}
getting
\begin{equation}
\label{standard}
\rho_{t} \simeq U_{t,t_{0}}e^{\gamma\mathfrak{L}\cdot(t-t_{0})}\rho_{t_{0}}U_{t,t_{0}}^{\dagger}.
\end{equation}
We notice that indeed in Eq. \eqref{standard} the two dynamics are independent.
\\

- \emph{Noisy gates approximation} - Also in this case the approximation is based on $\omega >> \gamma$, but we assume that $\gamma$ is not small enough to completely separate the dynamics. An example of this can be seen in the devices of IBM where the noise evolution can be influenced in a non-negligible manner by the pulse of the drive Hamiltonian \cite{carvalho2021error,alexander2020qiskit}. Thus we make a first order approximation over $\gamma$ in Eq. \eqref{formal}
\begin{equation}
\text{T}\biggl[e^{\gamma\int_{t_{0}}^{t} \D s\mathfrak{L}(s)}\biggr] \simeq 1 + \gamma \int_{t_{0}}^{t} \D s\mathfrak{L}(s),
\end{equation}
and we get
\begin{equation}
\label{ng_approx}
\rho_{t} \simeq U_{t,t_{0}}\biggl(1+\gamma\int_{t_{0}}^{t} \D s\mathfrak{L}(s)\biggr)\rho_{t_{0}}U_{t,t_{0}}^{\dagger}.
\end{equation}
In Eq. \eqref{ng_approx} the noise depends on the Hamiltonian dynamics through $\mathfrak{L}(s)$. We stress that the perturbative solution of the SDE in the noisy gates model reproduce density matrices of the form of Eq. \eqref{ng_approx}.

\section{Kraus maps used in Qiskit simulations}\label{Krausmaps}
The error channels that we included in the custom Qiskit noise model are a composition of depolarization and relaxation after the gates and bitflip before measurements. With relaxation we mean the amplitude and phase damping channel. In this appendix we show the corresponding Kraus maps for these channels that are used in Qiskit simulations through Alg. \ref{alqiskit} in section \ref{algorithms_comp}.

\subsection{State Preparation and measurement (SPAM)}\label{spamKraus}

This kind of error is usually described as a bit flip channel that acts on a single qubit \cite{benenti2019principles}. Hence, its Kraus representation reads:
\begin{equation}\label{krausbitflip}
\mathcal{E}(\rho) = (1-p) \rho + p X\rho X,
\end{equation}
where $\rho$ is the density matrix of a single qubit, $X$ is the x-Pauli matrix and $p$ is the probability of having a flip of the states of the computational basis. The probability $p$ that we used in the simulations is the readout error provided as a calibration parameter for IBM devices, see appendix \ref{Device_parameters}.

\subsection{Depolarization}\label{Depolarization}

Depolarization drives the qubit towards the maximally mixed state \cite{benenti2019principles} and models incoherent gate infidelities. Its Kraus representation reads:
\begin{equation}\label{krausdamping}
\mathcal{E}(\rho) = \big(1-\frac{3}{4} p\big) \rho + \frac{p}{4} X\rho X + \frac{p}{4} Y\rho Y + \frac{p}{4} Z\rho Z,
\end{equation}
where $\rho$ is the density matrix of a single qubit, $X,Y,Z$ are the Pauli matrices and $p/4$ is the equal probability of having a bit flip, a phase flip or a bit and phase flip of the states of the computational basis. The probability $p$ that we used in the simulations is the gate error provided as a calibration parameter for IBM devices, see appendix \ref{Device_parameters}.

\subsection{Amplitude and phase damping (Relaxation)}\label{relaxationKraus}
The amplitude-damping channel describes the decay $\ket{1}\rightarrow\ket{0}$ due to the interaction with the environment; on the other hand, phase-damping represents the process in which phase coherences decay over time. Here we briefly call relaxation the combination of both effects. The Kraus representation is given by \cite{benenti2019principles,nielsen2000quantum}
\begin{equation}
\label{relaxKraus}
\mathcal{E}(\rho) = K\rho K + p_{1}\sigma^{+} \rho \sigma^{-} + p_{z}\mathcal{P}_{1}\rho\mathcal{P}_{1},
\end{equation}
where we defined 
\begin{equation}
    K = \begin{pmatrix} 1 & 0 \\
0 & \sqrt{1 - p_{1} - p_{z}} \end{pmatrix};
\end{equation}
as usual, $\sigma^{+} = \ket{0}\bra{1}$, $\sigma^{-} = \ket{1}\bra{0}$ and $\mathcal{P}_{1} = \ket{1}\bra{1}$. Moreover, $p_{1} = 1 - e^{-t/T_{1}}$ is the probability of amplitude damping, $T_{1}$ being the relaxation time (the time it takes for the qubit to decay in the ground state), and $p_{z} = (1 - p_{1})p_{pd}$, where $p_{pd} = 1 - e^{-t/T_{pd}}$ and $T_{pd} = T_{1}T_{2} / (2T_{1} - T_{2})$, $T_{2}$ being the decoherence time. We mention that the time scales $T_{1}$ and $T_{2}$ are related as $T_{2} \le 2T_{1}$.
The times $T_1$ and $T_2$ that we used in the simulations are directly provided as calibration parameters for IBM devices, see appendix \ref{Device_parameters}.

\section{Noise gates for Spam and Relaxation on idle qubits}\label{spam_relax}
In this section we address SPAM and relaxation noises on idle qubits, where the corresponding noise gates can be derived exactly \cite{bassi2008noise,jacobs2010stochastic,jacobs1998linear}. We do not consider depolarization error on idle qubits, because this channel is used to model incoherent gate infidelities.

\subsection{Noise gate for SPAM}\label{spam}

The Kraus map of SPAM is in Eq.\eqref{krausbitflip} of appendix \ref{Krausmaps}.
Assuming a behaviour in time of the form $p = (1 - e^{-2t/T})/2$ for a characteristic time $T=\gamma^{-1}$, one gets the corresponding Lindblad master equation
\begin{equation}
\label{lindbladbitflip}
\frac{\D}{\D t} \rho_{t} = \gamma(X\rho_{t}X - \rho_{t}).
\end{equation}
The associated stochastic differential equation is
\begin{equation}
\label{SDEbitflip}
\D \ket{\psi_{t}} = \biggl[i\sqrt{\gamma}X\D W_{t} - \frac{\gamma}{2} \D t\biggr]\ket{\psi_{t}}.
\end{equation}
This equation is analytically solvable with standard methods \cite{gardiner1985handbook,jacobs2010stochastic}, and thus we can exactly evaluate the corresponding noise gate as 
\begin{equation}
\label{SPAMgate}
\begin{aligned}
&N^{\mbox{\tiny{\text{SPAM}}}}(t,t_{0}) = e^{i\sqrt{\gamma}X\bar{W}(t,t_{0})}, 
\end{aligned}
\end{equation}
where $\bar{W}(t,t_0):=\int_{t_0}^t\D W_s$. In this case, the noise gate happens to be unitary, thus we can interpret it as a stochastic Schrödinger evolution due to the presence of the Wiener process $\bar{W}(t,t_{0})$. In the simulations we can directly sample  $\bar{W}(t,t_{0})$ from a Gaussian distribution with mean $\mathbb{E}[\bar{W}(t,t_{0})] = 0$ and variance $\mathbb{E}[\bar{W}^{2}(t,t_{0})] = t - t_{0}$.

\subsection{Noise gate for amplitude and phase damping (Relaxation)}\label{relaxation}

The Kraus map of the amplitude and phase damping is in Eq.\eqref{krausdamping} of appendix \ref{Krausmaps}.
Defining $\gamma_{1} =1/T_{1}$, $\gamma_{pd} = 1/T_{pd}$, the corresponding Lindblad equation is
\begin{equation}
\label{lindbladrelaxation}
\frac{\D}{\D t} \rho_{t} = \gamma_{1}\sigma^{+}\rho_{t}\sigma^{-} - \frac{\gamma_{1}}{2} \{\mathcal{P}_{1},\rho_{t}\} + \frac{\gamma_{pd}}{4}(Z\rho_{t}Z - \rho_{t}),
\end{equation}
and the stochastic term of the relative It\^o equation reads:
\begin{equation}
\label{SDErelaxation}
\D \mathcal{W} = i\sqrt{\gamma}_{1}\sigma^{+}\D W_{t,1} - \frac{\gamma_{1}}{2}\mathcal{P}_{1} \D t + i\sqrt{\frac{\gamma_{pd}}{4}}Z \D W_{t,2} - \frac{\gamma_{pd}}{8}\D t.
\end{equation}
With this stochastic term the It\^o equation is analytically solvable \cite{arnold1974stochastic} and we get the following non-unitary noisy gate
\begin{equation}
\label{relaxationgate1}
N^{\mbox{\tiny{\text{relax}}}}(t,t_{0}) = 
\begin{pmatrix} e^{i\alpha\bar{W}_{2}(t,t_{0})} & iS(t,t_{0})e^{i\alpha\bar{W}_{2}(t,t_{0})}\\
0 & e^{-\frac{\gamma_{1}}{2}(t - t_{0})}e^{-i\alpha\bar{W}_{2}(t,t_{0})} \end{pmatrix},
\end{equation}
where we defined for simplicity $\alpha:=\sqrt{\gamma_{pd}/4}$, and
\begin{equation}
S(t,t_{0}) = \sqrt{\gamma_{1}}\int_{t_{0}}^{t} e^{-\frac{\gamma_{1}}{2}(s - t_{0})}e^{-2i\alpha \bar{W}_{2}(s,t_{0})} \D W_{s,1}
\end{equation}
is a complex stochastic It\^o process. In principle, such a term is problematic in view of a simulation, since it is not easy to sample. To understand this, look for instance at the real part,
\begin{equation}
S_{R}(t,t_{0}) = \sqrt{\gamma_{1}}\int_{t_{0}}^{t} e^{-\frac{\gamma_{1}}{2}(s - t_{0})}\cos\biggl(2\alpha \bar{W}_{2}(s,t_{0})\biggr) \D W_{s,1};
\end{equation}
this is an It\^o integral of a stochastic function, and it is not easy to derive its probability distribution; thus, sampling $S(t,t_0)$ may be problematic. We can avoid such a difficulty by adequately substituting $N^{\mbox{\tiny{\text{relax}}}}(t,t_{0})$ with some modified noisy gate, which is equivalent to the former once the average is carried out, in the sense that Eq. \eqref{relaxKraus} still holds even if the new noisy gate is not a solution of the unraveling \eqref{SDErelaxation} anymore. For instance, it is straightforward to verify that this holds for the following choice:
\begin{equation}
\label{relaxationgate2}
\tilde{N}^{\mbox{\tiny{\text{relax}}}}(t,t_{0}) = 
\begin{pmatrix} e^{i\alpha \bar{W}_{2}(t,t_{0})} & i\tilde{S}(t,t_{0})e^{-i\alpha \bar{W}_{2}(t,t_{0})}\\
0 & e^{-\frac{\gamma_{1}}{2}(t - t_{0})}e^{-i\alpha \bar{W}_{2}(t,t_{0})} \end{pmatrix},
\end{equation}
with the definition
\begin{equation}
\tilde{S}(t,t_{0}) = \sqrt{\gamma_{1}}\int_{t_{0}}^{t} e^{-\frac{\gamma_{1}}{2}(s - t_{0})} \D W_{s,1};
\end{equation}
i.e., one always has that
\begin{equation}
    \mathbb{E}\big[N^{\mbox{\tiny{\text{relax}}}}\ket{\psi}\bra{\psi}N^{\mbox{\tiny{\text{relax}}}\dag}\big]=\mathbb{E}\big[\tilde{N}^{\mbox{\tiny{\text{relax}}}}\ket{\psi}\bra{\psi}\tilde{N}^{\mbox{\tiny{\text{relax}}}\dag}\big].
\end{equation}
The difference is that now the process $\tilde{S}(t,t_{0})$ is just the It\^o integral of a deterministic function, hence we know that it must have a Gaussian statistics \cite{gardiner1985handbook}, which makes it more convenient for a simulation.

\section{Device parameters}\label{Device_parameters}
For the simulations in Sec. \ref{simulations} and in appendix \ref{GHZ} we used the device parameters provided by IBM. Here we report the average value of such parameters.

$\text{ibmq\_manila}$ contains $5$ fixed-frequency transmons qubits \cite{koch2007charge}, with median fundamental transition frequency of $4.962$ GHz and median anharmonicity of $-0.34358$ GHz. The median qubit lifetime $T_1$ of the qubits is $149.11\mu$s, the median coherence time $T_2$ is $44.43\mu$s and the median readout error is $0.0217$. The single qubit gate error varies between $1.975\times 10^{-4}$ and $6.138 \times 10^{-4}$, while the CNOT error varies between $7.072 \times 10^{-3}$ and $1.125 \times 10^{-2}$, depending on the specific connection. In the simulations that reproduce the Lindblad equations, parameters of qubits zero and one were used.
$\text{ibmq\_kolkata}$ contains $27$ fixed-frequency transmons qubits, with median fundamental transition frequency of $5.102$ GHz and median anharmonicity of $-0.34345$ GHz. The median qubit lifetime $T_1$ of the qubits is $127.39\mu$s, the median coherence time $T_2$ is $86.41\mu$s and the median readout error is $0.0132$. The single qubit gate error varies between $1.443\times 10^{-4}$ and $5.410\times 10^{-3}$, while the CNOT error varies between $4.214\times 10^{-3}$ and $1\times 10^{-2}$, depending on the specific connection. The qubits, which are used to run $\text{QFT}^{\dagger}$ algorithm, belong to the list $[$0,1,4,7,10,12,15,18,21,23,24,25,22,19,16,14,11,8,5,3,2$]$.
$\text{ibmq\_quito}$ contains $7$ fixed-frequency transmons qubits, with the median fundamental transition frequency of $5.164$ GHz and median anharmonicity of $-0.3315$ GHz. The median qubit lifetime $T_1$ of the qubits is $105.84\mu$s, the median coherence time $T_2$ is $84.05\mu$s and the median readout error is $0.044$. The single qubit gate error varies between $3.054\times 10^{-4}$ and $6.929\times 10^{-4}$, while the CNOT error varies between $9.682\times 10^{-3}$ and $1.463\times 10^{-2}$, depending on the specific connection. The qubits, which are used for CNOT gate sequence are $0$ and $1$.
$\text{ibmq\_oslo}$ contains $7$ fixed-frequency transmons qubits, with the median fundamental transition frequency of $5.046$ GHz and median anharmonicity of $-0.3429$ GHz. The median qubit lifetime $T_1$ of the qubits is $128.12\mu$s, the median coherence time $T_2$ is $58.57\mu$s and the median readout error is $0.0216$. The single qubit gate error varies between $1.648\times 10^{-4}$ and $6.698\times 10^{-4}$, while the CNOT error varies between $6.471\times 10^{-3}$ and $2.067\times 10^{-2}$, depending on the specific connection. The qubits, which are used to run $\text{GHZ}$ algorithm, belong to the list $[$0,1,3,5,4$]$.

\section{Plots of the fidelities of the X and CR gates Lindblad simulations}\label{fidelities}
\begin{figure*}[htp]
    \begin{minipage}{0.4\textwidth}
        \includegraphics[width=\textwidth]{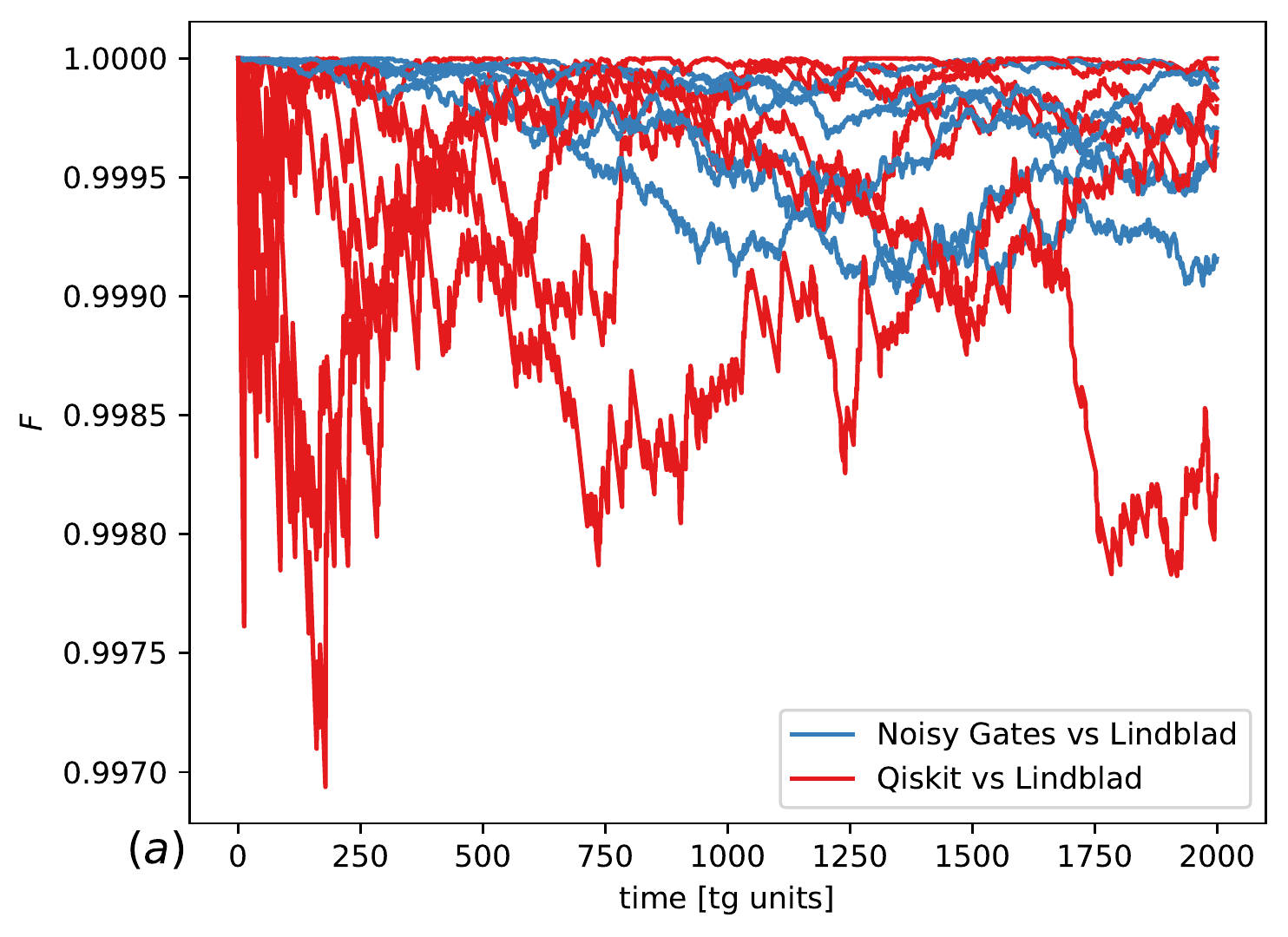}
    \end{minipage}%
    \begin{minipage}{0.4\textwidth}
        \includegraphics[width=\textwidth]{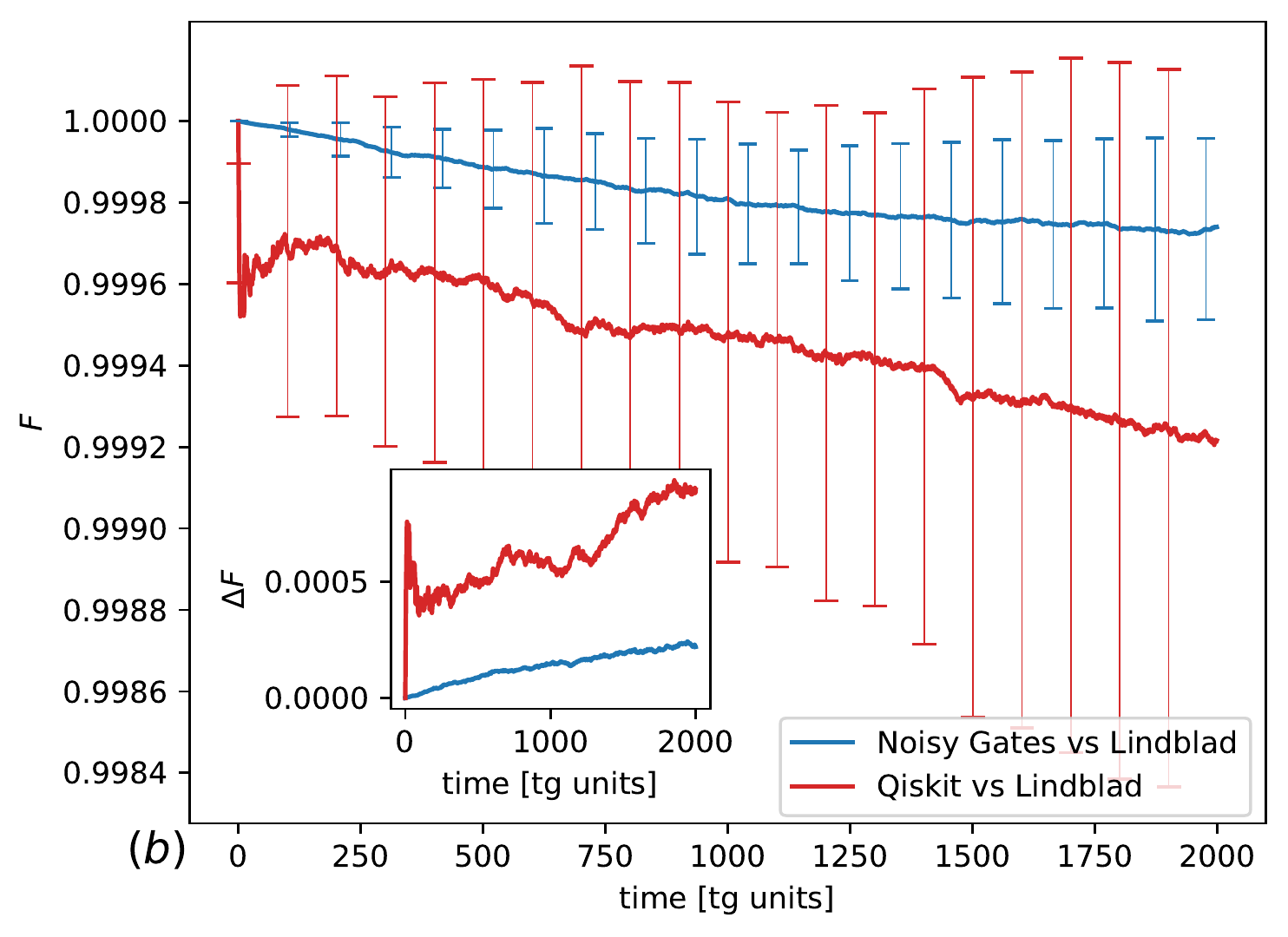}
    \end{minipage}
\caption{Fidelities $\cl{F}_{\sigma}^{\text{ng}}$, in blue, and $\cl{F}_{\sigma}^{\text{ibm}}$, in red, as a function of time, for a repetition of X gates. On panel (a), the fidelities obtained from $100$ independent runs of the two methods are pictured (for better readability only five are shown), where each simulation is obtained by averaging over $1000$ samples. On panel (b), the means $\Bar{\cl{F}}_{\sigma}^{\text{ng}}$, $\Bar{\cl{F}}_{\sigma}^{\text{ibm}}$ of the same simulations and their standard deviations $\Delta\cl{F}_{\sigma}^{\text{ng}}$, $\Delta\cl{F}_{\sigma}^{\text{ibm}}$ are displayed. The inset shows the standard deviations $\Delta\cl{F}_{\sigma}^{\text{ng}}$, $\Delta\cl{F}_{\sigma}^{\text{ibm}}$ as functions of time.}
\label{fidelities_X}
\end{figure*}

\begin{figure*}[htp]
    \begin{minipage}{0.4\textwidth}
        \includegraphics[width=\textwidth]{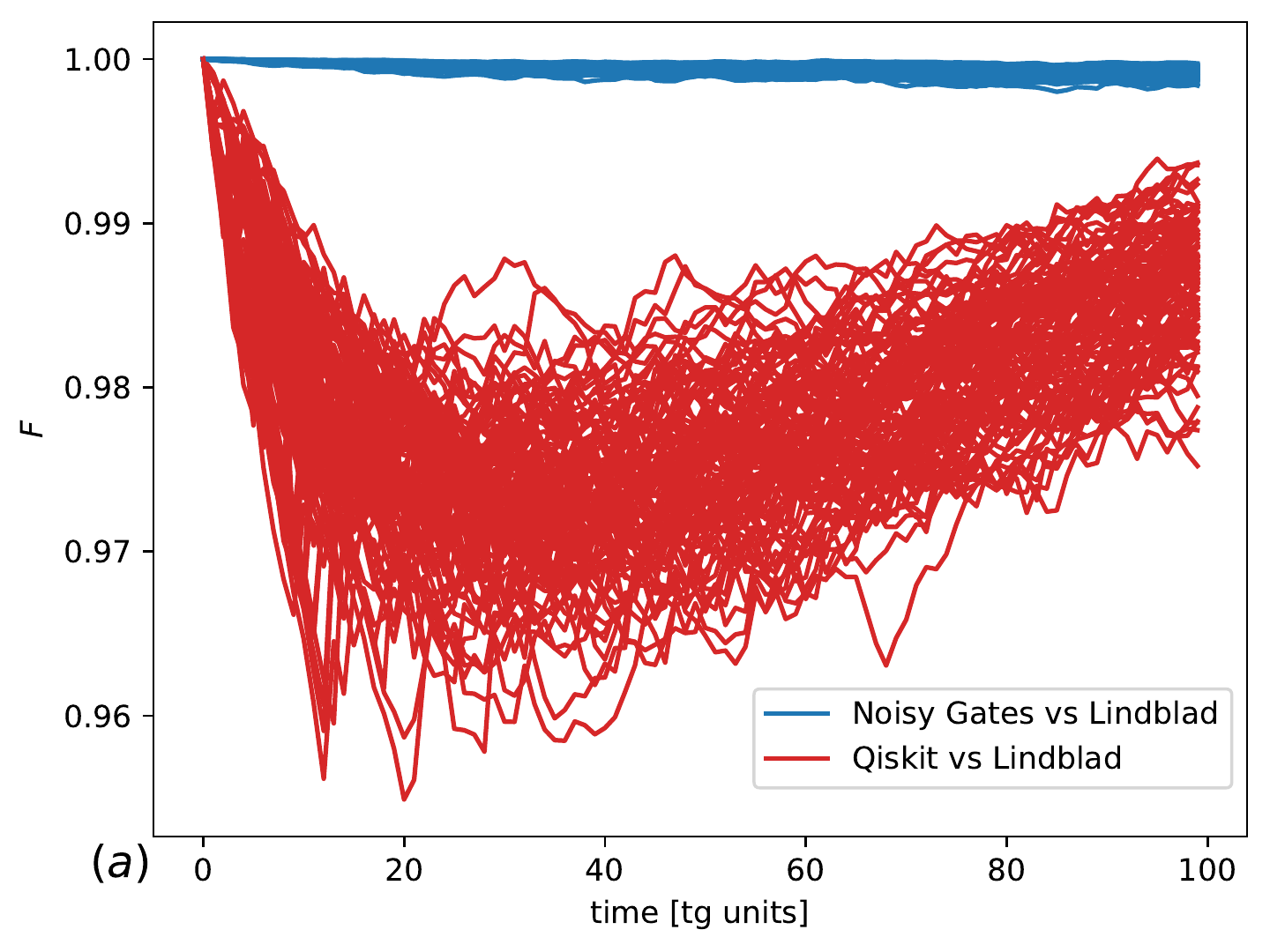}
    \end{minipage}%
    \begin{minipage}{0.4\textwidth}
        \includegraphics[width=\textwidth]{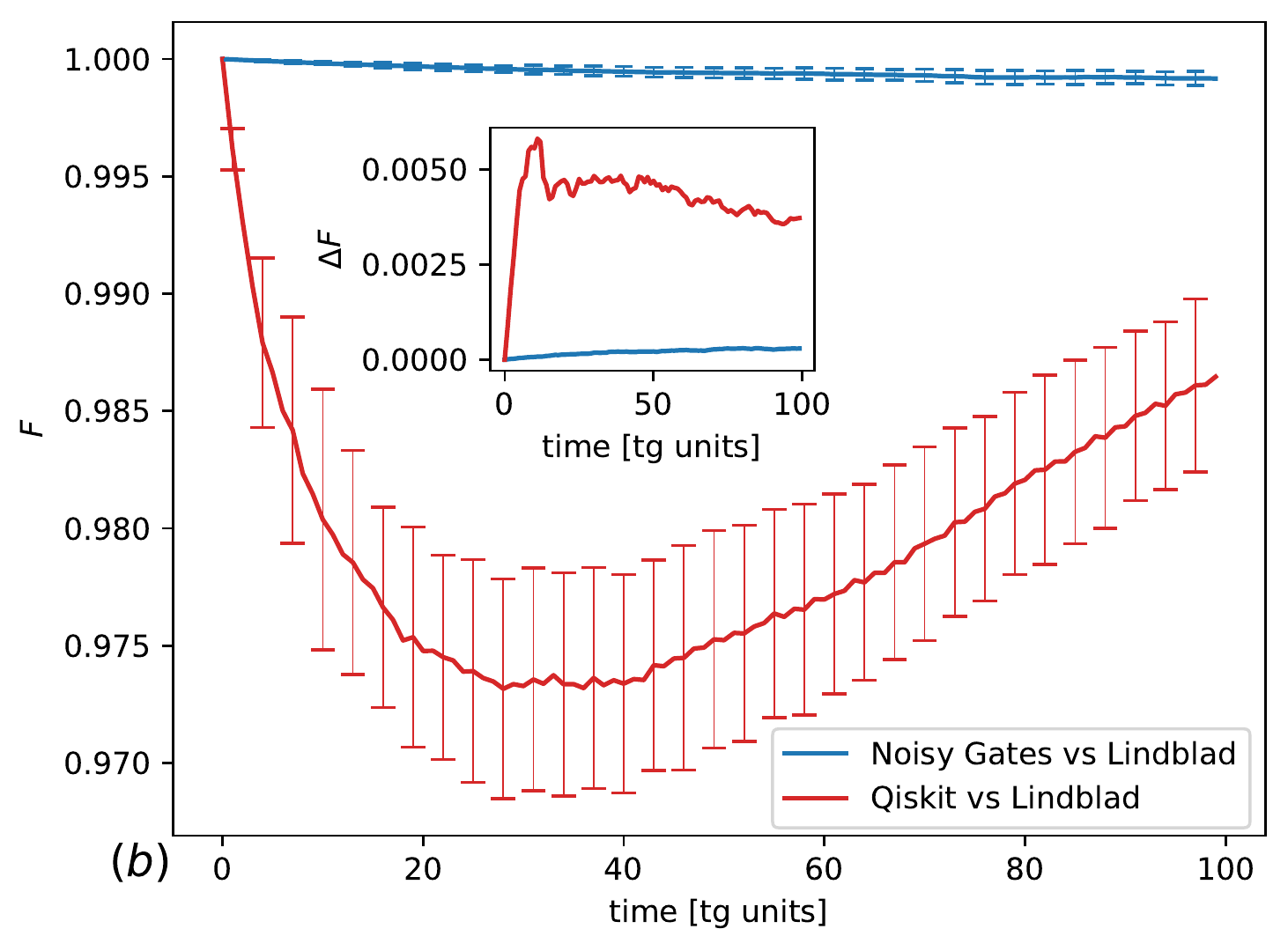}
    \end{minipage}
\caption{Fidelities $\cl{F}_{\sigma}^{\text{ng}}$, in blue, and $\cl{F}_{\sigma}^{\text{ibm}}$, in red, as a function of time, for a repetition of  CR gates. Panels (a) and (b) have the same meaning as for Fig.\ref{fidelities_X}.}
\label{CR_plot}
\end{figure*}

\begin{figure*}[htp]
    \begin{minipage}{0.4\textwidth}
        \includegraphics[width=\textwidth]{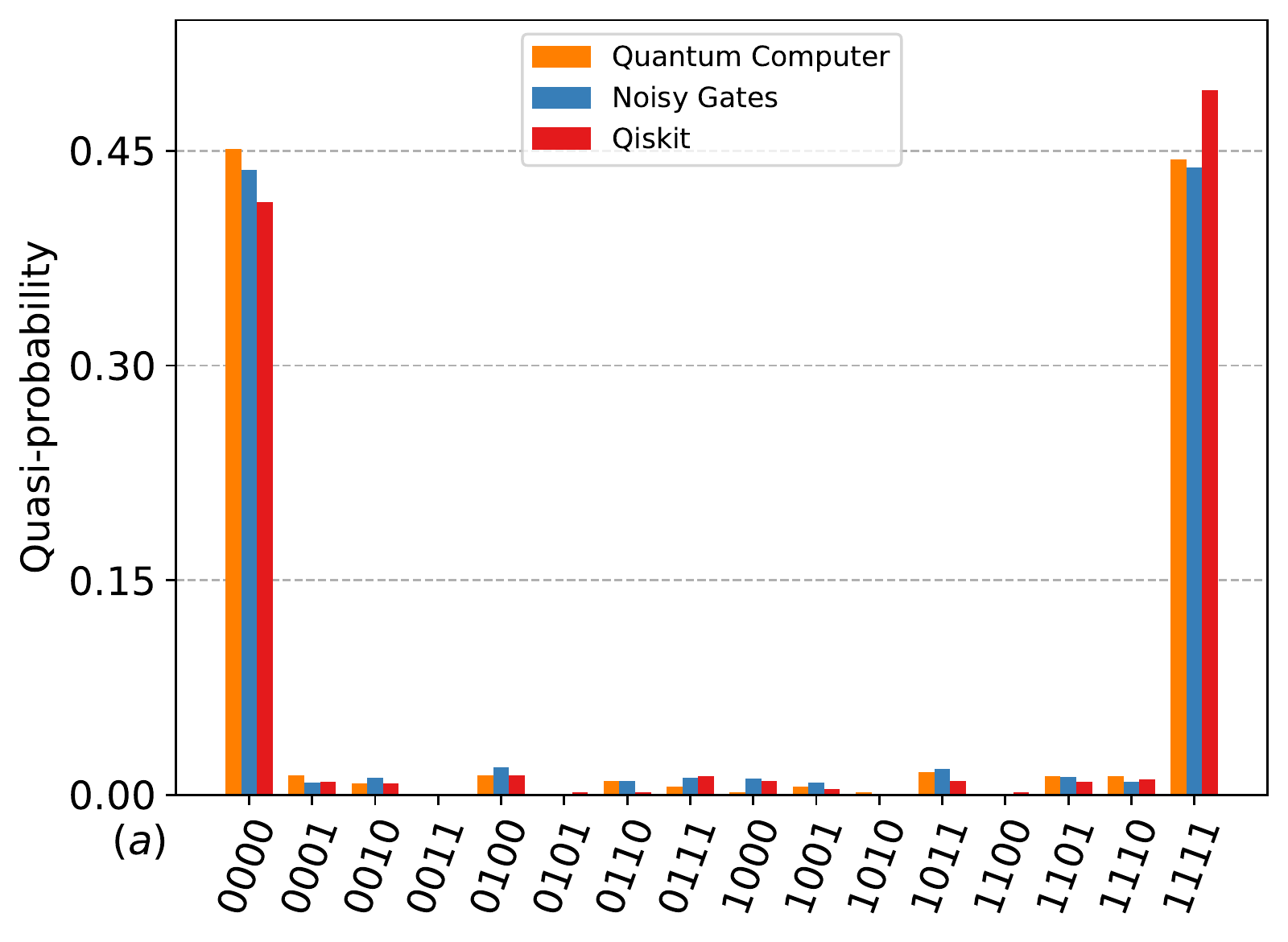}
    \end{minipage}%
    \begin{minipage}{0.4\textwidth}
        \includegraphics[width=\textwidth]{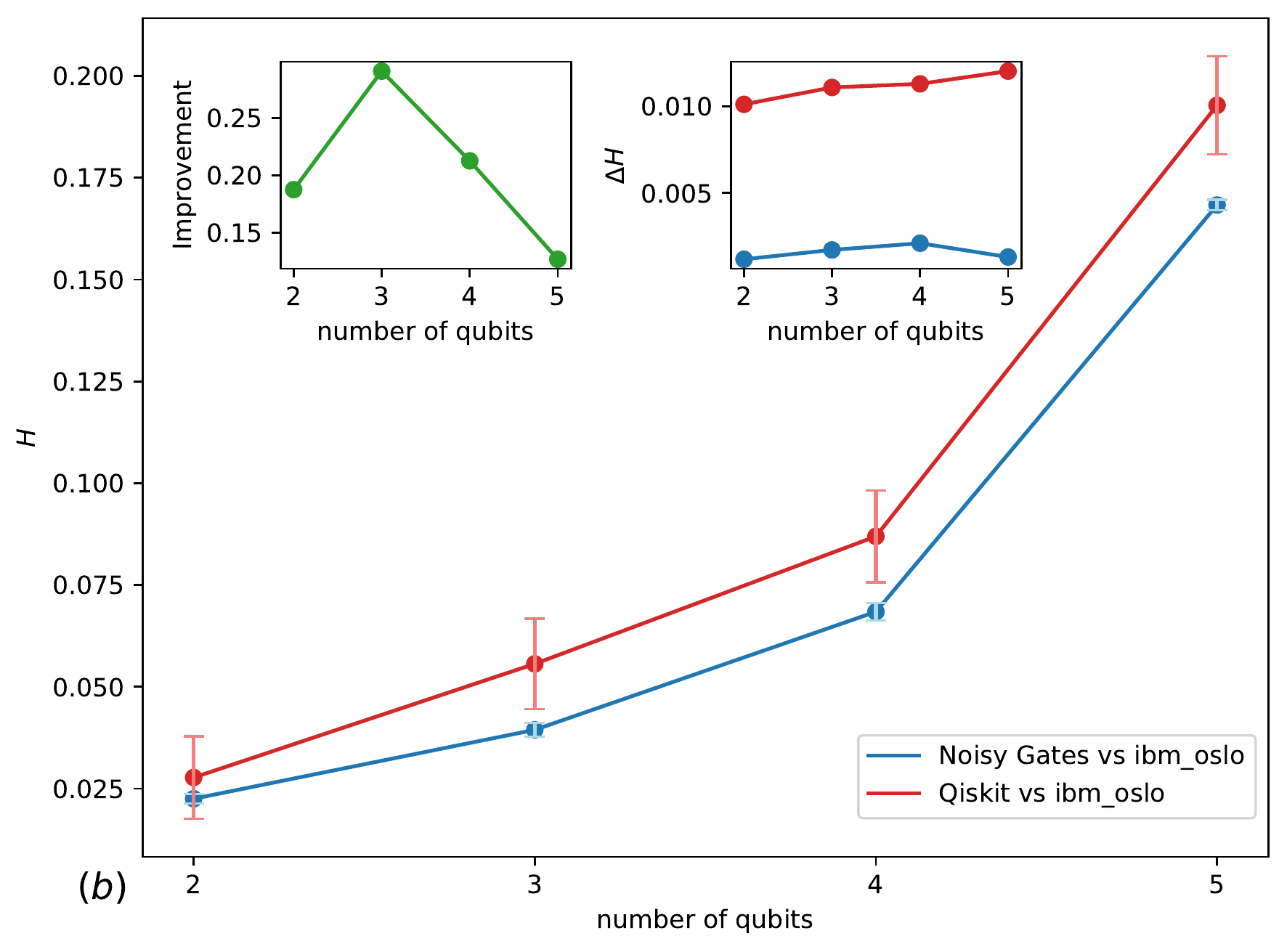}
    \end{minipage}
\caption{On panel (a), probabilities histograms for 4 qubits of a single independent simulation of the GHZ algorithm. In orange the results for $\text{ibmq\_oslo}$, in blue for the noisy gates and in red for the Qiskit simulator. On panel (b), Hellinger distance for the GHZ algorithm for $n = 2,\dots, 5$ qubits. Each value is the mean of $100$ independent simulations for the noisy gates, in blue, and for the Qiskit simulations, in red. The left inset shows the relative improvement, calculated as $|\Bar{\cl{H}}_{\sigma}^{\text{ibm}}-\Bar{\cl{H}}_{\sigma}^{\text{ng}}|/\Bar{\cl{H}}_{\sigma}{\text{ibm}}$, while the right inset shows the standard deviations as functions of the number of qubits.}
\label{hellinger_GHZ_plot}
\end{figure*}

Here in Fig \ref{fidelities_X} and Fig. \ref{CR_plot} we show the plots of the fidelities obtained from the simulations in Sec. \ref{simulations}. The fidelity is defined as:
\begin{equation}
    \cl{F}(\rho,\sigma)=\big(\rr{Tr}\sqrt{\sigma^{1/2}\rho\sigma^{1/2}}\big)^2\, .
\end{equation}
We notice that when one considers only diagonal density matrices, the fidelity is called Hellinger fidelity and it is related to the Hellinger distance as $\cl{F} = (1-\cl{H}^2)^2$. The Hellinger fidelity is not a proper mathematical distance, thus in the main text we used the Hellinger distance. As one can see the results are consistent with those in Sec. \ref{simulations}.

\section{GHZ simulations}\label{GHZ}
In this appendix we report the results of the analysis of the GHZ algorithm in order to inspect the performances of the noisy gates approach when trying to reproduce the behaviour of a real quantum computer.
We run GHZ for n = 2, ... , 5 on $\text{ibmq\_oslo}$ and we set as input the state $\ket{0}^{\otimes n}$. Runs on real quantum computer are performed by taking 1000
shots, i.e. measurements. We also run the corresponding classical simulations. We use the same custom noise model defined in Sec. \ref{simulations} in the $QFT^{\dagger}$ case. The resulting probability histograms for 4 qubits of a single independent simulation is reported in Fig. \ref{hellinger_GHZ_plot}.
We notice that, as for the QFT$^{\dagger}$ case, for every $n$ we get $\Bar{\cl{H}}^{\text{ng}} < \Bar{\cl{H}}^{\text{ibm}}$ and $\Delta\cl{H}^{\text{ng}} < \Delta\cl{H}^{\text{ibm}}$, see Fig. \ref{hellinger_GHZ_plot}.

\section{Comparison between relevant quantum computing frameworks on noisy simulations}\label{relevant_QC_frame}
In the following we report a table with a list of relevant quantum computing frameworks where we verify whether they support noise simulation (NS) and if so wheter they implement the approach descibed in Eq. \eqref{QuantumMaps} of Sec.
\ref{introduction} that we call standard approach (SA).

\begin{table}[htp]\label{table}
\vspace{5mm}
\centering
\begin{tabular}{|c|c|c|c|c|}
\hline
$\textbf{Company}$ & $\textbf{Name}$ & $\textbf{Ref.}$ & $\textbf{NS}$ & $\textbf{SA}$ \\
\hline 
IBM & Qiskit & \cite{anis2021qiskit} & Yes & Yes\\
\hline
Rigetti & pyQuil & \cite{rigetti_noise} & Yes & Yes\\
\hline
Quantinuum & t$\ket{ket}$ & \cite{sivarajah2020t} & Yes & Yes\\
\hline
Xanadu & Pennylane & \cite{bergholm2018pennylane} & Yes & Yes\\
\hline
Xanadu & Strawberry Field & \cite{killoran2019strawberry} & Yes & Yes\\
\hline
Microsoft & Azure Quantum & \cite{azurequantum} & No & -\\
\hline
Microsoft & LIQUI$\ket{}$ & \cite{wecker2014liqui} & Yes & Yes\\
\hline
Google & Cirq & \cite{cirq_noise} & Yes & Yes\\
\hline
Google & TensorFlow Quantum & \cite{broughton2020tensorflow} & Yes & Yes\\
\hline
Intel & Intel QS & \cite{guerreschi2020intel} & Yes & Yes\\
\hline
Baidu & Paddle Quantum & \cite{Paddlequantum_noise} & Yes & Yes\\
\hline
Amazon & Braket & \cite{amazon_braket} & Yes & Yes\\
\hline
- & ProjectQ & \cite{steiger2018projectq} & No & -\\
\hline
- & QiBO & \cite{efthymiou2021qibo} & Yes & Yes\\
\hline
- & QCL & \cite{omer2005classical} & No & -\\
\hline
- & Quipper & \cite{green2013quipper} & No & -\\
\hline
- & Quirk & \cite{qirk} & No & -\\
\hline
- & SilQ & \cite{bichsel2020silq} & No & -\\
\hline
\end{tabular}
\caption{List of relevant quantum computing frameworks. In the fourth column is specified whether the corresponding framework support noise simulation (NS) while in the fifth column we specify whether the noise simulation is based on the approach described in Eq. \eqref{QuantumMaps} of Sec.
\ref{introduction} that we call standard approach (SA).}
\end{table}

\newpage
\bibliography{biblio}

\end{document}